\documentclass[preprint,11pt]{elsarticle}
\usepackage[top=2cm, bottom=2cm, left=1.5cm, right=1.5cm]{geometry}

\usepackage{amssymb}
\usepackage{float}
\DeclareUnicodeCharacter{0394}{$\Delta$}
\usepackage{bigdelim}
\usepackage{fontawesome5}
\usepackage[ruled,vlined,linesnumbered]{algorithm2e}
\usepackage{amsmath}
\usepackage{subcaption}
\usepackage{graphicx}
\usepackage{tabularx}
\usepackage{array}
\usepackage{booktabs}
\usepackage{xcolor}
\usepackage{float}
\usepackage{tabularx}
\usepackage{pgfplots}
\pgfplotsset{compat=1.18}
\usepackage{multirow}
\usepackage{amsmath}
\usepackage{tikz}
\usetikzlibrary{matrix, calc, positioning, arrows.meta, shadows, fit, backgrounds,shapes.geometric, decorations.pathmorphing}
\usetikzlibrary{shadows.blur}
\usepackage{hyperref}

\journal{Name of Target Journal}

\begin{document}

\begin{frontmatter}

\title{A Topology-Aware Quantum-Classical Framework for Multi-Target Active Debris Removal using Gaussian Boson Sampling}

\author{Komal Pandya\corref{cor1}}
\ead{komalpandya@zohomail.in}

\author{Saiyam Sakhuja}
\ead{sakhujasaiyam@gmail.com}

\author{Nagaraj Kattaian}
\ead{nagaiaf@gmail.com}

\cortext[cor1]{Corresponding author}

\address{Independent Researchers}
\begin{abstract}

The continued growth of orbital debris poses a significant challenge to the
safety, sustainability, and operational reliability of near-Earth space
activities, making Active Debris Removal (ADR) an important component of
long-term debris mitigation. Selecting a servicing subset of size $k$ from a
candidate population of $N$ objects introduces a combinatorial search space of
size $\binom{N}{k}$. Directly coupling target selection with detailed mission
sequencing can therefore impose a substantial computational burden as the
candidate population grows.

To address this challenge, we present a hybrid quantum-classical ADR
framework that integrates orbital mechanics, graph theory, Gaussian Boson
Sampling (GBS), and multi-objective evolutionary optimization. Candidate
debris objects are represented as vertices of a weighted orbital-transfer
network, with edge weights constructed from transfer cost, orbital-plane
relationships, and estimated phasing requirements. Seven target-prioritization
strategies are considered to represent physical consequence, orbital
persistence, conjunction risk, transfer-network importance, orbital community
structure, inferred source-family information, and random selection. The
resulting weighted graphs are supplied to a GBS-based subset-generation stage
that samples candidate target sets according to the structural properties of
the encoded transfer network, avoiding exhaustive enumeration of the
$\binom{N}{k}$ candidate subsets.

The sampled subsets are subsequently evaluated using a PyGMO-based
multi-objective evolutionary framework to approximate the Pareto front of
servicing sequences that jointly minimize mission $\Delta v$ and operational
duration, subject to propulsion, timing, mass, and mission constraints. The
framework is evaluated across three benchmark environments, namely SOCRATES
Minefield, Collision Cloud, and Random, and across candidate-pool sizes
$N\in\{15,18,22,25\}$ and target cardinalities $k\in\{3,5\}$. The experimental
results characterize how benchmark construction, target-prioritization
strategy, and combinatorial scale affect GBS subset diversity, transfer-network
structure, and downstream mission performance. The resulting framework
provides a topology-aware mechanism for generating candidate ADR target
subsets before detailed trajectory optimization and illustrates a hybrid
quantum-classical approach to decomposing large combinatorial orbital-planning
problems.

\end{abstract}

\end{frontmatter}
\tableofcontents
\newpage
\section{Introduction}
\label{introsec}

\begin{figure}[htbp]
    \centering
    \includegraphics[width=1\linewidth]{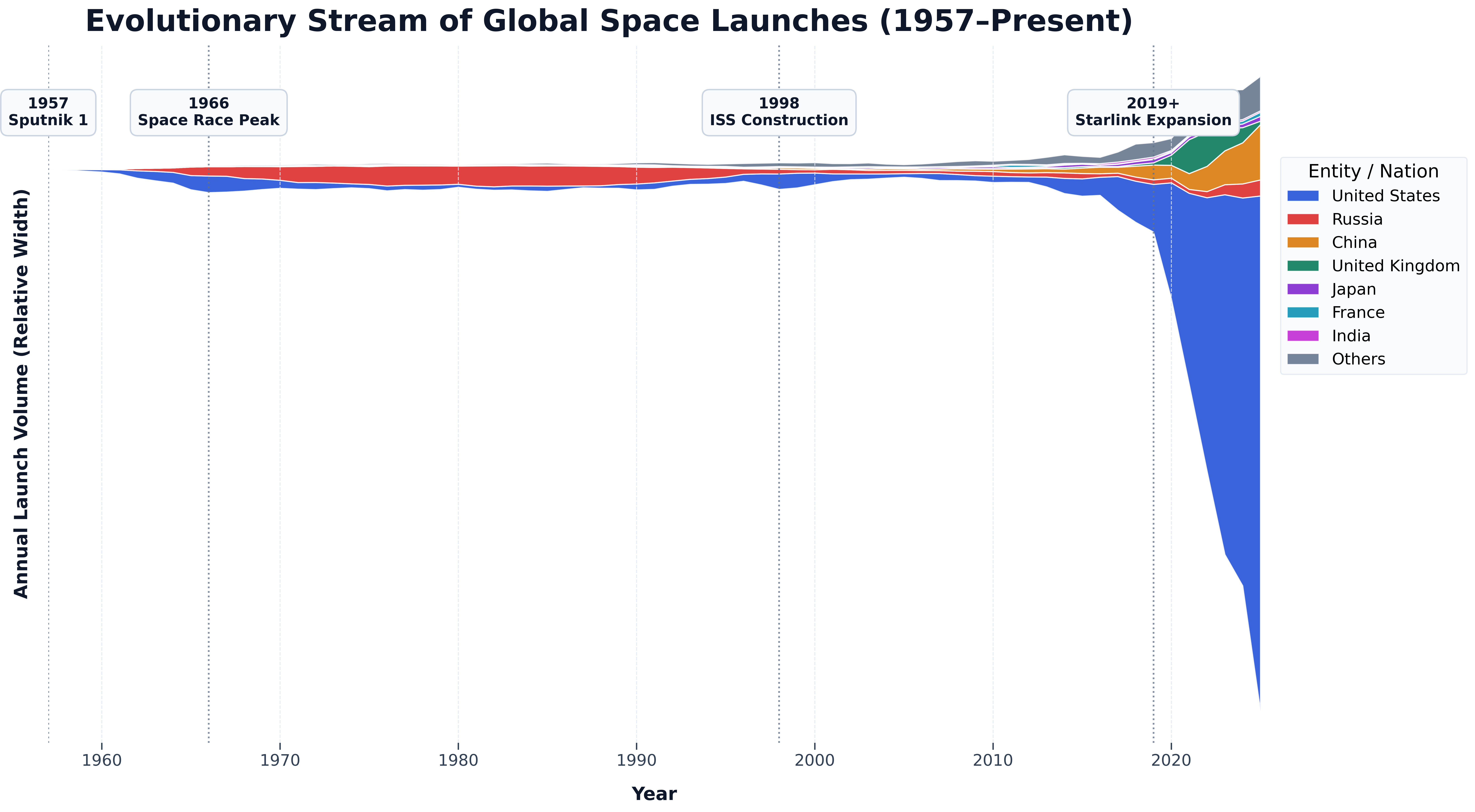}
    \caption{Historical evolution of annual space launch volume across the top
    seven spacefaring nations and aggregated other entities (1957-present).}
    \label{fig:launch_volume_top_countries_1957_present}
\end{figure}

The near-Earth space environment has evolved from a sparsely populated
operational domain into a densely congested and increasingly contested
ecosystem \footnote{\url{https://ourworldindata.org/grapher/yearly-number-of-objects-launched-into-outer-space}}. Since the launch of the first artificial satellite in 1957,
more than six decades of global space launch activity
(Figure~\ref{fig:launch_volume_top_countries_1957_present}), combined with
accidental explosions, on-orbit collisions, anti-satellite (ASAT) tests, and
routine spacecraft operations, have generated a large population of resident
space objects (RSOs) distributed across Low Earth Orbit (LEO), Medium Earth
Orbit (MEO), and Geostationary Earth Orbit (GEO)
\cite{belward_who_2015}.

While only a fraction of these objects remain operational, hundreds of
thousands of debris fragments larger than one centimeter and millions of
smaller particles continue to orbit the Earth at velocities exceeding
$7~\text{km/s}$. At such hypervelocity regimes, even small debris particles
can pose serious risks to spacecraft, while larger derelict satellites and
rocket bodies can retain sufficient kinetic energy to generate extensive
fragmentation following a collision
\cite{haridim_challenges_2026,shustov_space_2024,chen_space_2011}.\\

\begin{figure}[htbp]
    \centering
    \includegraphics[width=1\linewidth]{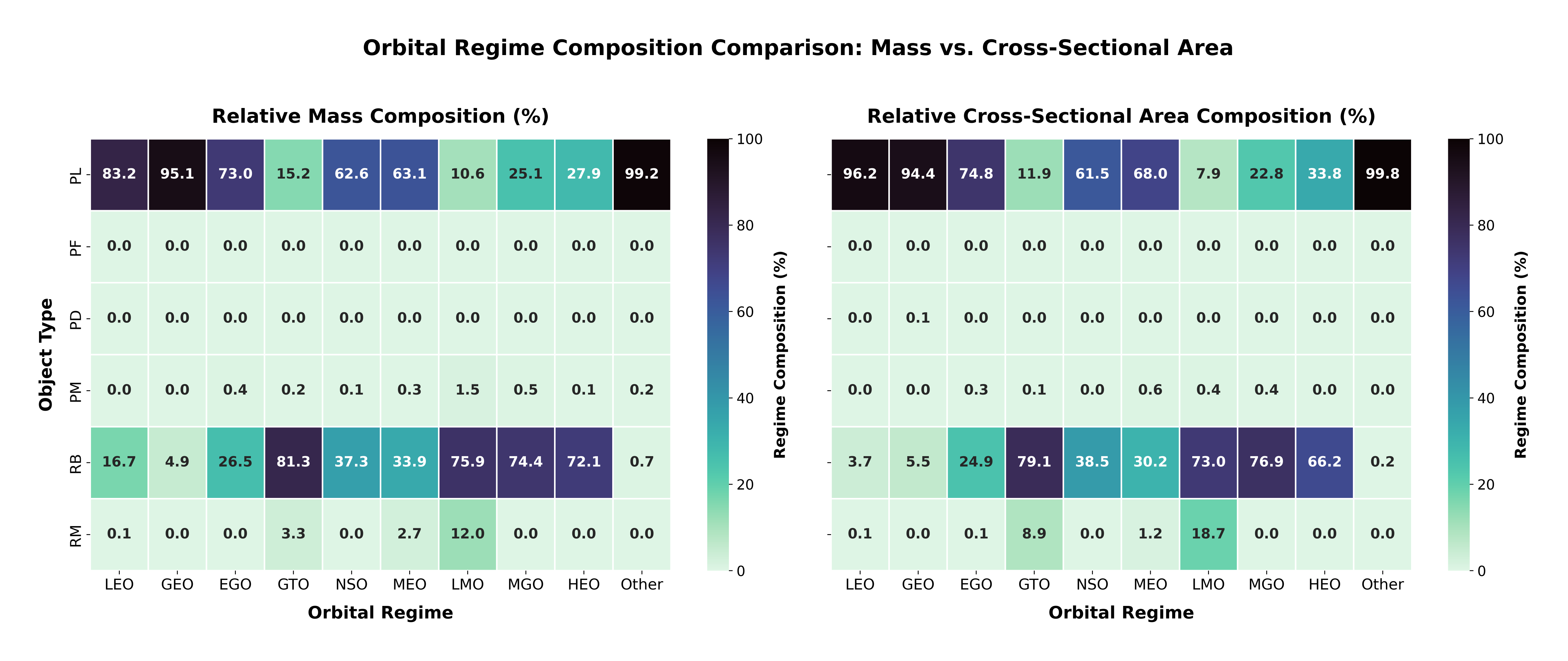}
    \caption{Side-by-side matrix heatmaps depicting relative mass composition
    (\%) and relative cross-sectional area composition (\%) across orbital
    regimes by object category.}
    \label{fig:mass_area_proportional_heatmap}
\end{figure}

A granular inspection of the spatial distribution and physical characteristics
of these objects yields three critical operational observations:

\begin{itemize}
    \item \textbf{Regime-Specific Compositional Fingerprints:}
    As mapped in Figure~\ref{fig:orbital_composition_comparison}, object
    distributions differ across orbital regimes. While LEO contains large
    numbers of active payloads and fragmentation debris, higher disposal and
    transfer regimes exhibit substantial populations of rocket bodies and
    unidentified fragments\footnote{\url{https://sdup.esoc.esa.int/discosweb/statistics/}}.

    \begin{figure}[htbp]
        \centering
        \begin{subfigure}[b]{0.38\textwidth}
            \centering
            \includegraphics[width=\textwidth]{chord_diagram_large_labels_big.png}
            \caption{Relational mapping between orbital regimes and object classes.}
            \label{fig:chord_diagram}
        \end{subfigure}
        \hfill
        \begin{subfigure}[b]{0.58\textwidth}
            \centering
            \includegraphics[width=\textwidth,height=8cm]{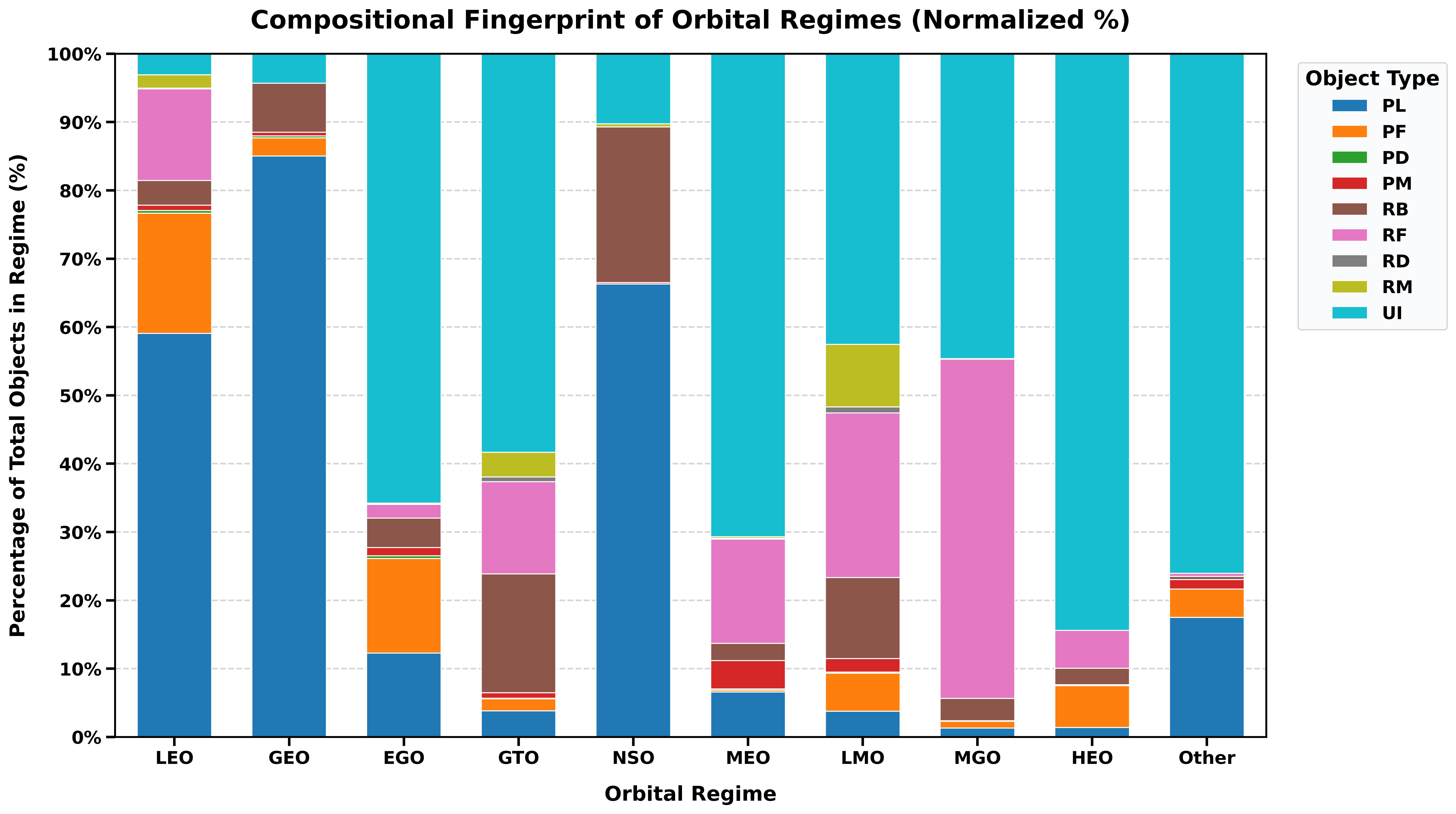}
            \caption{Proportional breakdown of object categories per orbital regime.}
            \label{fig:compositional_barchart}
        \end{subfigure}
        \caption{Comparative profile of space object distributions across
        orbital regimes, illustrating (a) cross-regime object interconnections
        and (b) regime-wise percentage compositions.}
        \label{fig:orbital_composition_comparison}
    \end{figure}

    \item \textbf{Mass and Area Concentration:}
    The comparative matrix heatmaps in Figure~\ref{fig:mass_area_proportional_heatmap}
    show that, despite the numerical dominance of small debris, large intact
    structures, particularly payloads (PL) and rocket bodies (RB), account for
    a substantial fraction of the cumulative orbital mass and cross-sectional
    area across the examined regimes.

    \item \textbf{Divergent Hazard Dynamics:}
    This structural imbalance creates distinct risk regimes. High object
    densities in LEO can increase conjunction activity, whereas substantial
    populations of massive objects in higher-altitude regimes can represent
    important long-term fragmentation hazards.
\end{itemize}

This increasing physical and spatial density has elevated the importance of
collision-risk management and orbital sustainability for civil, commercial,
and defense space missions
\cite{bradley_space_2009,olivieri_large_2020,filho_managing_2025}. Managing this structurally complex and spatially dense debris population
requires moving beyond passive observation and collision avoidance toward
physical remediation. Space Domain Awareness (SDA) systems provide essential
information for tracking objects and supporting conjunction assessment and
collision-avoidance operations, but these functions do not themselves remove
the underlying debris population. Similarly, post-mission disposal and other
debris-mitigation measures primarily address future debris generation and do
not directly remove legacy hazards accumulated over previous decades.
Long-term orbital sustainability requires complementary
approaches capable of physically reducing selected portions of the existing
debris population.\\

Among the mitigation strategies proposed to address this challenge, Active
Debris Removal (ADR) has emerged as an important component of long-term debris
mitigation
\cite{navaz_space_2026,suseela_space_2024,mark_review_2019}. Unlike passive
mitigation measures such as post-mission disposal and end-of-life deorbiting,
ADR seeks to actively remove existing objects before they participate in
future collision events. The importance of active intervention has been
emphasized by long-term debris-environment studies showing that collision
driven fragmentation can sustain or increase debris populations even under
substantially reduced launch activity
\cite{johnson_origin_2014,schafer_inter-agency_2005}. Selective
removal of particularly consequential or operationally important objects has
become an important research direction for long-term orbital remediation
\cite{buchs_collision_2021}.\\

Over the past decade, substantial research has been devoted to the development
of ADR technologies. Proposed capture and removal concepts include robotic
manipulators, harpoons, nets, electrodynamic tethers, drag augmentation
devices, ion-beam shepherds, laser momentum transfer systems, and autonomous
rendezvous-and-capture spacecraft
\cite{sizov_space_2021,ledkov_review_2022,zeng_concurrent_2024,
kaushik_active_nodate,bayin_review_2026,bigdeli_mechanics_2025,
fallahiarezoodar_review_2025,gao_dynamic_2025,arshad_emerging_2025}.
These concepts have progressed from theoretical investigations to mission
demonstrations involving rendezvous, inspection, servicing, and debris-removal
technologies. International agencies including ESA
\cite{biesbroek_e_2017,biesbroek_clearspace-1_2021,woicke_adrios_2025,
el-shawa_cradle_2021}, NASA
\cite{sullivan_securing_2023,liou_active_2011,colvin_cost_2023}, ISRO
\cite{anil_kumar_space_2025,sharma_space_2024}, and JAXA
\cite{kawamoto_current_2017,nakamura_study_2023,yamamoto_advancing_2025},
together with commercial operators, continue to investigate scalable servicing
and ADR architectures
\cite{forshaw_active_2020,taylor_remove_2018}. These developments reflect
the growing role of active servicing and removal in future orbital-sustainability
strategies.\\

Despite advances in capture technologies and servicing-spacecraft design, the
computational problem underlying multi-target ADR remains challenging. Given
a catalogue containing potentially large numbers of debris objects, mission
planners must determine which subset should be removed, how the selected
objects should be sequenced, and how the resulting mission can satisfy
propulsion, orbital-transfer, timing, launch-window, and mission-duration
requirements
\cite{wei_scheduling_2022,shen_simple_2021,nishida_space_2009}. These decisions
are coupled because the operational value of a target depends not only on its
individual characteristics but also on its transfer accessibility relative to
other selected targets and on the resulting mission architecture. ADR
planning therefore combines combinatorial target selection, orbital-transfer
modeling, network structure, environmental characterization, and
multi-objective mission design.\\

The design of ADR missions has traditionally been formulated as a trajectory
optimization problem in which a servicing spacecraft must rendezvous with and
remove a sequence of debris objects while minimizing propellant consumption,
mission duration, or other operational costs
\cite{medioni_trajectory_2023}. When represented as a graph whose vertices
correspond to debris objects and whose edge weights represent transfer
relationships, the resulting sequencing problem can be related to Traveling
Salesperson and selective-routing formulations
\cite{zhang_timeline_2022}. In general ADR formulations, servicing order can
affect the cumulative mission cost because orbital geometry and phasing evolve
with time. In the present framework, the baseline pairwise transfer matrices
used for graph construction are symmetric, while direction-dependent effects
are introduced during ordered route evaluation through time-dependent orbital
phasing, secular $J_2$ drift, waiting requirements, and the corresponding
active-transfer calculations. As a result, a large body of ADR research has
considered ATSP, selective ATSP, and generalized vehicle-routing formulations
for multi-target servicing
\cite{xia_task_2024,zhao_novel_2020,britto_enhanced_2021,li_optimization_2020}.\\

However, before trajectory or route optimization can be performed, a more
fundamental decision must be made: which debris objects should be serviced
in the first place. A servicing spacecraft has finite propellant, mission
duration, and operational resources, implying that only a limited subset of
the orbital population can be considered within a single mission campaign.
Let
\[
V=\{v_1,v_2,\ldots,v_N\}
\]
denote the candidate debris catalogue and let
\[
S\subseteq V,\qquad |S|=k,
\]
denote the subset selected for servicing, where typically $k\ll N$. The
selected subset subsequently forms the candidate target set for downstream
mission planning and route evaluation.\\

When target selection and route sequencing are considered jointly, the
combinatorial search space grows rapidly. The total number of ordered
target sequences is
\[
\binom{N}{k}k!
=
\frac{N!}{(N-k)!},
\]
where $\binom{N}{k}$ accounts for target-subset selection and $k!$ accounts
for the possible servicing orders. If an exact routing method based on
Held-Karp dynamic programming is then applied separately to every candidate
subset, the corresponding computational burden can be represented
schematically as
\[
\mathcal{O}
\left(
\binom{N}{k}k^2 2^k
\right).
\]
The dominant source of complexity is therefore the target-selection stage,
which grows combinatorially with $N$ even before detailed route evaluation
is performed. For example, selecting $k=10$ targets from $N=1000$ candidates
produces
\[
\binom{1000}{10}
\approx
2.63\times10^{23}
\]
possible target subsets before servicing order is considered. Such growth
makes exhaustive exploration impractical for large candidate populations.\\

The difficulty is further amplified because target utility is not solely an
intrinsic property of individual objects. Rather, mission utility emerges
from interactions among selected targets. The suitability of servicing a
particular object may depend on transfer accessibility relative to other
selected targets, environmental consequence, persistence, conjunction-risk
information, source-family relationships, and the resulting mission
architecture. A general target-selection formulation can therefore be written
as
\[
S^{*}
=
\arg\max_{S\subseteq V,\;|S|=k}
f(S),
\]
where $f(S)$ denotes a task-specific utility function incorporating relevant
environmental, astrodynamic, and mission-level criteria. The associated
selection-routing problem contains NP-hard combinatorial subproblems, and
exact methods can become computationally prohibitive as the candidate
population increases.\\

To address this challenge, a variety of classical optimization approaches
have been investigated in the ADR literature. Exact formulations based on
mixed-integer programming can become computationally expensive with
increasing problem size, motivating heuristic and metaheuristic methods such
as beam search, branch-and-bound pruning, ant-colony optimization, simulated
annealing, evolutionary algorithms, genetic algorithms, multi-objective
optimization, and learning-based mission planners
\cite{zhang_timeline_2022,medioni_trajectory_2023,xia_task_2024,
viavattene_design_2022,yang_application_2019,zhou_improved_2025}. These
methods can provide useful approximate solutions, but their computational
requirements remain influenced by the size of the candidate subset space and
the cost of downstream mission evaluation.\\

Recent work in quantum optimization has motivated alternative formulations
based on Quadratic Unconstrained Binary Optimization (QUBO) and related
Ising-model representations. ADR target selection and routing have been
studied using quantum annealing, gate-based variational algorithms, and
hybrid quantum-classical workflows
\cite{torta_quantum_2025,dubey_satellite_2023,swain_optimisation_2023,
gagliardi_quantum_2025,gagliardi_space_2025}. These approaches provide
alternative mechanisms for exploring combinatorial decision spaces, although
the underlying problem structure remains combinatorial and practical
performance continues to depend on encoding, hardware resources, and
algorithmic configuration.\\

In this work, we adopt a different decomposition of the ADR planning problem.
Rather than solving target selection and downstream mission optimization as a
single monolithic search, we explicitly separate candidate-subset generation
from mission-level route optimization. The central observation is that the
target-selection stage constitutes a large combinatorial component of the
overall planning problem. When debris objects are represented as vertices and
transfer accessibility is represented through weighted edges, candidate
selection can be formulated as a graph-structured sampling problem in which
the objective is to identify subsets associated with useful combinations of
target characteristics and transfer accessibility.\\

This observation motivates the use of Gaussian Boson Sampling (GBS), a
photonic quantum-computing paradigm that can sample combinatorial structures
encoded in weighted graphs without explicitly enumerating all
$\binom{N}{k}$ candidate subsets
\cite{hamilton_gaussian_2017,arrazola_quantum_2018,deshpande_quantum_2022,
noauthor_photonic_2019}. Given an adjacency matrix $A$ describing the encoded
orbital-transfer network, collision-free GBS probabilities are governed by
Hafnian-based quantities associated with induced subgraphs
\cite{gu_quantum_2019,bjorklund_faster_2019}. The weighted
graph structure influences which combinations of vertices are sampled. In
particular, subsets containing strongly weighted pairwise connections can
receive enhanced probability under the encoded graph, making GBS suitable as
a candidate-subset generation mechanism for topology-aware search
\cite{arrazola_quantum_2018,feige_dense_2001}.

Instead of explicitly enumerating
\[
\binom{N}{k}
\]
possible target subsets, the proposed framework generates a bounded collection
of candidate subsets
\[
M \ll \binom{N}{k},
\]
which are then forwarded to the downstream mission-optimization stage. Here,
$M$ denotes the number of retained candidate subsets rather than a new
theoretical complexity bound. In the implemented framework, the number of
retained subsets is explicitly capped, limiting the number of
candidate target sets requiring subsequent route evaluation.\\

The resulting decomposition can therefore be represented conceptually as
\[
\underbrace{\binom{N}{k}}_{\text{subset-selection space}}
\;\longrightarrow\;
\underbrace{M}_{\text{sampled candidate subsets}}
\]
followed by downstream route optimization for each retained candidate.
This does not eliminate the underlying combinatorial structure, but it reduces
the number of candidate subsets passed to the computationally more expensive
mission-evaluation stage. The approach therefore acts as a sampling-based
candidate-generation mechanism rather than as an exact replacement for the
full combinatorial optimization problem. \\ 

\begin{table*}[htbp]
\centering
\caption{Comparison of computational paradigms and mathematical complexity for multi-target Active Debris Removal planning.}
\label{tab:adr_complexity_grouped}
\renewcommand{\arraystretch}{1.08}
\setlength{\tabcolsep}{3pt}
\footnotesize
\begin{tabularx}{\textwidth}{@{} 
  >{\raggedright\arraybackslash}p{0.13\textwidth} 
  >{\raggedright\arraybackslash}p{0.16\textwidth} 
  >{\raggedright\arraybackslash}p{0.16\textwidth} 
  >{\raggedright\arraybackslash}X 
  >{\raggedright\arraybackslash}p{0.17\textwidth} 
@{}}
\toprule
\textbf{Method}
& \textbf{Optimization Scope}
& \textbf{Search Space}
& \textbf{Mathematical Complexity}
& \textbf{Primary Limitation}
\\
\midrule

\multicolumn{5}{@{}l}{\textbf{Classical Exact and Deterministic Formulations}} \\

Exhaustive ADR
& Target selection + sequencing
& $\displaystyle\binom{N}{k}k! = \frac{N!}{(N-k)!}$ ordered missions
& $\displaystyle\mathcal{O}\!\left(k\frac{N!}{(N-k)!}\right)$ for explicit enumeration and evaluation of all ordered missions
& Factorial growth with $N$
\\

ATSP / TSP
& Sequencing of a fixed target set
& $N!$ possible tours
& $\displaystyle\mathcal{O}(N^2 2^N)$ using Held-Karp dynamic programming
& Requires a predefined target set
\\

Selective ATSP
& Target selection + sequencing
& $\displaystyle\binom{N}{k}k!$ ordered candidate missions
& $\displaystyle\mathcal{O}\!\left(\binom{N}{k}k^2 2^k\right)$ when exact Held-Karp routing is applied to every selected subset
& Combinatorial growth in target selection
\\

Orienteering Problem
& Budget-constrained target selection + routing
& Up to $2^N$ candidate subsets
& NP-hard; exact formulations have exponential worst-case growth
& Problem-size dependent scalability
\\

MILP / MINLP
& Integrated discrete-continuous optimization
& Discrete-continuous combinatorial space
& NP-hard in general; worst-case solution effort can grow exponentially
& High computational cost for large formulations
\\

\midrule
\multicolumn{5}{@{}l}{\textbf{Approximate and Metaheuristic Methods}} \\

Beam / A* Search
& Pruned graph/tree search
& Branching search tree
& $\displaystyle\mathcal{O}(b d)$ for bounded beam width $b$ over search depth $d$; worst-case growth is exponential without effective pruning
& Heuristic-dependent search behavior
\\

Ant Colony Optimization
& Population-based routing search
& Permutation / graph search space
& $\displaystyle\mathcal{O}(I A k)$ under $\mathcal{O}(k)$ route evaluation per ant, where $I$ is the number of iterations and $A$ is the number of ants
& Parameter and iteration dependent
\\

Genetic Algorithms
& Evolutionary sequencing
& $k!$ possible permutations
& $\displaystyle\mathcal{O}(G P k)$ assuming $\mathcal{O}(k)$ fitness evaluation, where $G$ is the number of generations and $P$ is the population size
& Population and parameter dependent
\\

NSGA-II
& Multi-objective evolutionary optimization
& Population of candidate routes
& $\displaystyle\mathcal{O}\!\left(G\left(mP^2 + Pk\log k\right)\right)$ for $m$ objectives, including non-dominated sorting and random-key route decoding
& Population-size and fitness-evaluation dependent
\\

Reinforcement Learning
& Policy optimization
& State-action space
& Problem dependent; for $E_{\mathrm{tr}}$ training episodes and $T$ decision steps, a generic form is $\mathcal{O}(E_{\mathrm{tr}}T C_{\pi})$
& Training and representation dependent
\\

\midrule
\multicolumn{5}{@{}l}{\textbf{Quantum and Hybrid Architectures}} \\

QUBO / QAOA
& Binary optimization
& $2^N$ binary configurations
& Exact enumeration: $\mathcal{O}(2^N)$. Actual QAOA runtime depends on circuit depth, number of shots, and classical parameter optimization
& Qubit, circuit-depth, and encoding limitations
\\

Quantum Annealing
& Ising / QUBO optimization
& $2^N$ spin configurations
& Exact state-space enumeration: $\mathcal{O}(2^N)$. Physical annealing time is hardware- and problem-dependent
& Embedding and hardware constraints
\\

GBS Subset Discovery
& Candidate-subset generation
& $\displaystyle\binom{N}{k}$ possible subsets, without explicit enumeration
& $\mathcal{O}(N^3 + SN)$ for the dominant implemented preprocessing and sample-processing operations, where $S$ is the number of sampled threshold outcomes; the $N^3$ term arises from the adjacency eigendecomposition
& Sampling yield and simulator / hardware dependence
\\

\textbf{GBS + PyGMO (This Work)}
& Candidate-subset generation + multi-objective routing
& $\displaystyle M \ll \binom{N}{k}$ retained candidate subsets
& $\mathcal{O}(N^3 + SN) + \mathcal{O}\!\left(EMG\left(mP^2 + Pk\log k\right)\right)$
& Depends on graph construction, sampling, and evolutionary fitness evaluation
\\

\bottomrule
\end{tabularx}

\vspace{5pt}
\begin{minipage}{\textwidth}
\scriptsize
\noindent
Here, $N$ denotes the candidate-pool size, $k$ the number of serviced
targets, $M$ the number of retained candidate subsets, $S$ the number of
GBS threshold-sampling outcomes, $E$ the number of evaluated launch epochs,
$G$ the number of evolutionary generations, $P$ the evolutionary population
size, and $m$ the number of objectives. The complexity expression for the
proposed GBS-PyGMO framework represents the dominant computational
operations of the implemented workflow and is not intended as a claim of
asymptotic optimality or minimum complexity relative to all alternative ADR
methods.
\end{minipage}
\end{table*}
To implement this decomposition, we develop a hybrid quantum-classical ADR
planning framework that integrates orbital mechanics, graph theory, GBS, and
multi-objective evolutionary optimization into a unified pipeline. The
architecture comprises two principal stages:

\begin{enumerate}

    \item \textbf{Astrodynamic Network Construction:}
    Target RSOs are reconstructed from TLE data using SGP4 and represented as
    vertices of an orbital-transfer graph. Weighted edges encode transfer
    accessibility through the adopted transfer-cost model, orbital-plane
    relationships, secular $J_2$ evolution, phasing opportunities, and
    estimated $\Delta v$ requirements.

    \item \textbf{Multi-Criteria Target Weighting:}
    Six prioritization heuristics are used to modify the graph presented to the
    GBS stage. These heuristics represent physical consequence, orbital
    persistence, conjunction risk, transfer-network centrality, orbital
    community structure, and inferred source-family information. A random
    subset-selection procedure is additionally used as an experimental
    baseline.
\end{enumerate}

The resulting weighted orbital-transfer graph encodes both astrodynamic
accessibility and heuristic-specific target information. The graph is then
supplied to the GBS stage, where the weighted structure influences the
distribution of sampled target combinations through Hafnian-based sampling
dynamics. Candidate subsets are retained only when they satisfy the prescribed
target cardinality, after which they are forwarded to the downstream mission
optimization stage.\\

Each retained candidate subset is subsequently evaluated using a
PyGMO-based multi-objective evolutionary framework. The route representation
encodes servicing order, while candidate launch epochs are evaluated over a
predefined mission-design window. Mission objectives jointly minimize total
mission $\Delta v$ and operational duration subject to the modeled propulsion,
mass, timing, transfer, and mission constraints. The resulting non-dominated
solutions are used to characterize the approximate Pareto front of the
candidate mission architectures. Table~\ref{tab:adr_complexity_grouped} synthesizes the mathematical formulations, scalability constraints, and algorithmic trade-offs across candidate computational paradigms for multi-target ADR planning.\\

\begin{figure}[!h]
    \centering
    \includegraphics[width=1\linewidth]{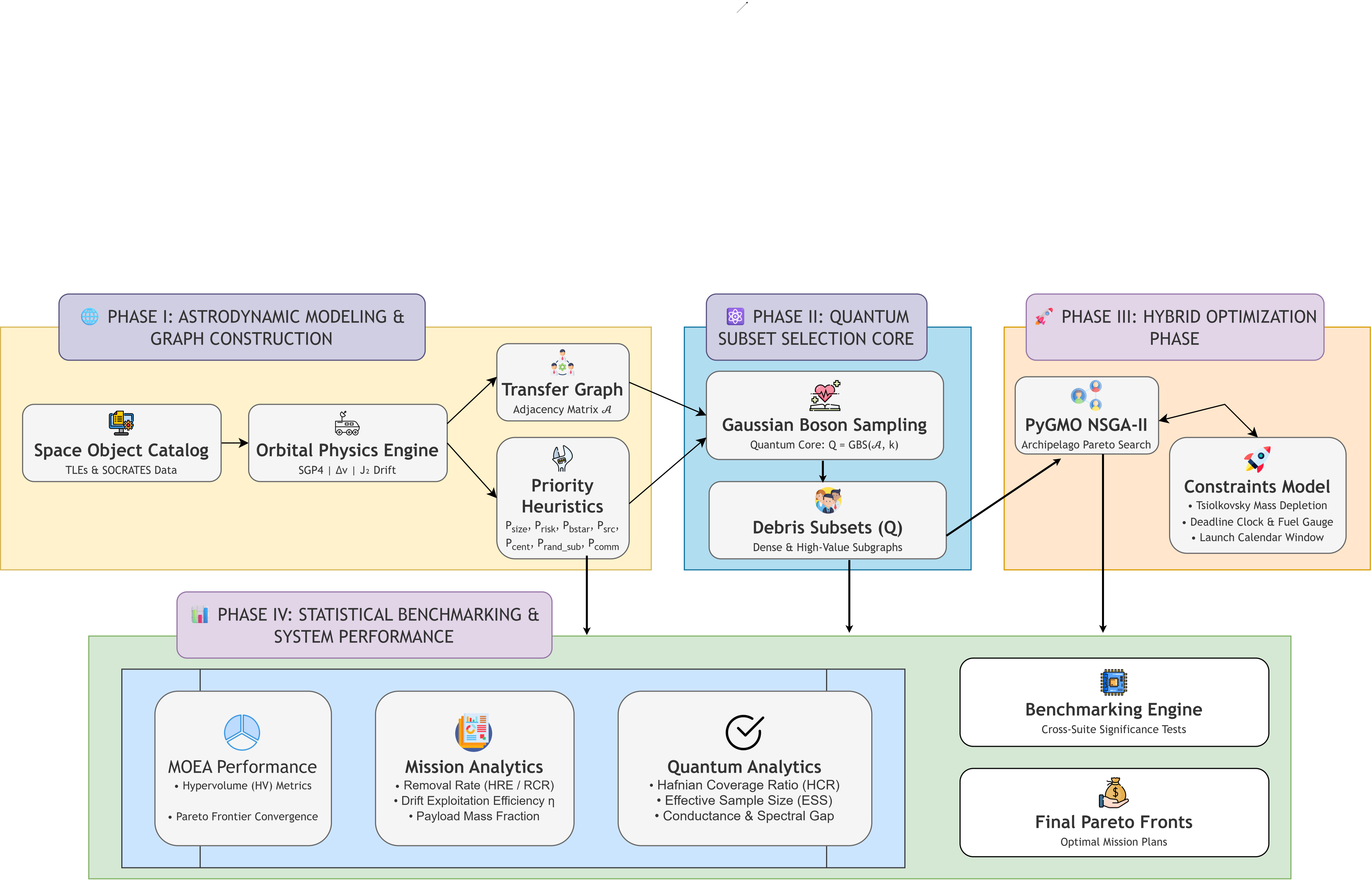}
    \caption{Overview of the experimental evaluation framework.}
    \label{fig:experimental_design}
\end{figure}

The framework is evaluated using three benchmark environments-SOCRATES Minefield, Collision Cloud, and Random-together with candidate-pool sizes $N\in\{15,18,22,25\}$ and target cardinalities $k\in\{3,5\}$. Figure~\ref{fig:experimental_design} depicts a high-level overview of the proposed architecture. This experimental design enables the influence of benchmark construction, target-prioritization strategy, and combinatorial scale to be examined within a common computational framework.

\subsection{Our Contributions}

This work introduces a topology-aware hybrid quantum-classical framework for
ADR candidate selection and mission planning. The principal contributions are
summarized as follows:

\begin{enumerate}

\item \textbf{A Topology-Aware Formulation of Active Debris Removal.}

We formulate ADR target selection using a weighted orbital-transfer network
in which vertices represent debris objects and edge weights encode
astrodynamic transfer accessibility through orbital geometry, phasing
opportunities, transfer cost, and related network structure. This formulation
provides a network-level representation in which target characteristics and
inter-target accessibility can jointly influence candidate-subset generation.

\item \textbf{A Hybrid GBS-PyGMO Architecture for ADR Mission Planning.}

We propose a hybrid quantum-classical framework that combines GBS-based
candidate-subset generation with multi-objective evolutionary mission
optimization. GBS generates candidate target subsets from weighted
orbital-transfer graphs, while PyGMO evaluates and refines the associated
servicing sequences. This decomposition separates combinatorial candidate
generation from downstream mission optimization and limits the number of
candidate subsets requiring detailed evaluation.

\item \textbf{Astrodynamically Informed Quantum Sampling.}

Unlike GBS applications in which the input graph is purely abstract, the
proposed framework constructs the sampling graph from orbital-transfer
information. Transfer accessibility, orbital geometry, estimated phasing
requirements, and target-prioritization criteria jointly influence the
weighted adjacency matrix supplied to the quantum sampler. The resulting
samples therefore reflect the encoded structure of the orbital environment.

\item \textbf{A Systematic Benchmarking Framework for Quantum-Assisted ADR.}

To evaluate topology-aware target selection, we construct three benchmark
environments, SOCRATES Minefield, Collision Cloud, and Random, together with
six heuristic prioritization strategies and a random subset baseline. The
framework is evaluated across multiple candidate-pool sizes and target
cardinalities under common mission-optimization settings.

\item \textbf{A Multi-Level Characterization of GBS Sampling Behavior.}

Beyond downstream mission performance, we characterize the internal behavior
of the GBS stage using graph-theoretic, statistical, and subset-sampling
metrics. This enables analysis of the relationship between encoded orbital
network structure, empirical subset distributions, and downstream mission
characteristics.

\item \textbf{Empirical Scaling Analysis of ADR Optimization.}

The study examines how variation in candidate-pool size $N$ and target
cardinality $k$ affects graph structure, candidate-subset generation,
sampling behavior, and downstream mission optimization. The resulting
analysis provides empirical insight into how the combinatorial and
astrodynamic components of the planning problem respond to changes in problem
size.

\end{enumerate}
\section{Mathematical and Technical Preliminaries}

To establish the theoretical foundation for the proposed quantum-classical ADR framework, this section outlines the underlying principles of orbital transfer mechanics, the graph-theoretic formulation of the space environment, the quantum probability distribution of GBS, and the mechanics of multi-objective evolutionary routing.

\subsection{Astrodynamics and Orbital Transfer Mechanics}

The transition between two debris objects is characterized primarily by the
required velocity increment, $\Delta v$, and the corresponding transfer and
waiting time. In the adopted ADR model, orbital states are reconstructed from
Two-Line Element (TLE) data, pairwise transfer costs are estimated using a
Hohmann-type surrogate, and time-dependent orbital evolution is incorporated
through secular $J_2$ nodal drift and orbital phasing.

\subsubsection{Orbit Reconstruction}

The initial debris population is obtained from publicly available orbital
catalogues containing TLE sets. For each object, the TLE is propagated using
the Simplified General Perturbation 4 (SGP4) model to obtain the Cartesian state
vector
\begin{equation}
\mathbf{x}_i=
(\mathbf{r}_i,\mathbf{v}_i).
\end{equation}

The resulting state is converted into the corresponding orbital quantities,
including semi-major axis $a_i$, eccentricity $e_i$, inclination $i_i$, right
ascension of the ascending node $\Omega_i$, and mean anomaly $M_i$. These
quantities provide the orbital information required for pairwise transfer
estimation and time-dependent route evaluation.

\subsubsection{Impulsive Velocity Change}

For an impulsive maneuver, the velocity change is represented by the vector
difference between the pre- and post-maneuver velocity states:
\begin{equation}
\Delta\mathbf{v}
=
\mathbf{v}_2-\mathbf{v}_1.
\end{equation}

Its scalar magnitude is
\begin{equation}
\Delta v
=
\left\|
\Delta\mathbf{v}
\right\|
=
\sqrt{
v_1^2+v_2^2-2v_1v_2\cos\alpha
},
\end{equation}
where $\alpha$ is the angle between the corresponding velocity vectors.

In the adopted transfer model, this relation is used to incorporate the
inclination-dependent component of the arrival maneuver.

\subsubsection{Tsiolkovsky Rocket Equation}

Propellant consumption is evaluated using the Tsiolkovsky rocket equation.
For an initial spacecraft mass $m_0$, specific impulse $I_{sp}$, and a maneuver
with velocity increment $\Delta v$, the remaining mass is
\begin{equation}
m_f
=
m_0
\exp
\left(
-\frac{\Delta v}
{I_{sp}g_0}
\right),
\end{equation}
where
\[
g_0=9.80665~\mathrm{m\,s^{-2}}.
\]

The mass state is propagated sequentially across servicing legs. A candidate
mission is rejected if the propagated mass falls below the prescribed dry-mass
floor.

\subsubsection{Hohmann-Type Transfer Surrogate}

For two circular orbits with radii corresponding to semi-major axes $a_i$ and
$a_j$, the intermediate transfer radius is defined as
\begin{equation}
a_{t,ij}
=
\frac{a_i+a_j}{2}.
\end{equation}

The corresponding circular velocities are
\begin{equation}
v_i=\sqrt{\frac{\mu}{a_i}},
\qquad
v_j=\sqrt{\frac{\mu}{a_j}},
\end{equation}
and the transfer-orbit velocities at the two apses are
\begin{equation}
v_{t,i}
=
\sqrt{
\mu
\left(
\frac{2}{a_i}
-
\frac{1}{a_{t,ij}}
\right)
},
\end{equation}

\begin{equation}
v_{t,j}
=
\sqrt{
\mu
\left(
\frac{2}{a_j}
-
\frac{1}{a_{t,ij}}
\right)
}.
\end{equation}

The departure velocity increment is evaluated as
\begin{equation}
\Delta v_{1,ij}
=
\left|
v_{t,i}-v_i
\right|.
\end{equation}

The arrival burn incorporates the orbital-plane separation:
\begin{equation}
\Delta v_{2,ij}
=
\sqrt{
v_j^2+v_{t,j}^2
-
2v_jv_{t,j}\cos(\Delta i_{ij})
},
\end{equation}
where
\[
\Delta i_{ij}=|i_i-i_j|.
\]

The adopted passive pairwise transfer cost is therefore
\begin{equation}
\Delta v^{\mathrm{passive}}_{ij}
=
\Delta v_{1,ij}
+
\Delta v_{2,ij}.
\end{equation}

Pairs exceeding the prescribed inclination-difference limit are treated as
inaccessible in the pairwise transfer matrices.

The corresponding Hohmann-type time of flight is
\begin{equation}
T^{\mathrm{TOF}}_{ij}
=
\frac{\pi}{\sqrt{\mu}}
a_{t,ij}^{3/2},
\end{equation}
with the implementation expressing this quantity in days.

\subsubsection{Secular $J_2$ Evolution and Orbital Phasing}

The secular evolution of the right ascension of the ascending node is modeled
using the standard $J_2$ approximation:
\begin{equation}
\dot{\Omega}
=
-\frac{3}{2}
J_2
n
\left(
\frac{R_E}
{a(1-e^2)}
\right)^2
\cos i,
\end{equation}
where $n$ is the mean motion, $R_E$ is the Earth reference radius, and $J_2$
is the second zonal harmonic coefficient.

For a transfer from object $i$ to object $j$, the relative nodal drift rate is
\begin{equation}
\Delta\dot{\Omega}_{ij}
=
\dot{\Omega}_i-\dot{\Omega}_j.
\end{equation}

The instantaneous RAAN values are propagated from the reference epoch
according to
\begin{equation}
\Omega_i(t)
=
\left(
\Omega_{i,0}
+
\dot{\Omega}_i t
\right)
\bmod 2\pi.
\end{equation}

When $\Delta\dot{\Omega}_{ij}\neq0$, the adopted RAAN waiting time is computed
using the periodic relative-drift relation
\begin{equation}
t_{\mathrm{RAAN},ij}
=
\left[
\frac{
\Omega_j(t)-\Omega_i(t)
}{
\Delta\dot{\Omega}_{ij}
}
\right]
\bmod
\left|
\frac{2\pi}
{\Delta\dot{\Omega}_{ij}}
\right|,
\end{equation}
subject to the model's prescribed waiting-time cap.

Orbital phase evolution is represented through the propagated mean anomalies:
\begin{equation}
M_i(t)
=
\left(
M_{i,0}+n_i t
\right)
\bmod 2\pi.
\end{equation}

The relative mean-motion difference is
\begin{equation}
\Delta n_{ij}
=
n_i-n_j,
\end{equation}
and the corresponding phasing wait is computed from the relative phase and
subject to the adopted phasing cap.

The total waiting estimate used for graph construction is a weighted
combination of the nodal-drift and mean-anomaly phasing times:
\begin{equation}
W_{ij}
=
\alpha t_{\mathrm{RAAN},ij}
+
(1-\alpha)t_{\mathrm{phase},ij},
\end{equation}
with $\alpha=0.7$ in the implemented model.

For ordered mission evaluation, the dynamic leg calculation also evaluates an
active-transfer alternative. The instantaneous orbital-plane separation is
obtained from the propagated RAAN and inclination values:
\begin{equation}
\cos\gamma_{ij}
=
\cos i_i\cos i_j
+
\sin i_i\sin i_j
\cos\left(
\Omega_j(t)-\Omega_i(t)
\right),
\end{equation}
where $\gamma_{ij}$ denotes the relative plane angle. The corresponding active
arrival burn is then computed from the velocity-vector magnitude relation.

\subsection{Graph-Theoretic Representation of the Environment}

The ADR target-selection problem can be naturally represented as a graph, where orbital objects correspond to vertices and feasible transfers correspond to weighted directed connections. This representation transforms mission planning from a purely astrodynamic routing problem into a combinatorial optimization problem defined over a network structure.\\

Formally, the debris environment can be modeled as a directed weighted graph
$G=(V,E)$, where each vertex $v_i \in V$ represents a candidate debris object and each directed edge $e_{ij} \in E$ encodes the transfer accessibility from object $v_i$ to object $v_j$. The directionality of the graph is useful when the transfer cost depends on the direction of traversal. In particular, the transfer cost from $v_i$ to $v_j$ need not equal the reverse transfer cost, i.e.,
\begin{equation}
\Delta v_{ij}\neq\Delta v_{ji},
\end{equation}
owing to differences in orbital geometry, relative phasing, inclination, and secular orbital evolution.\\

The directed transfer network can be represented by an $N\times N$ affinity matrix $\mathcal{C}$, whose elements are obtained by transforming the transfer cost through an exponentially decaying kernel:
\begin{equation}
C_{ij} =
\begin{cases}
\exp\left(-\frac{\Delta v_{ij}}{\sigma}\right), & i \neq j,\\
0, & i=j,
\end{cases}
\end{equation}
where $\Delta v_{ij}$ denotes the estimated transfer cost from object $i$ to object $j$, and $\sigma$ is a scaling parameter controlling the decay of transfer affinity. Since the transfer cost is directional, the resulting affinity matrix is generally asymmetric,
\begin{equation}
\mathcal{C}\neq\mathcal{C}^{T}.
\end{equation}

Under this construction, low-cost transfers yield affinity values close to unity, whereas energetically expensive transfers are assigned weights approaching zero. Highly accessible debris collections appear as regions containing strong directed connections. To illustrate this representation, consider a simplified catalog of four objects where $v_1$, $v_2$, and $v_3$ belong to the same fragmentation cloud, while $v_4$ occupies a dynamically distant orbital region. The corresponding directed affinity matrix may be written as
\begin{equation}
\mathcal{C}=
\begin{bmatrix}
0    & 0.88 & 0.72 & 0.01 \\
0.82 & 0    & 0.94 & 0.06 \\
0.84 & 0.88 & 0    & 0.02 \\
0.03 & 0.04 & 0.00 & 0
\end{bmatrix}.
\end{equation}

The upper-left block contains strong bidirectional connectivity, although the individual transfer affinities are not identical in opposite directions. In contrast, the connections involving $v_4$ are weak, indicating that transfers involving this object are less strongly connected to the remaining objects under the illustrative affinity model. Thus, a directed weighted graph provides a natural representation when transfer accessibility or cost depends on the direction of traversal.\\

\subsection{Gaussian Boson Sampling (GBS)}

GBS is a specialized paradigm of continuous-variable photonic quantum computing designed to generate samples from probability distributions associated with matrix Hafnians. Unlike universal gate-based architectures requiring fault-tolerant qubits, GBS operates using squeezed states of light, linear optical interference, and photon-number measurements \cite{zhong_experimental_2019, kruse_detailed_2019, phillips_benchmarking_2019}. Owing to its natural connection to matrix Hafnians, GBS provides a natural framework for graph-based combinatorial sampling \cite{gupt_classical_2020}.\\

The practical execution of GBS relies on three primary physical stages:
\begin{itemize}
\item \textbf{State Preparation:} Nonlinear optical elements generate squeezed vacuum states through spontaneous parametric down-conversion (SPDC) or four-wave mixing.
\item \textbf{Interferometric Evolution:} Squeezed optical modes propagate through a programmable linear optical network, producing multimode quantum interference.
\item \textbf{Photon Detection:} The output modes are measured using photon-number-resolving or threshold detectors, depending on the measurement architecture. Photon-number-resolving detectors record the number of photons detected in each mode, whereas threshold detectors record only whether a mode contains at least one photon.
\end{itemize}

To represent a graph within the standard GBS formalism, the matrix describing the graph is taken to be symmetric. When the underlying physical or application-specific network is represented by a directed affinity matrix, a symmetric representation can be constructed, for example, by symmetrizing the pairwise connectivity:
\begin{equation}
\mathcal{A}
=
\frac{1}{2}
\left(
\mathcal{C}+\mathcal{C}^{T}
\right).
\end{equation}

This transformation provides an undirected weighted representation of the pairwise connectivity while retaining the aggregate relationship between two objects. The resulting matrix is suitable for the symmetric-matrix formulation commonly used in GBS-based graph encodings. Direction-dependent information may be retained separately when it is required by a subsequent routing or scheduling formulation.\\

For the illustrative directed affinity matrix above, the corresponding symmetric matrix is
\begin{equation}
\mathcal{A}=
\begin{bmatrix}
0    & 0.85 & 0.78 & 0.02 \\
0.85 & 0    & 0.91 & 0.05 \\
0.78 & 0.91 & 0    & 0.01 \\
0.02 & 0.05 & 0.01 & 0
\end{bmatrix}.
\end{equation}

The resulting symmetric matrix encodes the structural connectivity of the graph for GBS sampling. Dense regions of $\mathcal{A}$ correspond to groups of objects exhibiting strong mutual connectivity and can therefore receive greater sampling weight under the GBS distribution. \\

For a complex symmetric matrix, the standard matrix factorization used in the Gaussian-state formulation is the Autonne--Takagi decomposition:
\begin{equation}
\mathcal{A}=U\Sigma U^{T},
\end{equation}
where $U$ is a unitary matrix and $\Sigma=\operatorname{diag}(\sigma_1,\sigma_2,\dots,\sigma_N)$ contains the non-negative Takagi singular values of $\mathcal{A}$. The Takagi decomposition is distinct from the ordinary eigendecomposition of a real symmetric matrix. In particular, for a real symmetric matrix one may write the spectral decomposition as
\begin{equation}
\mathcal{A}=Q\Lambda Q^{T},
\end{equation}
where $Q$ is orthogonal and $\Lambda$ contains the real eigenvalues. The Takagi representation instead uses non-negative singular values and is the form naturally associated with the Gaussian-state optical implementation.

The singular values determine the squeezing parameters according to
\begin{equation}
r_i = \operatorname{arctanh}(c \sigma_i),
\end{equation}
where $c$ is a scaling factor chosen such that $c\sigma_i<1$ for all $i$. The scaling controls the magnitude of the squeezing and consequently the mean photon generation rate of the GBS system.\\

After propagation through the interferometric network, the quantum state is registered by the detector array. For photon-number-resolving detection, the probability of a photon-number configuration associated with a matrix substructure is proportional to a Hafnian-dependent quantity. In the collision-free case, where at most one photon is detected in each occupied mode, this takes the form
\begin{equation}
\Pr(S) \propto
\left|\operatorname{Haf}(\mathcal{A}_S)\right|^2,
\end{equation}
where $\mathcal{A}_S$ denotes the submatrix associated with the detected modes. More generally, for photon-number patterns with occupation numbers $n_1,n_2,\ldots,n_N$, the probability contains the corresponding factorial normalization terms,
\begin{equation}
\Pr(S) \propto
\frac{\left|\operatorname{Haf}(\mathcal{A}_S)\right|^2}
{n_1! n_2! \cdots n_N!}.
\end{equation}

For threshold detection, the detector output is instead a binary click pattern indicating the set of modes in which one or more photons are detected. The associated graph-sampling probabilities are consequently defined through the threshold-detection formulation of GBS rather than directly by the photon-number-resolving expression above. The distinction is important because threshold detectors discard the exact photon occupation number while retaining the information about which modes register a click.\\

The Hafnian of a symmetric matrix is defined as
\begin{equation}
\operatorname{Haf}(\mathcal{A}_S) =
\sum_{M \in \mathcal{M}} \prod_{(u,v) \in M} \mathcal{A}_{u,v},
\end{equation}
where $\mathcal{M}$ denotes the set of all perfect matchings of the graph induced by $S$. Intuitively, the Hafnian aggregates the products of edge weights over all possible perfect matchings of the selected subgraph. Because exact computation of Hafnians is \#P-hard, classical evaluation becomes computationally difficult as the matrix dimension increases \cite{grier_complexity_2022,olson_role_2018}.\\

In the context of ADR, the core challenge is selecting a subset of $k$ target objects from a candidate catalog of size $N$. The number of possible target combinations grows combinatorially:
\begin{equation}
\binom{N}{k},
\end{equation}
making exhaustive evaluation computationally prohibitive prior to trajectory optimization.\\

The graph-structured sampling property of GBS provides a mechanism for generating candidate subsets that reflect the connectivity encoded in the graph. When the orbital environment is mapped to the weighted symmetric matrix $\mathcal{A}$, subsets containing stronger collections of pairwise connections can produce larger Hafnian magnitudes and consequently higher sampling probabilities under the corresponding GBS distribution. This property motivates the use of GBS as a graph-based sampling mechanism for identifying structurally cohesive subsets prior to subsequent optimization.\\

To illustrate this sampling bias, consider two candidate target pairs from the symmetrized four-node graph:
\begin{equation}
S_1 = \{v_1, v_2\}, \qquad S_2 = \{v_3, v_4\}.
\end{equation}
The induced submatrices are:
\begin{equation}
\mathcal{A}_{S_1} = \begin{bmatrix} 0 & 0.85 \\ 0.85 & 0 \end{bmatrix}, \qquad
\mathcal{A}_{S_2} = \begin{bmatrix} 0 & 0.01 \\ 0.01 & 0 \end{bmatrix}.
\end{equation}

For a $2 \times 2$ matrix, the Hafnian reduces to the off-diagonal element ($\operatorname{Haf}(\mathcal{A}_{S_1}) = 0.85$ and $\operatorname{Haf}(\mathcal{A}_{S_2}) = 0.01$). Under the collision-free photon-number-resolving formulation, the relative probability contribution between these two subgraphs is
\begin{equation}
\frac{\Pr(S_1)}{\Pr(S_2)}
=
\frac{|0.85|^2}{|0.01|^2}
=
7225.
\end{equation}
Thus, in this illustrative example, the first pair has a sampling-probability ratio of $7225$ relative to the second pair. The example demonstrates how the quadratic dependence on the Hafnian magnitude can produce large differences in relative sampling probability for differently weighted subgraphs.\\

For the full four-node subgraph, the total Hafnian integrates all perfect-matching contributions:
\begin{equation}
\operatorname{Haf}(\mathcal{A})
=
(0.85)(0.01)
+
(0.78)(0.05)
+
(0.02)(0.91)
=
0.0657.
\end{equation}
The largest individual contribution arises from the matching $(1,3)$ and $(2,4)$, which contributes $0.0390$ to the total Hafnian. These matching contributions collectively determine the Hafnian associated with the selected subgraph. In GBS, the corresponding sampling probability is governed by these Hafnian-dependent contributions, such that graph structures containing strong weighted matchings can receive greater sampling weight. The photonic sampling process can therefore generate candidate subsets exhibiting strong internal connectivity, which may subsequently be evaluated for their suitability in application-specific optimization problems.\\
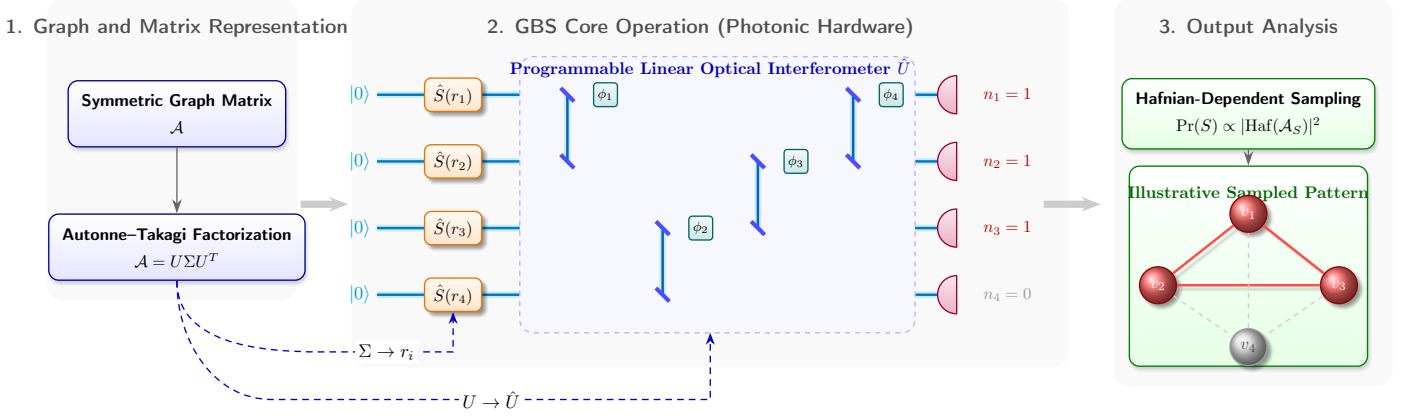
\begin{figure*}[t]
\centering
\resizebox{\textwidth}{!}{
\begin{tikzpicture}[
    font=\small\sffamily,
    arrow/.style={->, >={Stealth[length=3mm,width=2mm]}, thick, draw=black!60},
    control/.style={->, >={Stealth[length=2.5mm]}, thick, dash pattern=on 4pt off 3pt, draw=blue!80!black},
    transition/.style={->, >={Stealth[length=4mm, width=3mm]}, line width=1.5mm, draw=black!20},
    %
    block/.style={
        rectangle, draw, thick, rounded corners=6pt,
        align=center, inner sep=8pt, 
        blur shadow={shadow blur steps=5, shadow opacity=20, shadow xshift=1pt, shadow yshift=-2pt}
    },
    blueBlock/.style={block, draw=blue!60!black, top color=white, bottom color=blue!10, minimum width=4.0cm},
    greenBlock/.style={block, draw=green!50!black, top color=white, bottom color=green!10, minimum width=4.0cm},
    %
    waveguide/.style={line width=1.5pt, draw=cyan!60!blue, preaction={draw=cyan!40, opacity=0.4, line width=4pt}},
    squeezer/.style={rectangle, rounded corners=4pt, draw=orange!90!black, top color=orange!5, bottom color=orange!20, thick, minimum width=1.2cm, minimum height=0.7cm, align=center, blur shadow},
    pnr/.style={semicircle, draw=purple!80!black, top color=purple!10, bottom color=purple!30, thick, minimum width=0.8cm, shape border rotate=90},
    bs/.style={line width=2.5pt, draw=blue!70!white},
    ps/.style={rectangle, rounded corners=2pt, draw=teal!80!black, top color=teal!5, bottom color=teal!20, thick, minimum width=0.5cm, minimum height=0.5cm, inner sep=1pt, font=\scriptsize},
    %
    mathnode/.style={circle, thick, minimum size=22pt, inner sep=0pt, blur shadow},
    activeNode/.style={mathnode, draw=red!60!black, ball color=red!60},
    inactiveNode/.style={mathnode, draw=black!40, ball color=gray!30, text=black!70},
    zoneLabel/.style={font=\large\bfseries\sffamily, text=black!70}
]

\node[zoneLabel] (z1) at (0, 2.2) {1. Graph and Matrix Representation};

\node[blueBlock] (adj) at (0, 0.4) 
{\textbf{Symmetric Graph Matrix}\\[1.5mm] $\mathcal A$};

\node[blueBlock] (eig) at (0, -2.4) 
{\textbf{Autonne--Takagi Factorization}\\[1.5mm] $\mathcal A = U\Sigma U^{T}$};

\draw[arrow] (adj) -- (eig);

\draw[transition] (2.6, -1.5) -- (3.6, -1.5);

\node[zoneLabel] (z2) at (11.0, 2.2) {2. GBS Core Operation (Photonic Hardware)};

\def\yA{0.8} \def\yB{-0.6} \def\yC{-2.0} \def\yD{-3.4}

\foreach \y in {\yA, \yB, \yC, \yD} {
    \draw[waveguide] (4.2, \y) -- (16.2, \y);
}

\foreach \y in {\yA, \yB, \yC, \yD} {
    \node[left, font=\normalsize, text=cyan!80!black] at (4.2, \y) {$|0\rangle$};
}

\foreach \y/\sub in {\yA/1, \yB/2, \yC/3, \yD/4} {
    \node[squeezer] (sq\sub) at (5.8, \y) {};
    \node[] at (5.8, \y) {$\hat{S}(r_\sub)$};
}

\draw[fill=blue!3, draw=blue!30, dashed, thick, rounded corners=8pt] 
(7.2, 1.6) rectangle (15.5, -4.2);

\node[anchor=north, font=\small\bfseries, text=blue!70!black] 
at (11.2, 1.7) {Programmable Linear Optical Interferometer $\hat{U}$};


\draw[waveguide, draw=cyan!40!blue] (8.2, \yA) -- (8.2, \yB); 
\draw[bs, rotate around={-45:(8.2, \yA)}] (8.0, \yA) -- (8.4, \yA);
\draw[bs, rotate around={-45:(8.2, \yB)}] (8.0, \yB) -- (8.4, \yB);
\node[ps] at (9.0, \yA) {$\phi_1$};

\draw[waveguide, draw=cyan!40!blue] (10.2, \yC) -- (10.2, \yD); 
\draw[bs, rotate around={-45:(10.2, \yC)}] (10.0, \yC) -- (10.4, \yC);
\draw[bs, rotate around={-45:(10.2, \yD)}] (10.0, \yD) -- (10.4, \yD);
\node[ps] at (11.0, \yC) {$\phi_2$};

\draw[waveguide, draw=cyan!40!blue] (12.2, \yB) -- (12.2, \yC); 
\draw[bs, rotate around={-45:(12.2, \yB)}] (12.0, \yB) -- (12.4, \yB);
\draw[bs, rotate around={-45:(12.2, \yC)}] (12.0, \yC) -- (12.4, \yC);
\node[ps] at (13.0, \yB) {$\phi_3$};

\draw[waveguide, draw=cyan!40!blue] (14.2, \yA) -- (14.2, \yB); 
\draw[bs, rotate around={-45:(14.2, \yA)}] (14.0, \yA) -- (14.4, \yA);
\draw[bs, rotate around={-45:(14.2, \yB)}] (14.0, \yB) -- (14.4, \yB);
\node[ps] at (15.0, \yA) {$\phi_4$};

\foreach \y in {\yA, \yB, \yC, \yD} {
    \node[pnr] at (16.2, \y) {};
}

\node[right, font=\small\bfseries, text=red!70!black] (out1) at (16.8, \yA) {$n_1 = 1$};
\node[right, font=\small\bfseries, text=red!70!black] (out2) at (16.8, \yB) {$n_2 = 1$};
\node[right, font=\small\bfseries, text=red!70!black] (out3) at (16.8, \yC) {$n_3 = 1$};
\node[right, font=\small\bfseries, text=black!40]      (out4) at (16.8, \yD) {$n_4 = 0$};

\draw[control] 
(eig.south) to[out=270, in=180] (3.0, -4.6) 
-- node[fill=white, inner sep=2pt, font=\bfseries] {$\Sigma \to r_i$} 
(5.8, -4.6) -- (sq4.south);

\draw[control] 
(eig.south) to[out=270, in=180] (2.0, -5.6) 
-- node[fill=white, inner sep=2pt, font=\bfseries] {$U \to \hat{U}$} 
(11.2, -5.6) -- (11.2, -4.2);

\draw[transition] (18.2, -1.5) -- (19.4, -1.5);

\node[zoneLabel] (z3) at (22.5, 2.2) {3. Output Analysis};

\node[greenBlock] (haf) at (22.5, 0.4) 
{\textbf{Hafnian-Dependent Sampling}\\[1.5mm] 
$\Pr(S) \propto |\mathrm{Haf}(\mathcal A_S)|^2$};

\node[greenBlock, minimum height=4.2cm, minimum width=5.0cm] 
(subbox) at (22.5, -2.8) {};

\node[anchor=north, font=\small\bfseries, text=green!40!black] 
at (22.5, -1.0) {Illustrative Sampled Pattern};

\node[activeNode, text=white] (v1) at (22.5, -1.7) {$v_1$};
\node[activeNode, text=white] (v2) at (20.6, -3.2) {$v_2$};
\node[activeNode, text=white] (v3) at (24.4, -3.2) {$v_3$};
\node[inactiveNode]           (v4) at (22.5, -4.5) {$v_4$};

\draw[line width=1.5pt, draw=red!70, blur shadow={shadow blur steps=3, shadow opacity=30}] 
(v1) -- (v2);

\draw[line width=1.5pt, draw=red!70, blur shadow={shadow blur steps=3, shadow opacity=30}] 
(v1) -- (v3);

\draw[line width=1.5pt, draw=red!70, blur shadow={shadow blur steps=3, shadow opacity=30}] 
(v2) -- (v3);

\draw[dashed, thick, draw=black!20] (v1) -- (v4);
\draw[dashed, thick, draw=black!20] (v2) -- (v4);
\draw[dashed, thick, draw=black!20] (v3) -- (v4);

\draw[arrow] (haf.south) -- (subbox.north);

\begin{pgfonlayer}{background}

    \coordinate (bg1_tl) at (-2.6, 2.7); 
    \coordinate (bg1_br) at (2.4, -3.4);
    \node[fill=gray!5, rounded corners=10pt, fit=(bg1_tl)(bg1_br)] {};

    \coordinate (bg2_tl) at (3.8, 2.7);  
    \coordinate (bg2_br) at (18.6, -4.8);
    \node[fill=gray!5, rounded corners=10pt, fit=(bg2_tl)(bg2_br)] {};

    \coordinate (bg3_tl) at (19.8, 2.7); 
    \coordinate (bg3_br) at (25.4, -5.2);
    \node[fill=gray!5, rounded corners=10pt, fit=(bg3_tl)(bg3_br)] {};

\end{pgfonlayer}

\end{tikzpicture}
} 

\caption{
Conceptual workflow of Gaussian Boson Sampling for a weighted graph. 
The symmetric graph matrix is factorized using the Autonne-Takagi decomposition, 
with the singular values determining the squeezing parameters and the unitary factor 
defining the linear optical transformation. Photon-number-resolving (PNR) detection 
produces photon-count patterns whose probabilities are governed by Hafnian-dependent 
expressions. The illustrative four-node graph continues the numerical example 
introduced in the text and shows one possible sampled pattern.
}
\label{fig:gbs_pure_integrated}
\end{figure*}

Figure~\ref{fig:gbs_pure_integrated} summarizes the conceptual GBS architecture corresponding to the matrix representation introduced above. The graph matrix is first transformed into the form required for the Gaussian optical realization. Its matrix decomposition determines the squeezing parameters and the interferometric transformation, after which photon detection generates graph-structured samples. The numerical example shown in the figure continues the illustrative matrix construction and demonstrates how a strongly connected subset can emerge from the sampled graph representation.

\subsection{Multi-Objective Evolutionary Optimization}

ADR mission planning naturally constitutes a multi-objective optimization problem. A servicing spacecraft must simultaneously minimize propellant expenditure while minimizing mission duration, two objectives that are fundamentally conflicting due to the underlying mechanics of orbital transfers. The optimization process seeks not a single optimal solution, but a family of trade-off solutions representing different mission strategies.\\

Formally, a multi-objective optimization problem (MOP) may be written as:
\begin{equation}
\min_{\mathbf{x}\in\Omega} \mathbf{f}(\mathbf{x}) = \left( f_1(\mathbf{x}), f_2(\mathbf{x}), \dots, f_M(\mathbf{x}) \right),
\end{equation}
where $\mathbf{x}$ denotes a decision vector within the feasible search space $\Omega$, and $M$ represents the number of objectives.\\

Because the objectives are generally conflicting, optimality is defined through Pareto dominance. A solution $\mathbf{x}$ is said to dominate another solution $\mathbf{y}$, denoted $\mathbf{x} \prec \mathbf{y}$, if:
\begin{equation}
\forall i,\; f_i(\mathbf{x}) \leq f_i(\mathbf{y}) \quad\land\quad \exists j,\; f_j(\mathbf{x}) < f_j(\mathbf{y}).
\end{equation}
The collection of all non-dominated solutions forms the Pareto-optimal set $\mathcal{P}^{*}$, while its image in objective space constitutes the Pareto front.

\subsubsection{ADR as a Bi-Objective Optimization Problem}

For ADR mission design, the decision vector $\mathbf{x}$ represents the servicing order through a continuous random-key representation. The optimization problem is formulated using two primary objectives:
\begin{equation}
f_1(\mathbf{x}) = \Delta v_{\mathrm{tot}}
\end{equation}
and
\begin{equation}
f_2(\mathbf{x}) = T_{\mathrm{tot}},
\end{equation}
where $\Delta v_{\mathrm{tot}}$ denotes the cumulative mission velocity increment and $T_{\mathrm{tot}}$ denotes the total mission duration.\\

The conflict between these objectives arises directly from orbital mechanics. Orbital-plane alignment may be achieved through active propulsive maneuvers, reducing transfer time while increasing fuel expenditure. Alternatively, secular $J_2$ perturbations may be exploited to passively align orbital planes through differential nodal drift, substantially reducing $\Delta v$ requirements while increasing mission duration. As a result, ADR planning naturally produces a continuum of Pareto-optimal trade-offs ranging from rapid, fuel-intensive interventions to slower but highly fuel-efficient servicing campaigns.\\

For a selected subset containing $k$ debris objects, the implementation uses
a continuous random-key decision vector
\begin{equation}
\mathbf{x}
=
(x_1,x_2,\ldots,x_k),
\qquad
x_i\in[0,1].
\end{equation}

The servicing permutation is obtained through rank-order decoding:
\begin{equation}
\pi
=
\operatorname{argsort}(\mathbf{x}).
\end{equation}

Thus, the evolutionary optimizer searches a continuous hypercube while the
ordering of the selected debris objects is determined by the relative values
of the random keys. Transfer costs, orbital phasing, waiting time, and
mass evolution are subsequently computed inside the mission fitness
evaluation rather than being represented as independent decision variables.\\

For a decoded permutation
\[
\pi=(\pi_1,\ldots,\pi_k),
\]
the route is evaluated sequentially across the $k-1$ inter-target legs.

At each leg, the dynamic transfer model returns the passive transfer cost, the
corresponding waiting time, and an active-transfer alternative:
\begin{equation}
\left(
\Delta v^{\mathrm{passive}}_{ab},
W_{ab},
\Delta v^{\mathrm{active}}_{ab}
\right).
\end{equation}

The passive arrival time is first evaluated. When a target is associated with a
conjunction-assessment deadline and passive arrival occurs after that
deadline, the active-transfer alternative is evaluated instead. If the active
alternative also arrives late, a lateness penalty is incorporated into the
objective evaluation.

For each leg,
\begin{equation}
T_{\mathrm{leg}}
=
W_{ab}
+
T^{\mathrm{TOF}}_{ab}
+
T_{\mathrm{service}},
\end{equation}
where $T_{\mathrm{service}}$ is the adopted service-duration allowance.

The cumulative mission duration is therefore
\begin{equation}
T_{\mathrm{tot}}
=
\sum_{\ell=1}^{k-1}
T_{\mathrm{leg},\ell},
\end{equation}
while the cumulative velocity expenditure is
\begin{equation}
\Delta v_{\mathrm{tot}}
=
\sum_{\ell=1}^{k-1}
\Delta v_{\mathrm{leg},\ell}.
\end{equation}

After every maneuver, the spacecraft mass is updated using the rocket
equation. Candidate solutions that violate the modeled leg-level transfer
limit or fall below the dry-mass floor are assigned a large infeasibility
penalty and are excluded from subsequent mission analysis.

\subsubsection{PyGMO and the Archipelago Paradigm}

The ADR search space is highly non-linear, discontinuous, and multimodal. Multiple target orderings, transfer opportunities, and phasing strategies can generate numerous local optima. Conventional gradient-based optimization methods are therefore poorly suited to this problem.\\

To efficiently explore this complex design space, the framework employs the Parallel Global Multiobjective Optimizer (PyGMO), an evolutionary optimization platform that implements a distributed island-model architecture known as the \emph{Archipelago} \cite{biscani_parallel_2020}. Rather than evolving a single population, PyGMO evolves multiple populations concurrently and exchanges solutions among them according to the specified migration topology. Mathematically, the Archipelago may be represented as a directed graph:
\begin{equation}
\mathcal{G}_{\mathrm{arch}} = (\mathcal{I}, \mathcal{E}_{\mathrm{mig}}),
\end{equation}
where $\mathcal{I}=\{I_1, I_2, \ldots, I_K\}$ denotes a collection of optimization islands and $\mathcal{E}_{\mathrm{mig}}$ defines the migration topology connecting them. Each island maintains an independent population:
\begin{equation}
P_k(t) \subset \Omega,
\end{equation}
which evolves according to an assigned evolutionary operator $\mathcal{O}_k$. During an evolutionary epoch $\tau$, the population updates as:
\begin{equation}
P_k(t+\tau) = \mathcal{O}_k \left( P_k(t) \right).
\end{equation}
At suitable intervals during archipelago evolution, solutions may be exchanged between connected islands according to the specified migration topology. The heterogeneous island structure is intended to combine complementary search behaviors while allowing information exchange between populations.

\subsubsection{NSGA-II for Pareto Front Exploration}

\begin{figure*}[t]
\centering
\resizebox{0.85\textwidth}{!}{
\begin{tikzpicture}[
    font=\small\sffamily,
    population/.style={rectangle, draw=blue!70!black, top color=white, bottom color=blue!10, thick, rounded corners=4pt, inner sep=6pt, align=center, blur shadow={shadow blur steps=5, shadow opacity=20}},
    frontBlock/.style={rectangle, draw=red!60!black, top color=white, bottom color=red!10, thick, rounded corners=3pt, minimum width=2.2cm, minimum height=0.8cm, align=center, drop shadow={opacity=0.15}},
    operatorBox/.style={rectangle, draw=orange!80!black, fill=orange!5, thick, rounded corners=6pt, inner sep=8pt, align=center, blur shadow={shadow blur steps=3, shadow opacity=20}},
    main_arrow/.style={->, >={Stealth[length=3mm, width=2mm]}, thick, draw=black!70},
    sort_arrow/.style={->, >={Stealth[length=2.5mm]}, thick, draw=blue!70!black},
    reject_arrow/.style={->, >={Stealth[length=2.5mm]}, thick, dashed, draw=red!70},
    mathLabel/.style={font=\small\bfseries, text=black!80}
]

\node[population, minimum width=3cm, minimum height=1cm] (Pt) at (-2.5, 4) {\textbf{Parent Population}\\[1mm] $P_t$ (Size: $P$)};
\node[population, minimum width=3cm, minimum height=1cm] (Qt) at (2.5, 4) {\textbf{Offspring Population}\\[1mm] $Q_t$ (Size: $P$)};

\node[population, minimum width=6cm, minimum height=1.2cm] (Rt) at (0, 2) {\textbf{Combined Population}\\[1mm] $R_t = P_t \cup Q_t$ (Size: $2P$)};

\draw[main_arrow] (Pt.south) -- (-1.5, 2.6);
\draw[main_arrow] (Qt.south) -- (1.5, 2.6);

\node[operatorBox, minimum width=5cm] (SortBox) at (0, 0) {\textbf{Fast Non-Dominated Sorting}\\Assigns dominance ranks ($\mathcal{F}_k$)};
\draw[main_arrow] (Rt.south) -- (SortBox.north);

\node[frontBlock] (F1) at (-4.5, -2) {Front $\mathcal{F}_1$};
\node[frontBlock] (F2) at (-1.5, -2) {Front $\mathcal{F}_2$};
\node[frontBlock] (F3) at (1.5, -2) {Front $\mathcal{F}_3$};
\node[frontBlock, draw=gray, bottom color=gray!10] (F4) at (4.5, -2) {Front $\mathcal{F}_4 \dots \mathcal{F}_k$};

\draw[sort_arrow] (SortBox.south) to[out=270, in=90] (F1.north);
\draw[sort_arrow] (SortBox.south) -- (F2.north);
\draw[sort_arrow] (SortBox.south) -- (F3.north);
\draw[sort_arrow] (SortBox.south) to[out=270, in=90] (F4.north);

\node[population, minimum width=7cm, minimum height=1.5cm, label={[font=\bfseries, text=black!70]below:New Parent Population $P_{t+1}$ (Size: $P$)}] (Pt1) at (-1.5, -5.5) {};

\node[frontBlock, minimum width=2cm, minimum height=1.2cm, fill=red!15] (fill1) at (-3.5, -5.5) {$\mathcal{F}_1$};
\node[frontBlock, minimum width=2cm, minimum height=1.2cm, fill=red!15] (fill2) at (-1.2, -5.5) {$\mathcal{F}_2$};
\node[frontBlock, minimum width=1.5cm, minimum height=1.2cm, fill=red!5, draw=orange, dashed, thick] (fill3) at (0.9, -5.5) {$\mathcal{F}_3$};

\draw[main_arrow] (F1.south) -- (fill1.north);
\draw[main_arrow] (F2.south) -- (fill2.north);

\node[operatorBox, draw=orange!90!black, minimum width=3cm] (CrowdBox) at (1.5, -3.8) {\textbf{Crowding Distance} ($d_i$)\\Sort to fill remaining slots};

\draw[sort_arrow] (F3.south) -- (CrowdBox.north);
\draw[main_arrow] (CrowdBox.south) to[out=270, in=90] (fill3.north);

\node[font=\small\bfseries, text=red!70] (Reject) at (4.5, -4.5) {Discarded};
\draw[reject_arrow] (CrowdBox.east) to[out=0, in=90] (Reject.north);
\draw[reject_arrow] (F4.south) -- (Reject.north);

\node[operatorBox, minimum width=4cm, minimum height=1cm] (EvoBox) at (-6.5, -0.5) {\textbf{Evolutionary Operators}\\Tournament Selection $\prec_n$\\SBX Crossover \& PM};

\draw[main_arrow, rounded corners=10pt] (Pt1.west) -- (-6.5, -5.5) -- (EvoBox.south);

\draw[main_arrow, rounded corners=10pt] (EvoBox.north) -- (-6.5, 4) -- node[above, font=\bfseries] {Generates $Q_{t+1}$} (Pt.west);

\end{tikzpicture}
} 

\caption{Execution architecture of the NSGA-II algorithm deployed within the Archipelago framework. The combined population $R_t$ undergoes fast non-dominated sorting to establish Pareto fronts. Elitism is enforced by retaining complete higher-ranked fronts ($\mathcal{F}_1,\mathcal{F}_2$) in the subsequent generation $P_{t+1}$. For the marginal front ($\mathcal{F}_3$) that exceeds the population limit $P$, the crowding distance operator ($d_i$) is calculated, systematically preserving individuals located in less densely populated regions of the objective space while discarding redundant clustered solutions.}
\label{fig:nsga2_flow}
\end{figure*}
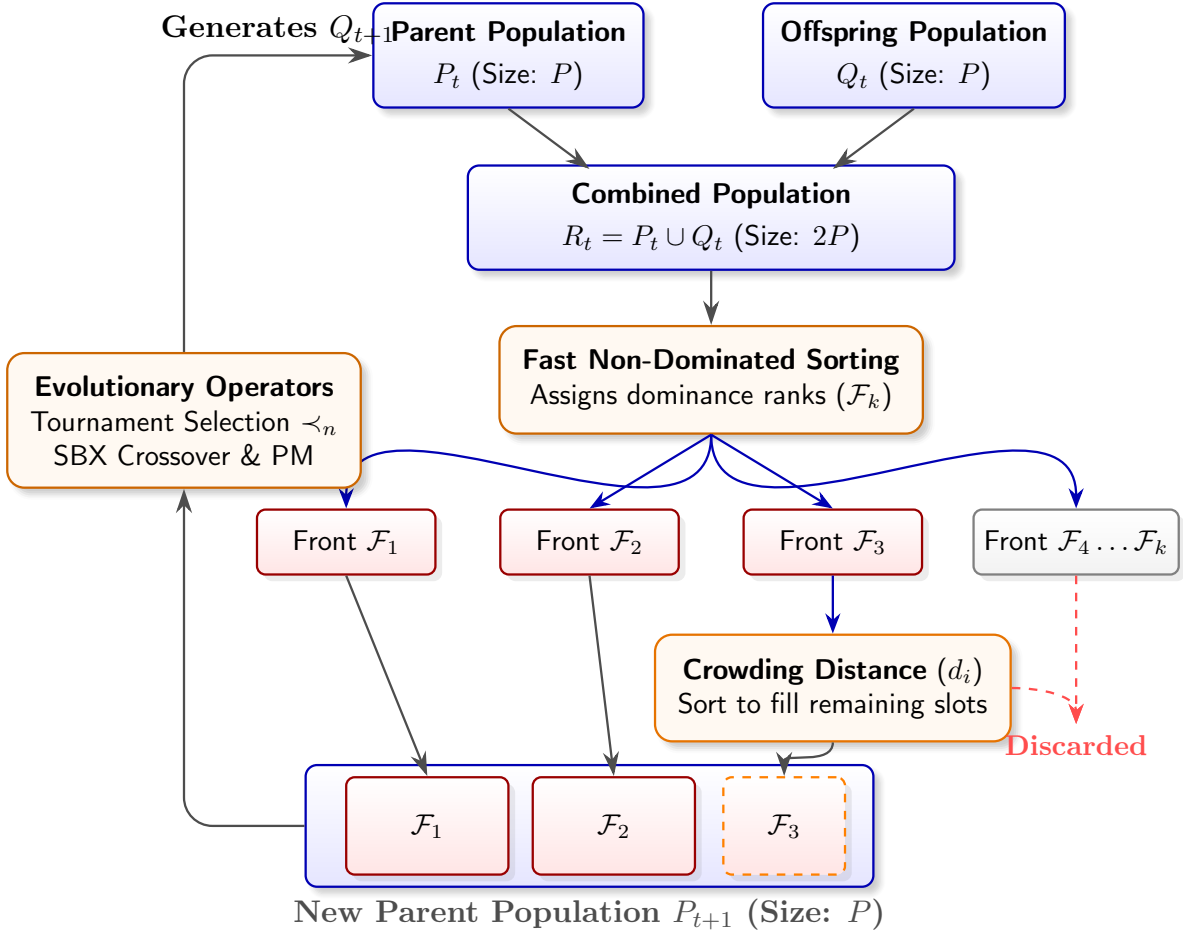

Among the algorithms available within PyGMO, the Non-dominated Sorting Genetic Algorithm II (NSGA-II) is particularly effective for exploring complex Pareto fronts. Within the proposed ADR optimization framework, NSGA-II is employed to explore the trade-offs among competing mission objectives and construct a diverse approximation of the Pareto-optimal solution set. As a multi-objective evolutionary algorithm, NSGA-II combines offspring generation, non-dominated sorting, elitist population selection, and crowding-distance preservation to simultaneously improve objective-space convergence and maintain diversity among competing mission architectures \cite{deb_fast_2002}. The overall execution cycle employed in the proposed framework is illustrated in Figure~\ref{fig:nsga2_flow}.

At each generation, the current parent population $P_t$ is evolved to generate an offspring population $Q_t$ using tournament selection, simulated binary crossover (SBX), and polynomial mutation. The parent and offspring populations are subsequently combined to form
\begin{equation}
R_t = P_t \cup Q_t,
\end{equation}
where $R_t$ contains twice the prescribed population size. The combined population is then evaluated and classified into successive non-dominated fronts according to Pareto dominance. Solutions belonging to the first front $\mathcal{F}_1$ are not dominated by any other solution in the combined population, while subsequent fronts contain solutions with progressively lower dominance ranks. The resulting sequence of fronts,
\begin{equation}
\mathcal{F}_1,\mathcal{F}_2,\ldots,\mathcal{F}_k,
\end{equation}
provides the basis for elitist environmental selection.\\

The fronts are considered sequentially during selection, with complete higher-ranked fronts preferentially retained in the next parent population until the prescribed population capacity is approached. When the inclusion of an entire front would exceed the available population size, crowding distance is used to determine which solutions from the marginal front are retained. For solution $i$, the crowding distance is given by:
\begin{equation}
d_i = \sum_{m=1}^{M}
\frac{f_m^{(i+1)}-f_m^{(i-1)}}
{f_m^{\max}-f_m^{\min}}.
\end{equation}
Solutions located in sparsely populated regions of the objective space receive larger crowding distances and are preferentially retained, whereas solutions located in densely populated regions are more likely to be discarded when the population capacity is exceeded. This mechanism preserves a broad distribution of solutions along the Pareto front rather than allowing the population to collapse toward a narrow region. The resulting population $P_{t+1}$ is subsequently used as the parent population for generating $Q_{t+1}$, completing the evolutionary cycle.\\

Within the ADR framework, each individual represents the servicing sequence through a continuous random-key vector, while the associated transfer and phasing quantities are computed during fitness evaluation. Different individuals may represent fundamentally different servicing strategies. One solution may prioritize rapid intervention through aggressive plane-change maneuvers, whereas another may exploit longer-duration $J_2$ drift opportunities to reduce propellant expenditure. The evolutionary operators therefore generate alternative mission architectures while the selection mechanism progressively retains solutions exhibiting favorable objective trade-offs.\\

NSGA-II consequently provides the evolutionary mechanism for systematically exploring these competing mission architectures. Non-dominated sorting identifies solutions representing different levels of Pareto quality, while elitist selection preserves the highest-ranked solutions across generations. Within the marginal front, crowding-distance preservation maintains solutions distributed across less densely populated regions of the objective space. The iterative interaction between offspring generation, population combination, non-dominated sorting, elitist selection, and diversity preservation enables the optimizer to progressively construct a broad approximation of the ADR Pareto front rather than converging toward a single mission configuration, as summarized in Figure~\ref{fig:nsga2_flow}.

\subsubsection{MOEA/D for Localized Pareto Convergence}

While NSGA-II excels at global exploration, rapid convergence toward highly optimized solutions can be achieved using the Multi-Objective Evolutionary Algorithm based on Decomposition (MOEA/D) \cite{zhang_moead_2007}. Within the proposed ADR optimization framework, MOEA/D is employed as a complementary optimizer to refine promising mission architectures and accelerate convergence toward specific regions of the Pareto front. Rather than treating the optimization problem as a single multi-objective task, MOEA/D decomposes it into a collection of scalar optimization subproblems, each associated with a distinct preference direction in the objective space.

The decomposition begins by generating a set of weight vectors
\begin{equation}
\boldsymbol{\lambda}^{1},\boldsymbol{\lambda}^{2},\ldots,
\boldsymbol{\lambda}^{P},
\end{equation}
where each vector defines a particular scalarization of the original multi-objective problem. An ideal reference point $\mathbf{z}^{*}$ is initialized from the best objective values observed in the current population. Using a Tchebycheff scalarization, the $i$-th subproblem minimizes:
\begin{equation}
g^{\mathrm{te}}(\mathbf{x}\mid\boldsymbol{\lambda}^{i},\mathbf{z}^{*})
=
\max_{1\leq m\leq M}
\left\{
\lambda_m^{i}
\left|
f_m(\mathbf{x})-z_m^{*}
\right|
\right\},
\end{equation}
where $\boldsymbol{\lambda}^{i}$ denotes the weight vector associated with subproblem $i$, $\mathbf{z}^{*}$ is the ideal reference point, and $f_m(\mathbf{x})$ represents the $m$-th objective function. Thus, each weight vector defines a distinct scalar optimization problem within the decomposed objective space.\\

In the ADR context, each weight vector corresponds to a particular mission philosophy. Weight vectors emphasizing $\Delta v$ generate highly fuel-efficient servicing campaigns, whereas vectors emphasizing mission duration favor rapid debris-removal operations. Intermediate vectors produce balanced trade-off solutions. The resulting subproblems are optimized concurrently, allowing different regions of the objective space to be explored according to their associated preference directions rather than forcing all candidate solutions toward a common scalar objective.\\

For each subproblem, a candidate solution $\mathbf{x}^{i}$ represents a specific servicing sequence through the random-key representation and is evolved using variation operators such as crossover and mutation. Information exchange is restricted to neighboring subproblems in the weight-vector space, forming a localized neighborhood registry. A subproblem primarily interacts with solutions associated with nearby weight vectors rather than with the entire population. This localized cooperation allows solutions corresponding to similar objective preferences to exchange information while limiting unnecessary interactions between fundamentally different regions of the Pareto front.\\

Following variation and evaluation, the resulting candidate is compared with the current solution of the corresponding scalar subproblem. If the new candidate provides an improved value of the scalarized objective, the current solution is replaced by the new candidate; the improved solution may also be used to update neighboring subproblems. The ideal reference point is updated whenever improved objective values are encountered. This iterative decomposition, neighborhood-based information exchange, and scalar replacement process is repeated across generations.\\

Within the ADR framework, each individual is represented by a continuous random-key vector whose rank ordering determines the servicing sequence. Transfer costs, waiting times, phasing effects, and mass evolution are then evaluated from the decoded sequence within the mission fitness function. MOEA/D therefore operates on the same decision representation while decomposing the resulting bi-objective problem into scalar subproblems. Weight vectors emphasizing propellant expenditure refine fuel-efficient architectures, whereas those emphasizing mission duration refine faster servicing strategies, with intermediate vectors maintaining balanced solutions.\\

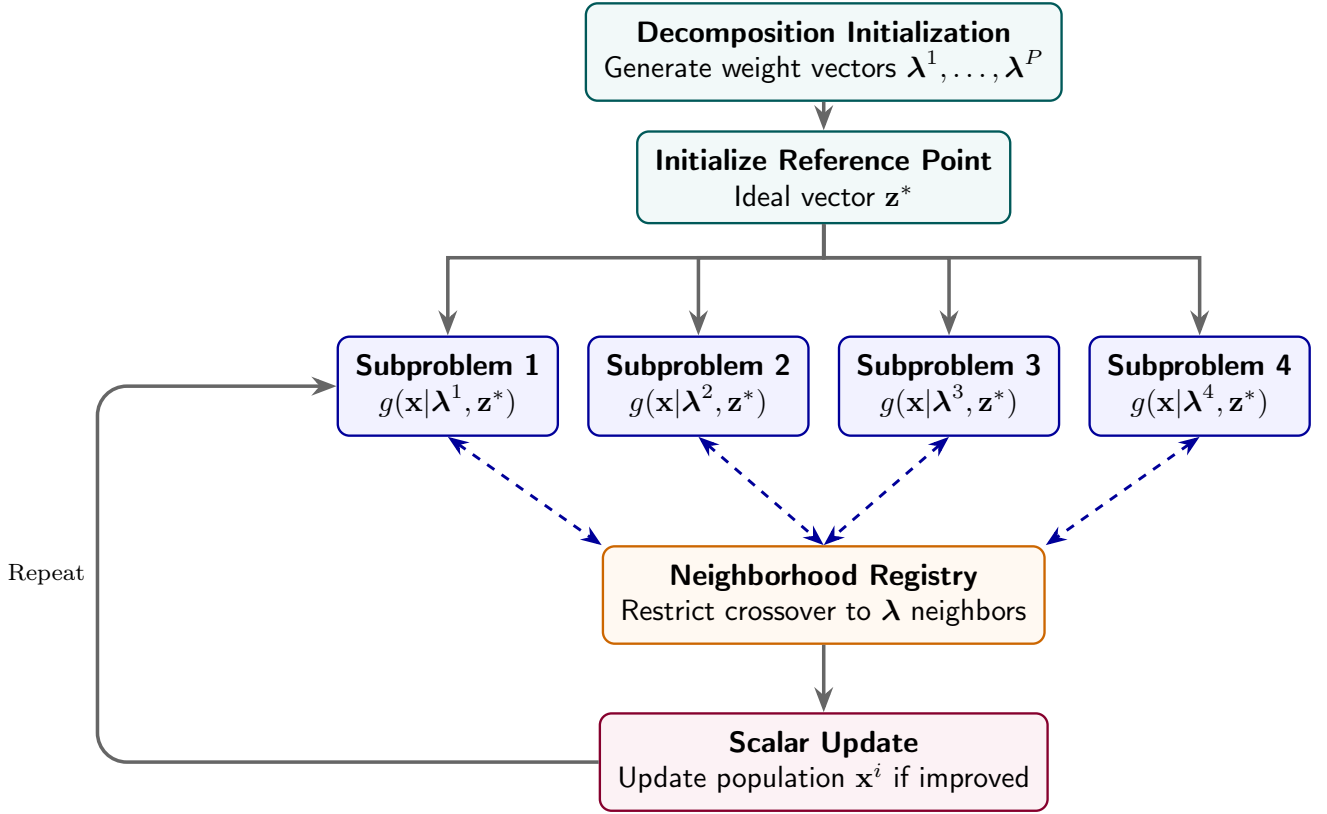
\begin{figure*}[t]
\centering
\resizebox{0.95\textwidth}{!}{
\begin{tikzpicture}[
    font=\small\sffamily,
    weightBlock/.style={rectangle, draw=teal!70!black, fill=teal!5, thick, rounded corners=4pt, inner sep=6pt, align=center},
    subpBlock/.style={rectangle, draw=blue!60!black, fill=blue!5, thick, rounded corners=4pt, inner sep=6pt, align=center},
    neighborBox/.style={rectangle, draw=orange!80!black, fill=orange!5, thick, rounded corners=4pt, inner sep=6pt, align=center},
    flowArrow/.style={->, >={Stealth[length=3mm]}, line width=1.2pt, draw=black!60},
    neighborArrow/.style={<->, >={Stealth[length=2.5mm]}, line width=1pt, draw=blue!60!black, dashed}
]

\node[weightBlock] (Weights) at (0, 4) {\textbf{Decomposition Initialization}\\Generate weight vectors $\boldsymbol{\lambda}^1, \dots, \boldsymbol{\lambda}^P$};
\node[weightBlock] (Ideal) at (0, 2.5) {\textbf{Initialize Reference Point}\\Ideal vector $\mathbf{z}^*$};

\draw[flowArrow] (Weights.south) -- (Ideal.north);

\node[subpBlock] (SP1) at (-4.5, 0) {\textbf{Subproblem 1}\\$g(\mathbf{x}|\boldsymbol{\lambda}^1, \mathbf{z}^*)$};
\node[subpBlock] (SP2) at (-1.5, 0) {\textbf{Subproblem 2}\\$g(\mathbf{x}|\boldsymbol{\lambda}^2, \mathbf{z}^*)$};
\node[subpBlock] (SP3) at (1.5, 0) {\textbf{Subproblem 3}\\$g(\mathbf{x}|\boldsymbol{\lambda}^3, \mathbf{z}^*)$};
\node[subpBlock] (SP4) at (4.5, 0) {\textbf{Subproblem 4}\\$g(\mathbf{x}|\boldsymbol{\lambda}^4, \mathbf{z}^*)$};

\draw[flowArrow] (Ideal.south) -- ++(0,-0.4) -| (SP1.north);
\draw[flowArrow] (Ideal.south) -- ++(0,-0.4) -| (SP2.north);
\draw[flowArrow] (Ideal.south) -- ++(0,-0.4) -| (SP3.north);
\draw[flowArrow] (Ideal.south) -- ++(0,-0.4) -| (SP4.north);

\node[neighborBox] (Neigh) at (0, -2.5) {\textbf{Neighborhood Registry}\\Restrict crossover to $\boldsymbol{\lambda}$ neighbors};

\draw[neighborArrow] (SP1.south) -- (Neigh.north west);
\draw[neighborArrow] (SP2.south) -- (Neigh.north);
\draw[neighborArrow] (SP3.south) -- (Neigh.north);
\draw[neighborArrow] (SP4.south) -- (Neigh.north east);

\node[subpBlock, draw=purple!70!black, fill=purple!5] (Update) at (0, -4.5) {\textbf{Scalar Update}\\Update population $\mathbf{x}^i$ if improved};
\draw[flowArrow] (Neigh.south) -- (Update.north);

\draw[flowArrow, rounded corners=10pt] (Update.west) -- ++(-6,0) -- node[left, font=\scriptsize] {Repeat} ++(0, 4.5) -- (SP1.west);

\end{tikzpicture}
}
\caption{Architecture of the MOEA/D algorithm. Unlike sorting-based methods, MOEA/D decomposes the multi-objective space into $P$ scalar optimization subproblems defined by weight vectors $\boldsymbol{\lambda}$. Evolution is localized: each subproblem interacts only with neighboring weight vectors in the objective-space decomposition, allowing different regions of the Pareto front to be explored through their associated scalarization functions.}
\label{fig:moead_flow}
\end{figure*}

This decomposition strategy is particularly advantageous for orbital trajectory optimization because small adjustments in transfer timing, waiting duration, or phasing can produce significant changes in the resulting mission objectives. By decomposing the global multi-objective problem into coordinated scalar subproblems and restricting evolutionary cooperation to neighboring weight vectors, MOEA/D provides a fine-grained trajectory optimization mechanism that refines mission architectures across different regions of the Pareto front. The complete decomposition, neighborhood interaction, scalar update, and iterative evolution process is summarized in Figure~\ref{fig:moead_flow}.

\subsubsection{Complementary Search Dynamics}

The heterogeneous combination of NSGA-II and MOEA/D within the PyGMO Archipelago is particularly well suited to ADR mission planning because the optimization problem contains both discrete combinatorial decisions and continuous astrodynamic decisions. At the combinatorial level, the optimizer must identify effective target sequences from a large set of candidate debris objects. At the astrodynamic level, it must determine the precise transfer timings, phasing opportunities, and maneuver schedules that minimize mission cost. To address these complementary requirements, the archipelago distributes the optimization process across heterogeneous islands, with NSGA-II and MOEA/D operating as distinct evolutionary search mechanisms.\\

Within this framework, NSGA-II primarily serves as a global exploration mechanism. By preserving diversity across the population through non-dominated sorting and crowding-distance selection, NSGA-II continuously explores fundamentally different mission architectures. In the ADR context, this corresponds to discovering alternative debris-removal campaigns involving different target combinations, routing strategies, and fuel-time trade-offs. MOEA/D provides a complementary localized search mechanism by decomposing the objective space into scalar subproblems and refining solutions associated with specific regions of the Pareto front. The heterogeneous island structure therefore allows broad exploration and localized exploitation to proceed independently within the same optimization environment.\\

Each island maintains its own population and independently evolves its candidate mission architectures for a number of generations using the evolutionary mechanism associated with that island. In the illustrated archipelago, separate islands employ NSGA-II or MOEA/D, allowing the same ADR objective space to be explored through different search dynamics. Within each island, candidate solutions are evaluated according to the mission objectives and progressively evolved toward improved Pareto regions. This isolated evolution allows each algorithm to develop solutions according to its intrinsic exploration or decomposition strategy before information is exchanged with neighboring islands.\\

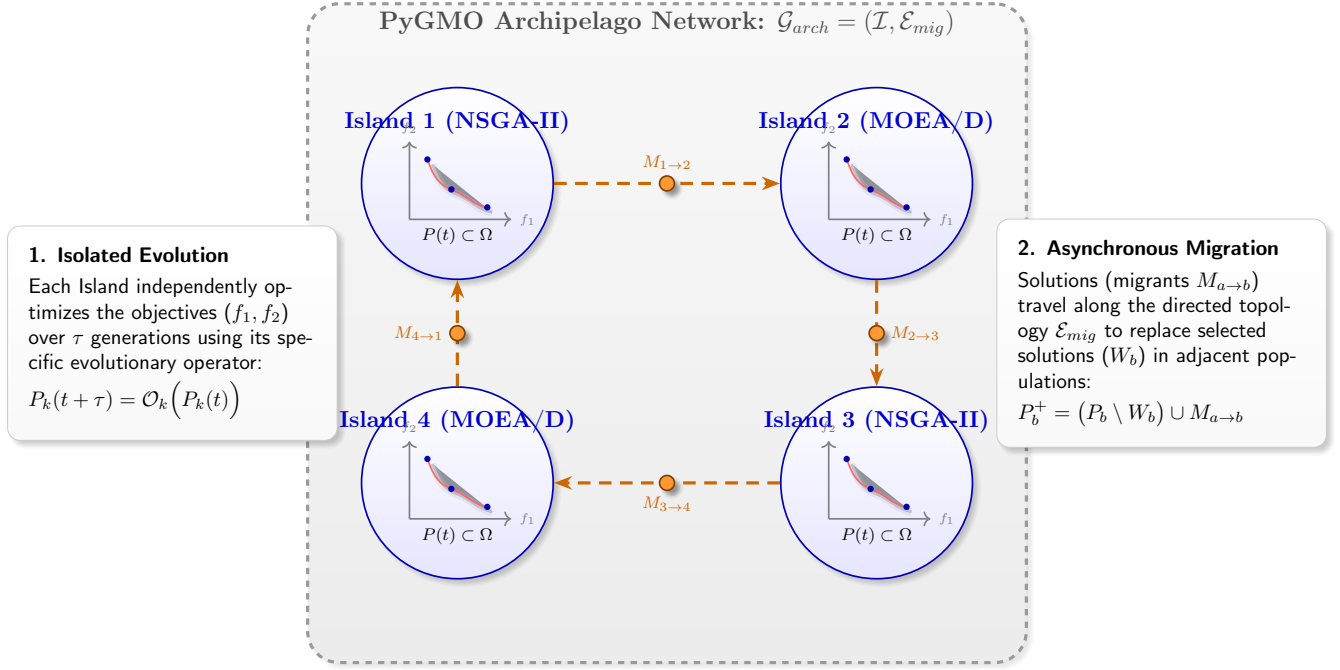
\begin{figure*}[t]
\centering
\resizebox{0.95\textwidth}{!}{
\begin{tikzpicture}[
    font=\small\sffamily,
    archipelago/.style={rounded corners=15pt, draw=black!40, dashed, line width=1.5pt, top color=gray!5, bottom color=gray!15, inner sep=15pt, blur shadow={shadow blur steps=5, shadow opacity=10}},
    island/.style={circle, draw=blue!70!black, top color=white, bottom color=blue!10, thick, minimum size=3.2cm, blur shadow={shadow blur steps=5, shadow opacity=20}},
    migr_path/.style={->, >={Stealth[length=3mm]}, line width=1.5pt, draw=orange!80!black, dash pattern=on 6pt off 4pt},
    migrant/.style={circle, fill=orange!80, draw=orange!50!black, thick, inner sep=2.5pt, blur shadow={shadow blur steps=3, shadow opacity=40}},
    infobox/.style={rectangle, draw=black!20, top color=white, bottom color=gray!5, rounded corners=6pt, blur shadow={shadow blur steps=5, shadow opacity=20}, inner sep=10pt, text width=4.8cm}
]

\node[archipelago, minimum width=12cm, minimum height=11cm] (container) at (0,0) {};
\node[anchor=north, font=\large\bfseries, text=black!70] at (container.north) {PyGMO Archipelago Network: $\mathcal{G}_{arch} = (\mathcal{I}, \mathcal{E}_{mig})$};

\newcommand{\drawIsland}[4]{
    \node[island] (#1) at (#2, #3) {};
    \node[anchor=south, font=\bfseries, text=blue!80!black] at ([yshift=0.7cm]#1.center) {#4};

    \draw[->, thick, black!50] ([xshift=-0.8cm, yshift=-0.6cm]#1.center) -- ([xshift=0.9cm, yshift=-0.6cm]#1.center) node[right, font=\tiny] {$f_1$};
    \draw[->, thick, black!50] ([xshift=-0.8cm, yshift=-0.6cm]#1.center) -- ([xshift=-0.8cm, yshift=0.7cm]#1.center) node[above, font=\tiny] {$f_2$};

    \draw[thick, draw=red!60, blur shadow={shadow blur steps=2}] 
        ([xshift=-0.5cm, yshift=0.4cm]#1.center) to[out=-70, in=160] 
        ([xshift=-0.1cm, yshift=-0.1cm]#1.center) to[out=-20, in=170] 
        ([xshift=0.5cm, yshift=-0.4cm]#1.center);

    \fill[blue!70!black] ([xshift=-0.5cm, yshift=0.4cm]#1.center) circle (1.5pt);
    \fill[blue!70!black] ([xshift=-0.1cm, yshift=-0.1cm]#1.center) circle (1.5pt);
    \fill[blue!70!black] ([xshift=0.5cm, yshift=-0.4cm]#1.center) circle (1.5pt);

    \node[anchor=north, font=\scriptsize] at ([yshift=-0.6cm]#1.center) {$P(t) \subset \Omega$};
}

\drawIsland{I1}{-3.5}{2.5}{Island 1 (NSGA-II)}
\drawIsland{I2}{3.5}{2.5}{Island 2 (MOEA/D)}
\drawIsland{I3}{3.5}{-2.5}{Island 3 (NSGA-II)}
\drawIsland{I4}{-3.5}{-2.5}{Island 4 (MOEA/D)}

\draw[migr_path] (I1.east) -- (I2.west);
\node[migrant] at (0, 2.5) {};
\node[above, font=\scriptsize\bfseries, text=orange!80!black] at (0, 2.6) {$M_{1 \to 2}$};

\draw[migr_path] (I2.south) -- (I3.north);
\node[migrant] at (3.5, 0) {};
\node[right, font=\scriptsize\bfseries, text=orange!80!black] at (3.6, 0) {$M_{2 \to 3}$};

\draw[migr_path] (I3.west) -- (I4.east);
\node[migrant] at (0, -2.5) {};
\node[below, font=\scriptsize\bfseries, text=orange!80!black] at (0, -2.6) {$M_{3 \to 4}$};

\draw[migr_path] (I4.north) -- (I1.south);
\node[migrant] at (-3.5, 0) {};
\node[left, font=\scriptsize\bfseries, text=orange!80!black] at (-3.6, 0) {$M_{4 \to 1}$};


\node[infobox, anchor=east] at (-5.5, 0) {
    \textbf{1. Isolated Evolution}\\
    \vspace{1mm}
    Each Island independently optimizes the objectives ($f_1, f_2$) over $\tau$ generations using its specific evolutionary operator:\\[1mm]
    $P_k(t+\tau) = \mathcal{O}_k \Big( P_k(t) \Big)$
};

\node[infobox, anchor=west] at (5.5, 0) {
    \textbf{2. Asynchronous Migration}\\
    \vspace{1mm}
    Solutions (migrants $M_{a \to b}$) travel along the directed topology $\mathcal{E}_{mig}$ to replace selected solutions ($W_b$) in adjacent populations:\\[1mm]
    $P_b^+ = \big( P_b \setminus W_b \big) \cup M_{a \to b}$
};

\end{tikzpicture}
}
\caption{Advanced structural diagram of the PyGMO Archipelago architecture. Each island independently maintains and evolves a population $P_k$ for the bi-objective optimization problem. Solutions are exchanged across the directed migration topology, with the illustrated heterogeneous configuration pairing NSGA-II and MOEA/D islands.}
\label{fig:pygmo_advanced_archipelago}
\end{figure*}

The Archipelago framework subsequently introduces directed migration between islands, as demonstrated in Figure~\ref{fig:pygmo_advanced_archipelago}. Candidate solutions generated by one island are transferred as migrants $M_{a\rightarrow b}$ through the migration topology and made available to another island, where they can replace inferior candidate solutions. In this manner, the populations evolve independently for successive generations while exchanging information across the heterogeneous search landscape. The resulting interaction can be represented as
\begin{equation}
P_b^{+}
=
\left(P_b\setminus W_b\right)
\cup M_{a\rightarrow b},
\end{equation}
where $P_b$ denotes the receiving population, $W_b$ represents solutions selected for replacement, and $M_{a\rightarrow b}$ denotes the incoming migrants.\\

This migration mechanism establishes a feedback pathway between global exploration and localized exploitation. An NSGA-II island may discover a previously unexplored routing architecture that is subsequently available to a MOEA/D island for further refinement, while solutions refined by MOEA/D can in turn provide high-quality regions of the objective space to other evolutionary populations. Through this heterogeneous interaction, the archipelago reduces the likelihood that individual islands become trapped within isolated regions of the search space while preserving the distinct search behavior of the underlying algorithms.\\

This synergy is especially valuable for large-scale ADR campaigns. The graph-theoretic and GBS stages first reduce the original combinatorial search space to promising debris subsets, after which NSGA-II explores alternative servicing architectures and MOEA/D refines the associated orbital-transfer parameters. The resulting archipelago therefore combines independent island evolution with cross-island exchange of solutions, simultaneously supporting mission diversity, localized trajectory refinement, and continued exploration of alternative Pareto regions. The final Pareto archive consequently integrates mission architectures generated through complementary evolutionary search dynamics, providing a diverse set of fuel-time trade-offs for multi-target debris-removal operations.

\section{Benchmark Environments}

A rigorous evaluation of an ADR planning framework requires testing across
orbital populations exhibiting different structural and operational
characteristics. Performance measured on a single debris catalogue may
conflate algorithmic behavior with properties specific to the sampled
population. To address this issue, three benchmark suites are constructed
from the adopted orbital catalogue and conjunction-risk data. The suites are
designed to expose the proposed framework to risk-focused, provenance-focused,
and randomly sampled target populations.

The benchmark environments are evaluated through the weighted orbital-transfer
network
\[
A\in\mathbb{R}^{N\times N},
\]
where $N$ denotes the candidate target-pool size. Because the graph-based
prioritization strategies and the GBS stage operate on this adjacency matrix,
differences in the orbital composition of the benchmark populations induce
differences in transfer accessibility, graph structure, and consequently in
the observed GBS subset distributions. Evaluating the framework across
multiple benchmark constructions therefore provides an empirical assessment
of its behavior across different orbital environments.

\subsection{Benchmark Design Philosophy}

The three benchmark suites represent complementary target-population
constructions and associated ADR scenarios:

\begin{itemize}
    \item \textbf{SOCRATES Minefield} represents a conjunction-risk-focused
    environment constructed from objects associated with high reported
    conjunction probabilities.

    \item \textbf{Collision Cloud} represents a provenance-focused population
    constructed from objects belonging to the dominant identifiable inferred
    source family in the catalogue.

    \item \textbf{Random Baseline} represents a reference population obtained
    through uniform random sampling from the available catalogue without
    benchmark-specific structural filtering.
\end{itemize}

Together, these suites provide distinct experimental environments for
examining how target-selection strategies interact with differences in
conjunction-risk concentration, source-family composition, and sampled orbital
structure. The benchmark construction itself does not impose a predetermined
graph topology. Instead, the transfer-network structure is determined by the
orbital states of the objects selected for each suite and random seed.

\subsection{SOCRATES Minefield}

\subsubsection{Data Source and Construction}

The SOCRATES Minefield benchmark is derived from conjunction information
obtained through the CelesTrak SOCRATES database
\footnote{\url{https://celestrak.org/SOCRATES/}}. For each object appearing in
the conjunction dataset, the maximum reported conjunction probability across
the available encounter records is retained as its conjunction-risk score.
The corresponding catalogue objects are then ranked in descending order of
this score.

Rather than selecting the first $N$ objects directly, the implementation first
retains the highest-risk 200 objects as a candidate tier. For each benchmark
configuration and experimental seed, $N$ objects are then sampled from this
200-object tier. The suite preserves a strong emphasis on
conjunction-associated objects while retaining stochastic variation among
benchmark instances.

\subsubsection{Topological Characteristics}

The orbital-transfer structure of the resulting population is determined by
the orbital states of the sampled high-risk objects. Since conjunction risk
can be concentrated among objects occupying similar or repeatedly intersecting
orbital regimes, the resulting transfer network may contain regions of
enhanced connectivity or localized transfer structure. However, such
structures are empirical properties of the sampled population rather than
explicit constraints imposed during benchmark construction.

\subsubsection{Expected Sampling Behavior}

The GBS sampling behavior is determined by the weighted adjacency matrix
constructed from the sampled population and the selected prioritization
heuristic. When the induced transfer network contains strongly connected
regions, GBS may assign greater probability to subsets associated with those
regions through the corresponding graph and Hafnian structure. The degree of
concentration and subset diversity is therefore expected to vary with the
realized orbital configuration and the applied prioritization strategy.

\subsubsection{Operational Interpretation}

This suite provides a risk-focused benchmark for examining the behavior of the
ADR framework on a population selected from objects with elevated reported
conjunction probabilities. Performance in this environment characterizes how
different target-prioritization strategies affect subset generation and
downstream mission planning for conjunction-risk-focused populations.

\subsection{Collision Cloud}

\subsubsection{Data Source and Construction}

Debris populations may contain groups of objects associated with common
launch programs, satellite constellations, or fragmentation events. To
construct a provenance-focused benchmark, source-family labels are inferred
from catalogue object names using recurring naming patterns. Objects that
cannot be assigned to an identifiable family are labeled
\texttt{UNKNOWN}.

The dominant identifiable family is then determined from the catalogue by
selecting the family with the largest number of associated objects. For each
benchmark configuration, $N$ objects are sampled from this dominant family
using the corresponding experimental seed. The family assignment is therefore
used as a catalogue-derived provenance proxy rather than as an independently
reconstructed fragmentation genealogy.

\subsubsection{Topological Characteristics}

The transfer-network structure of the resulting population depends on the
orbital characteristics of the objects belonging to the selected family.
Objects originating from related deployment or fragmentation histories may
exhibit similar orbital characteristics, which can produce relatively strong
transfer accessibility within portions of the population. Nevertheless, the
actual connectivity, community structure, and transfer-cost distribution are
computed from the orbital states of the sampled objects rather than imposed
directly by the benchmark construction.

\subsubsection{Expected Sampling Behavior}

When the selected family contains groups of mutually accessible objects, the
corresponding weighted transfer graph may contain dense local structures.
Under such conditions, the GBS stage can generate subsets associated with
these structurally connected regions. Conversely, greater orbital dispersion
within the sampled family can produce a more heterogeneous subset
distribution. The observed sampling behavior is therefore evaluated
empirically for each benchmark instance.

\subsubsection{Operational Interpretation}

This suite provides a benchmark for examining the ability of the framework to
exploit populations with common inferred provenance. It emphasizes the
interaction between source-family structure, transfer accessibility, graph
weighting, and downstream multi-target mission optimization.

\subsection{Random Baseline}

\subsubsection{Data Source and Construction}

To establish a reference population without benchmark-specific selection
criteria, a Random benchmark is generated through uniform random sampling of
$N$ objects from the complete available catalogue. The sampling is performed
using the corresponding experimental seed. No filtering is applied based on
conjunction probability, source-family membership, orbital similarity, or
other benchmark-specific structural characteristics.

\subsubsection{Topological Characteristics}

Unlike the SOCRATES Minefield and Collision Cloud suites, the Random benchmark
does not intentionally concentrate objects according to conjunction risk or
inferred source family. The resulting transfer-network structure is therefore
determined by the orbital characteristics of the randomly sampled population.
Different seeds may consequently produce different levels of connectivity,
dispersion, and community structure.

\subsubsection{Expected Sampling Behavior}

Because no benchmark-specific structural property is imposed, the GBS subset
distribution is determined by the weighted transfer graph induced by the
randomly sampled objects and by the selected prioritization strategy. The
result may exhibit either concentrated or dispersed sampling depending on the
connectivity realized for a particular benchmark instance.

\subsubsection{Operational Interpretation}

The Random benchmark provides an unstructured reference case for evaluating
how the prioritization strategies and downstream mission-optimization
framework behave when the candidate population is obtained without a
benchmark-specific selection criterion.

\subsection{Benchmark Summary}

Table~\ref{tab:benchmark_environments} summarizes the three benchmark suites
used in the experimental evaluation. Each suite is generated independently for
the prescribed candidate-pool sizes and experimental seeds, after which the
same orbital-transfer, target-prioritization, GBS, and trajectory-optimization
procedures are applied as appropriate.

\begin{table}[htbp]
\centering
\caption{Summary of the benchmark environments used for evaluation.}
\label{tab:benchmark_environments}
\renewcommand{\arraystretch}{1.3}
\begin{tabularx}{\textwidth}{
>{\raggedright\arraybackslash}p{3.0cm}
>{\raggedright\arraybackslash}X
>{\raggedright\arraybackslash}X
>{\raggedright\arraybackslash}X}
\toprule
\textbf{Environment} &
\textbf{Network Characteristics} &
\textbf{Construction Principle} &
\textbf{Operational Role} \\
\midrule

\textbf{SOCRATES Minefield}
&
Risk-focused population whose transfer-network structure is determined by the
sampled high-conjunction-risk objects.
&
Objects are ranked by maximum reported conjunction probability; the highest-risk
200 objects form the candidate tier, from which $N$ objects are sampled using
the experimental seed.
&
Evaluates framework behavior for conjunction-risk-focused target populations.
\\
\addlinespace

\textbf{Collision Cloud}
&
Population drawn from a dominant inferred source family, with network structure
determined by the orbital characteristics of the sampled family members.
&
The dominant identifiable source family is inferred from catalogue naming
patterns, followed by seeded sampling of $N$ objects from that family.
&
Evaluates framework behavior for provenance-focused and potentially clustered
orbital populations.
\\
\addlinespace

\textbf{Random Baseline}
&
Network structure determined by the orbital characteristics of the randomly
sampled population.
&
Uniform random sampling of $N$ objects from the complete catalogue without
benchmark-specific filtering.
&
Provides an unstructured reference case for evaluating the effect of
target-selection strategies.
\\

\bottomrule
\end{tabularx}
\end{table}
\section{Target Prioritization Framework}

A fundamental challenge in ADR is determining which
objects should be serviced when the available mission capacity is substantially
smaller than the size of the orbital debris population. Let
\[
V=\{v_1,v_2,\ldots,v_N\}
\]
denote the candidate debris catalog and let
\[
S\subseteq V,\qquad |S|=k,
\]
represent the subset selected for removal during a single remediation campaign,
where typically $k\ll N$.

The quality of an ADR mission depends both on the individual characteristics of
the selected targets and on the collective structure of the selected subset.
An object may possess high environmental consequence, long orbital persistence,
or substantial conjunction risk, while remaining operationally difficult to
service if it is poorly connected to other accessible targets. Conversely, a
group of highly accessible objects may support efficient multi-target servicing
while providing a smaller environmental contribution. Target selection in the
proposed framework therefore considers complementary dimensions including
environmental consequence, orbital persistence, conjunction risk, astrodynamic
accessibility, transfer-network structure, and provenance-related information.

Within the proposed framework, target prioritization is separated from
trajectory optimization. The prioritization stage does not determine the
servicing sequence directly. Instead, it modifies the weighted orbital-transfer
network presented to the GBS stage, thereby influencing the distribution from
which candidate target subsets are sampled. Different prioritization strategies
therefore induce different weighted graph representations of the same debris
environment. The resulting candidate subsets are subsequently passed to the
trajectory-optimization stage, where detailed mission-level objectives are
evaluated independently of the prioritization score.

To investigate how different target-selection criteria affect the resulting
mission architectures, seven selection strategies are considered. Six are
heuristic prioritization strategies and one is a random baseline. The
heuristics span three complementary categories. The first captures
object-level environmental and operational characteristics, the second
captures collective astrodynamic structure, and the third incorporates
provenance-related information. The random baseline provides a reference in
which no heuristic-specific target weighting is used.

\begin{enumerate}

    \item \textbf{Physical and Environmental Characteristics}
    \begin{itemize}
        \item \textbf{Physical Consequence Index}
        \par
        \emph{Which objects have the largest surrogate mass and therefore
        potentially contribute greater consequences in the event of
        fragmentation?}

        \item \textbf{Persistence Index}
        \par
        \emph{Which objects are associated with longer estimated orbital
        persistence under the adopted drag-related surrogate?}

        \item \textbf{Operational Threat Index}
        \par
        \emph{Which objects combine fragmentation consequence, persistence,
        and conjunction-risk information into a composite hazard score?}
    \end{itemize}

    \item \textbf{Astrodynamic Network Structure}
    \begin{itemize}
        \item \textbf{Transfer-Network Importance}
        \par
        \emph{Which objects occupy structurally important positions within
        the weighted orbital-transfer network?}

        \item \textbf{Orbital Cluster Membership}
        \par
        \emph{Which objects belong to large and densely connected communities
        of transfer-accessible debris?}
    \end{itemize}

    \item \textbf{Provenance-Aware Prioritization}
    \begin{itemize}
        \item \textbf{Source-Family Prioritization}
        \par
        \emph{Which objects belong to larger inferred source or fragmentation
        families represented in the catalog?}
    \end{itemize}

    \item \textbf{Baseline Control}
    \begin{itemize}
        \item \textbf{Random Subset Selection}
        \par
        \emph{What mission performance is obtained when target subsets are
        generated without heuristic-specific prioritization?}
    \end{itemize}

\end{enumerate}

The seven strategies emphasize different characteristics of the debris
environment and are not mutually independent. For example, the Operational
Threat Index combines surrogate mass, persistence, and conjunction-risk
information into a single composite score, while the Transfer-Network
Importance and Orbital Cluster Membership strategies both operate on the same
underlying transfer network but emphasize different structural properties.
Accordingly, the purpose of the experimental comparison is not to treat the
heuristics as independent estimators, but to examine how alternative
prioritization mechanisms alter subset generation and downstream mission
characteristics.

Each heuristic-specific strategy is evaluated using the same subsequent
mission-optimization procedure. This provides a controlled basis for
quantifying how the choice of prioritization strategy affects the sampled
target subsets, their graph-theoretic properties, and the resulting
multi-objective ADR mission architectures. For each strategy, four aspects are
described below: (i) operational motivation, (ii) available catalog
information, (iii) mathematical formulation, and (iv) mission-level
significance.

\subsection{Heuristic-Dependent Graph Weighting}

Let $P_h(v_i)$ denote the priority assigned to debris object $v_i$ under
heuristic $h$. The priority scores are used to modify the weighted
orbital-transfer network before GBS sampling.

Let
\[
A^{\mathrm{base}}=[A^{\mathrm{base}}_{ij}]
\]
denote the baseline astrodynamic affinity matrix derived from the pairwise
transfer model. For a heuristic $h$, the priority-dependent component of the
edge weighting can be represented conceptually as
\begin{equation}
A^{(h)}_{ij}
\propto
A^{\mathrm{base}}_{ij}
\left(P_h(v_i)P_h(v_j)\right)^2.
\end{equation}

In the implementation, the baseline accessibility component is constructed
from a composite transfer cost
\begin{equation}
C_{ij}
=
\Delta V^{\mathrm{passive}}_{ij}
+
0.2\,\Delta i_{ij}^{\circ}
+
0.05\,\Delta\Omega_{ij}^{\circ}
+
0.01\,W_{ij},
\end{equation}
where $\Delta V^{\mathrm{passive}}_{ij}$ denotes the passive transfer
velocity increment, $\Delta i_{ij}^{\circ}$ and
$\Delta\Omega_{ij}^{\circ}$ are the inclination and RAAN differences in
degrees, respectively, and $W_{ij}$ denotes the estimated waiting time
associated with orbital phasing.

A characteristic transfer-cost scale is defined by
\begin{equation}
\sigma
=
\operatorname{median}
\left\{
\Delta V^{\mathrm{passive}}_{ij}:
\Delta V^{\mathrm{passive}}_{ij}<\infty
\right\}.
\end{equation}

The corresponding accessibility term is
\begin{equation}
B_{ij}
=
\exp\left(-\frac{C_{ij}}{\sigma}\right).
\end{equation}

For a heuristic $h$, the unnormalized edge weight is then constructed as
\begin{equation}
\widetilde A^{(h)}_{ij}
=
B_{ij}
\left(P_h(v_i)P_h(v_j)\right)^2.
\end{equation}

For the Source-Family Prioritization strategy, an additional multiplicative
factor of five is applied when both endpoints belong to the same identified
non-unknown source family:
\begin{equation}
\widetilde A^{(\mathrm{src})}_{ij}
=
\begin{cases}
5\,B_{ij}
\left(P_{\mathrm{src}}(v_i)P_{\mathrm{src}}(v_j)\right)^2,
&
F(v_i)=F(v_j)\neq\mathrm{UNKNOWN},
\\[4pt]
B_{ij}
\left(P_{\mathrm{src}}(v_i)P_{\mathrm{src}}(v_j)\right)^2,
&
\text{otherwise}.
\end{cases}
\end{equation}

A small multiplicative stochastic perturbation is subsequently introduced
through
\begin{equation}
\widetilde A^{(h)}_{ij}
\leftarrow
\widetilde A^{(h)}_{ij}\,\xi_{ij},
\qquad
\xi_{ij}
=
\max\left(0.5,\mathcal N(1,0.05^2)\right),
\end{equation}
and the resulting symmetric adjacency matrix is spectrally normalized before
being supplied to the GBS procedure.

The heuristic-dependent graph can be decomposed into a baseline
astrodynamic accessibility component and a priority-dependent component.
Let $A^{\mathrm{base}}$ denote the baseline adjacency associated with the
astrodynamic accessibility model, and define
\[
D_h=
\operatorname{diag}
\left(
P_h(v_1)^2,\ldots,P_h(v_N)^2
\right).
\]
If the additional source-family factor, stochastic perturbation, and
subsequent normalization are temporarily omitted, the priority-weighted
adjacency can be written as
\begin{equation}
A^{(h)}_{\mathrm{prio}}
=
D_h A^{\mathrm{base}} D_h.
\end{equation}
Element-wise, this gives
\begin{equation}
A^{(h)}_{\mathrm{prio},ij}
=
A^{\mathrm{base}}_{ij}
\left(P_h(v_i)P_h(v_j)\right)^2,
\end{equation}
which is the priority-weighting component used in the implemented graph
construction.

For a subset $S$, the corresponding diagonal transformation satisfies
\begin{equation}
\operatorname{Haf}
\left(
D_{h,S}A^{\mathrm{base}}_S D_{h,S}
\right)
=
\left(
\prod_{v_i\in S}P_h(v_i)^2
\right)
\operatorname{Haf}
\left(A^{\mathrm{base}}_S\right).
\end{equation}
Thus, the matrix identity describes the priority-only component of the
graph, whereas the implemented adjacency additionally includes the
source-family factor for the \texttt{SOURCE} heuristic, stochastic edge
jitter, and spectral normalization.

For a collision-free GBS outcome associated with a vertex subset $S$, the
Hafnian obeys the diagonal scaling identity
\begin{equation}
\operatorname{Haf}
\left(D_{h,S}A^{\mathrm{base}}_S D_{h,S}\right)
=
\left(
\prod_{v_i\in S}P_h(v_i)^2
\right)
\operatorname{Haf}
\left(A^{\mathrm{base}}_S\right).
\end{equation}


In the present implementation, threshold detection is used and samples are
retained when exactly $k$ distinct modes are active. The
empirical subset distribution is determined directly from the simulated GBS
outcomes rather than being replaced by a simplified closed-form product of
individual priority values. The graph weighting should therefore be
interpreted as a mechanism for modifying the combinatorial structure presented
to the quantum sampler, rather than as an independent probability model for
the final subsets.

\subsection{Physical Consequence Index (\texttt{SIZE})}

Large debris objects, particularly spent upper stages and defunct spacecraft,
can represent substantial fragmentation consequences because a breakup event
can generate large numbers of secondary fragments. The Physical Consequence
Index therefore uses surrogate object mass as a proxy for the potential
environmental consequence associated with removing a given object.

\textbf{Available Data:}
Public orbital catalogs provide object classification information through
\texttt{OBJECT\_TYPE} together with approximate size information through
\texttt{RCS\_SIZE}. Because direct object mass is generally unavailable from a
standard TLE record, a surrogate mass is assigned from these catalog
attributes.

\textbf{Mathematical Formulation:}

A surrogate mass estimate is defined as
\begin{equation}
\hat m_i
=
f(\texttt{OBJECT\_TYPE}_i,\texttt{RCS\_SIZE}_i),
\end{equation}
with the adopted lookup values
\begin{equation}
\hat m_i =
\begin{cases}
8000 & \text{Rocket Body, Large}\\
3000 & \text{Rocket Body, Medium}\\
1000 & \text{Rocket Body, Small}\\
2000 & \text{Payload, Large}\\
700 & \text{Payload, Medium}\\
200 & \text{Payload, Small}\\
500 & \text{Debris, Large}\\
100 & \text{Debris, Medium}\\
10 & \text{Debris, Small}.
\end{cases}
\end{equation}

The corresponding priority score is
\begin{equation}
P_{\mathrm{size}}(v_i)=\sqrt{\hat m_i}.
\end{equation}

The square-root transformation preserves the ordering induced by the
surrogate mass while reducing the numerical dominance of the largest objects.

\textbf{Operational Significance:}

This heuristic emphasizes objects with larger estimated physical consequence.
It does not by itself incorporate orbital transfer cost, conjunction risk, or
orbital persistence, which are represented separately by the other
prioritization strategies.

\subsection{Persistence Index (\texttt{BSTAR})}

Objects in low Earth orbit experience atmospheric drag, with the magnitude of
the effect depending strongly on orbital altitude and the adopted drag-related
parameter. The Persistence Index therefore constructs a relative persistence
indicator from quantities available in the TLE catalog.

\textbf{Available Data:}

The metric utilizes the drag-related parameter \texttt{BSTAR} and the mean
motion $n_i$ obtained from the TLE.

\textbf{Mathematical Formulation:}

The semi-major axis is estimated from the mean motion, expressed in revolutions
per day, as
\begin{equation}
a_i
=
\left(
\frac{\mu}
{\left(2\pi n_i/86400\right)^2}
\right)^{1/3},
\end{equation}
where $\mu$ denotes Earth's gravitational parameter. The corresponding orbital
altitude is
\begin{equation}
h_i=a_i-R_E,
\end{equation}
where $R_E$ is Earth's reference radius.

A relative persistence indicator is then defined as
\begin{equation}
L_i
=
\frac{h_i^{3/2}}
{|B_i^*|},
\end{equation}
where $B_i^*$ denotes the TLE BSTAR parameter.

Because this quantity is a surrogate indicator rather than a full atmospheric
lifetime model, it is used after logarithmic scaling:
\begin{equation}
P_{\mathrm{bstar}}(v_i)
=
\log_{10}(1+L_i).
\end{equation}

The quantity $L_i$ should therefore be interpreted as a relative persistence
measure within the adopted computational model, rather than as a direct
prediction of re-entry time.

\textbf{Operational Significance:}

This heuristic emphasizes objects associated with larger values of the adopted
persistence surrogate and thus provides a long-term environmental perspective
that is distinct from the immediate conjunction-risk metric.

\subsection{Operational Threat Index (\texttt{SOCRATES\_RISK})}

The Physical Consequence and Persistence indices characterize environmental
impact from different perspectives but do not directly incorporate conjunction
risk. The Operational Threat Index combines these dimensions into a composite
object-level score.

\textbf{Available Data:}

The metric uses the surrogate mass estimate $\hat m_i$, the BSTAR-derived
persistence indicator $L_i$, and a conjunction-risk quantity $P_{c,i}$ obtained
from the adopted conjunction-assessment data source.

\textbf{Mathematical Formulation:}

The operational threat score is defined as
\begin{equation}
P_{\mathrm{risk}}(v_i)
=
\hat m_i
\sqrt{P_{c,i}}
\log_{10}(1+L_i).
\end{equation}

The three factors respectively represent surrogate physical consequence,
orbital persistence, and conjunction-risk information. The square-root
transformation applied to $P_{c,i}$ moderates the numerical contribution of
the conjunction-risk term within the composite score.

\textbf{Operational Significance:}

The heuristic combines environmental and conjunction-related information into
a single priority value. It therefore provides a composite target-selection
perspective rather than isolating any single physical property.

\subsection{Transfer-Network Importance (\texttt{CENTRALITY})}

The operational relevance of a debris object also depends on its position
within the orbital transfer network. Objects that are connected to many other
accessible targets may influence the structure of multi-target servicing
campaigns even when their object-level physical characteristics are moderate.
The Transfer-Network Importance heuristic therefore uses graph centrality to
characterize structural importance within the transfer network.

\textbf{Available Data:}

The heuristic uses the pairwise astrodynamic transfer information from which
the weighted orbital-transfer graph is constructed.

\textbf{Mathematical Formulation:}

Let
\[
G=(V,E)
\]
denote the weighted orbital-transfer graph. Three centrality measures are
computed:

\begin{itemize}
    \item PageRank ($PR$), representing global network influence;
    \item Eigenvector Centrality ($EV$), representing connectivity to other
    structurally important objects;
    \item Betweenness Centrality ($BC$), representing participation in
    shortest-path transfer corridors.
\end{itemize}

The composite centrality score is defined as
\begin{equation}
P_{\mathrm{cent}}(v_i)
=
4PR(v_i)
+
3EV(v_i)
+
2BC(v_i).
\end{equation}

The coefficients specify the relative contribution assigned to the three
centrality components in the computational heuristic.

\textbf{Operational Significance:}

The resulting score identifies objects that are structurally important within
the weighted transfer network. It therefore emphasizes network position
rather than direct environmental consequence.

\subsection{Orbital Cluster Membership (\texttt{COMMUNITY})}

Orbital debris populations can contain groups of objects that are strongly
connected under the adopted transfer-cost metric. The Orbital Cluster
Membership heuristic identifies such communities using graph partitioning and
assigns higher priority to objects belonging to larger and more internally
connected communities.

\textbf{Available Data:}

The heuristic uses the weighted orbital-transfer graph constructed from the
pairwise transfer model.

\textbf{Mathematical Formulation:}

Communities are identified using Louvain modularity maximization, yielding a
partition
\begin{equation}
V
=
\bigcup_{r=1}^{K}\mathcal C_r,
\end{equation}
where the communities $\mathcal C_r$ are the graph partitions obtained from
the weighted transfer network.

For each community, an average internal accessibility density is calculated as
\begin{equation}
\rho_r
=
\frac{
\sum_{(i,j)\in E(\mathcal C_r)}w_{ij}
}{
\binom{|\mathcal C_r|}{2}
}.
\end{equation}

The corresponding object-level priority is
\begin{equation}
P_{\mathrm{comm}}(v_i)
=
|\mathcal C_r|\rho_r,
\qquad
v_i\in\mathcal C_r.
\end{equation}

Thus, objects belonging to larger communities with greater internal
accessibility density receive larger priority values.

\textbf{Operational Significance:}

This heuristic emphasizes local network structure rather than global
centrality. It provides a mechanism for identifying target populations that
are concentrated within densely connected regions of the transfer graph.

\subsection{Source-Family Prioritization (\texttt{SOURCE})}

Debris populations may contain objects associated with recurring launch
programs, satellite constellations, or fragmentation families. Provenance
information can therefore be incorporated as an additional target-selection
criterion. In the present framework, source-family information is used as a
catalog-derived proxy rather than as a reconstructed physical genealogy.

\textbf{Available Data:}

Source families are inferred from catalog naming and identifier information,
including recurring object-name patterns and related available metadata.
Objects for which a family cannot be identified are assigned the label
\texttt{UNKNOWN}.

\textbf{Mathematical Formulation:}

Let
\[
F_r\subseteq V
\]
denote an inferred source family and let
\[
n_r=|F_r|
\]
represent the number of catalog objects assigned to that family. The
source-family priority is defined as
\begin{equation}
P_{\mathrm{src}}(v_i)
=
n_r,
\qquad
v_i\in F_r.
\end{equation}

In addition to this object-level priority, the graph-construction procedure
applies an additional multiplicative factor to edges whose endpoints belong to
the same non-unknown family.

\textbf{Operational Significance:}

This heuristic emphasizes provenance structure within the benchmark catalog.
It enables experiments in which target selection is influenced by the
concentration of objects within inferred debris families. Because the family
assignment is a catalog-based proxy, the resulting interpretation is likewise
limited to the provenance information encoded by the adopted classification
procedure.

\subsection{Random Subset Selection (\texttt{RANDOM\_SUBSET})}

A random baseline is included to provide a reference against which the
heuristic-driven target-selection strategies can be evaluated. The baseline
removes heuristic-specific prioritization from the subset-generation stage and
therefore isolates the effect of the selection strategy from the common
downstream mission-optimization procedure.

\textbf{Available Data:}

No object-level prioritization information is required.

\textbf{Mathematical Formulation:}

For a benchmark containing $N$ candidate objects, a subset of cardinality $k$
is sampled directly from the candidate pool according to the random-selection
procedure. Thus,
\begin{equation}
S_{\mathrm{rand}}
\sim
\operatorname{Uniform}
\left(
\left\{
S\subseteq V:\,|S|=k
\right\}
\right).
\end{equation}

Unlike the heuristic-driven strategies, \texttt{RANDOM\_SUBSET} bypasses the
GBS sampling stage and directly generates candidate $k$-subsets. These subsets
are then passed to the same downstream trajectory-optimization and
mission-assessment procedures used for the other methods.

\textbf{Operational Significance:}

The random baseline establishes a reference level for evaluating the effect of
heuristic-guided candidate selection. Differences observed downstream can
therefore be examined in the context of whether the candidate targets were
generated using object-level, network-level, provenance-aware, or random
selection mechanisms.

\subsection{Summary of Target-Prioritization Strategies}

For each heuristic $h$, the corresponding weighted graph is constructed
according to the procedure above and propagated through the GBS-based subset
generation stage. The resulting candidate subsets are subsequently passed to
the common trajectory-optimization framework. The \texttt{RANDOM\_SUBSET}
baseline follows the same downstream optimization and evaluation pipeline but
bypasses GBS during subset generation.

This design maintains a common mission-evaluation framework while varying the
mechanism used to construct candidate target subsets. The resulting
experimental comparisons characterize how alternative prioritization
strategies influence graph structure, GBS sampling behavior, target-subset
composition, and downstream ADR mission objectives. Table \ref{tab:heuristic_summary} summarizes the seven heuristics and their formulations.

\begin{table}[!h]
\centering
\caption{Summary of the target-selection strategies used in this study.}
\label{tab:heuristic_summary}
\renewcommand{\arraystretch}{1.5}
\resizebox{\textwidth}{!}{
\begin{tabular}{@{} p{3.5cm} p{4.8cm} p{3.0cm} p{4.0cm} p{4.8cm} @{}}
\toprule
\textbf{Heuristic} &
\textbf{Motivation} &
\textbf{Available Data} &
\textbf{Mathematical Formulation} &
\textbf{Operational Significance} \\
\midrule

\multicolumn{5}{@{}l}{\textit{\textbf{Physical and Environmental Characteristics}}} \\
\cmidrule(lr){1-5}

\hspace{0.3cm}\textbf{Physical Consequence}
&
Prioritize objects according to their surrogate physical consequence.
&
\texttt{OBJECT\_TYPE}, \texttt{RCS\_SIZE}
&
$P_{\mathrm{size}}(v_i)=\sqrt{\hat m_i}$
&
Emphasizes objects with larger estimated mass and corresponding fragmentation consequence.
\\

\hspace{0.3cm}\textbf{Persistence Index}
&
Prioritize objects according to the adopted orbital-persistence surrogate.
&
\texttt{BSTAR}, $n_i$
&
$P_{\mathrm{bstar}}(v_i)=\log_{10}(1+L_i)$
&
Emphasizes objects associated with larger values of the adopted persistence indicator.
\\

\hspace{0.3cm}\textbf{Operational Threat}
&
Combine physical consequence, persistence, and conjunction-risk information.
&
$\hat m_i$, $B_i^*$, $P_{c,i}$
&
$P_{\mathrm{risk}}(v_i)=\hat m_i\sqrt{P_{c,i}}\log_{10}(1+L_i)$
&
Provides a composite target-selection criterion incorporating environmental and conjunction-risk information.
\\

\addlinespace[0.3cm]

\multicolumn{5}{@{}l}{\textit{\textbf{Astrodynamic Network Structure}}} \\
\cmidrule(lr){1-5}

\hspace{0.3cm}\textbf{Transfer-Network Importance}
&
Identify structurally important objects within the weighted transfer network.
&
Pairwise transfer information
&
$P_{\mathrm{cent}}(v_i)=4PR+3EV+2BC$
&
Emphasizes global and corridor-level network structure.
\\

\hspace{0.3cm}\textbf{Orbital Cluster Membership}
&
Identify large and densely connected transfer communities.
&
Weighted transfer graph
&
$P_{\mathrm{comm}}(v_i)=|\mathcal C_r|\rho_r$
&
Emphasizes locally dense transfer regions.
\\

\addlinespace[0.3cm]

\multicolumn{5}{@{}l}{\textit{\textbf{Provenance-Aware Strategy}}} \\
\cmidrule(lr){1-5}

\hspace{0.3cm}\textbf{Source-Family Prioritization}
&
Incorporate inferred source-family structure into target selection.
&
Catalog naming and identifier metadata
&
$P_{\mathrm{src}}(v_i)=n_r$
&
Emphasizes larger inferred debris families and applies an additional same-family graph-weighting factor.
\\

\addlinespace[0.3cm]

\multicolumn{5}{@{}l}{\textit{\textbf{Baseline Control}}} \\
\cmidrule(lr){1-5}

\hspace{0.3cm}\textbf{Random Subset}
&
Provide a reference without heuristic-specific target prioritization.
&
None
&
$S_{\mathrm{rand}}\sim\operatorname{Uniform}\{S\subseteq V:|S|=k\}$
&
Provides a baseline for quantifying the effect of heuristic-guided subset generation.
\\

\bottomrule
\end{tabular}}
\end{table}
\section{Methodology}
\subsection{Problem Formulation}

ADR mission planning involves two coupled decision processes: selecting a
set of debris objects for servicing and determining the visitation order
through which those objects are serviced. Let

\begin{equation}
V=\{v_1,v_2,\ldots,v_N\}
\end{equation}

denote a catalog containing $N$ candidate debris objects. Each object is
associated with an orbital state and a set of mission-priority attributes
derived from the available catalog and conjunction data.

The proposed framework decomposes the ADR planning problem into two
hierarchical stages. The first stage performs combinatorial target-subset
discovery, while the second stage performs route evaluation and
multi-objective mission optimization for the selected subsets. This
decomposition avoids direct enumeration of the complete target-selection
space while retaining explicit astrodynamic and spacecraft feasibility
checks in the downstream mission layer.

\subsubsection{ADR Subset-Selection Problem}

Let

\begin{equation}
S\subseteq V,\qquad |S|=k
\end{equation}

denote a candidate debris subset containing exactly $k$ targets.

Each target is characterized by mission-relevant attributes including
estimated physical size or mass, orbital persistence, conjunction-related
risk, inferred source-family information, and graph-theoretic importance.
These attributes are converted into normalized heuristic priorities and
combined with pairwise orbital transfer accessibility.
The debris catalog is represented as a weighted orbital transfer graph

\begin{equation}
G=(V,E,W),
\end{equation}

where vertices correspond to debris objects and weighted edges represent
candidate transfer relationships between pairs of objects. The resulting
weighted adjacency matrix incorporates both pairwise transfer
accessibility and target-level priority information.

The direct combinatorial formulation of subset selection would require
consideration of

\begin{equation}
\binom{N}{k}
\end{equation}

possible target subsets. Rather than exhaustively evaluating this space,
the proposed framework uses GBS as a
probabilistic subset-discovery mechanism. The weighted orbital graph is
encoded into a Gaussian photonic state, and threshold-detection outcomes
are converted into candidate target subsets. The resulting GBS distribution preferentially samples graph configurations
with larger encoded connectivity weights. GBS acts as a
probabilistic reduction mechanism that supplies the downstream mission
optimizer with a limited collection of candidate target subsets.

Let

\begin{equation}
\mathcal{Q}
=
\{S_1,S_2,\ldots,S_m\}
\end{equation}

denote the candidate-subset pool produced after cardinality filtering and
duplicate removal. The resulting subsets are subsequently evaluated by
the astrodynamic mission-optimization layer.

\subsubsection{Multi-Objective Routing Problem}

For a selected subset

\begin{equation}
S=\{v_1,v_2,\ldots,v_k\},
\end{equation}

the second stage determines a visitation sequence

\begin{equation}
\pi
=
(v_{\pi_1},v_{\pi_2},\ldots,v_{\pi_k}),
\end{equation}

where $\pi$ is a permutation of the selected objects and every target is
visited exactly once.

The mission planner simultaneously minimizes the aggregate maneuvering
requirement and total mission duration. The first objective is

\begin{equation}
f_1(\pi)
=
\sum_{i=1}^{k-1}
\Delta v_{\pi_i,\pi_{i+1}},
\end{equation}

where $\Delta v_{\pi_i,\pi_{i+1}}$ denotes the modeled transfer cost of
the corresponding route leg.

The second objective is

\begin{equation}
f_2(\pi)
=
\sum_{i=1}^{k-1}
\left(
t_{\mathrm{wait},i}
+
TOF_i
+
t_{\mathrm{service}}
\right),
\end{equation}

where $t_{\mathrm{wait},i}$ is the passive phasing delay,
$TOF_i$ is the transfer time-of-flight, and $t_{\mathrm{service}}$ is
the fixed servicing allowance associated with each inter-target leg.

The resulting routing problem is

\begin{equation}
\min_{\pi}
\left(
f_1(\pi),f_2(\pi)
\right).
\label{eq:routing_problem}
\end{equation}

The objective pair is evaluated using Pareto dominance rather than a
single scalarized objective. A candidate solution is considered
non-dominated when no other feasible candidate has no larger value in both
objectives and a strictly smaller value in at least one.

The optimization is subject to spacecraft mass, per-leg maneuver,
and conjunction-timing conditions. These constraints are evaluated during
route fitness computation and during post-optimization mission scoring.

\subsection{Orbital-State Reconstruction and Transfer Modeling}

The orbital states of the catalog objects are reconstructed from their
TLE records. Each TLE pair is propagated using the
Simplified General Perturbations 4 (SGP4) model to obtain the initial
position and velocity state. The corresponding osculating orbital state
is then represented using the astrodynamic orbital model employed by the
implementation. For each object, the framework stores the semimajor axis $a$, inclination
$i$, RAAN $\Omega$, eccentricity $e$, orbital velocity, initial mean
anomaly, and secular RAAN-drift rate. These quantities form the basis of
the pairwise transfer and phasing calculations.

\subsubsection{Pairwise Transfer Model}

For every object pair $(v_i,v_j)$, the framework first evaluates whether
the inclination difference is within the prescribed mission bound:

\begin{equation}
\Delta i_{ij}
=
|i_i-i_j|.
\end{equation}

Pairs violating the inclination constraint are assigned infinite transfer
cost and are therefore excluded from the weighted transfer graph.

For admissible pairs, an intermediate transfer orbit is defined through

\begin{equation}
a_t
=
\frac{a_i+a_j}{2}.
\end{equation}

The first transfer impulse is computed from the corresponding
Hohmann-type energy transfer,

\begin{equation}
v_{t,1}
=
\sqrt{
\mu
\left(
\frac{2}{a_i}-\frac{1}{a_t}
\right)
},
\end{equation}

with associated first-burn magnitude

\begin{equation}
\Delta v_{1,ij}
=
|v_{t,1}-v_i|.
\end{equation}

The second transfer component includes the orbital-plane separation and
is computed using the velocity-triangle relation

\begin{equation}
\Delta v_{2,ij}
=
\sqrt{
v_j^2
+
v_{t,2}^2
-
2v_jv_{t,2}\cos(\Delta i_{ij})
},
\end{equation}

where

\begin{equation}
v_{t,2}
=
\sqrt{
\mu
\left(
\frac{2}{a_j}-\frac{1}{a_t}
\right)
}.
\end{equation}

The passive-transfer estimate is then

\begin{equation}
\Delta v^{\mathrm{passive}}_{ij}
=
\Delta v_{1,ij}
+
\Delta v_{2,ij}.
\end{equation}

The corresponding transfer time-of-flight is approximated by

\begin{equation}
TOF_{ij}
=
\frac{\pi\sqrt{a_t^3/\mu}}{86400},
\end{equation}

expressed in days.

This model is a Hohmann-type surrogate transfer model rather than a
full Lambert trajectory solution.

\subsubsection{Secular $J_2$ Phasing Model}

The time dependence of orbital-plane geometry is incorporated through
the secular nodal precession produced by Earth's oblateness. The
implementation uses the second zonal harmonic $J_2$ to calculate the
secular RAAN-drift rate

\begin{equation}
\dot{\Omega}_i
=
-\frac{3}{2}
\frac{
J_2\sqrt{\mu}R_E^2\cos i_i
}{
a_i^{7/2}(1-e_i^2)^2
}.
\end{equation}

For a given evaluation epoch, the RAAN of each object is propagated
according to its secular drift rate. The relative drift between objects
$i$ and $j$ is

\begin{equation}
\Delta\dot{\Omega}_{ij}
=
\dot{\Omega}_i-\dot{\Omega}_j.
\end{equation}

The next RAAN-alignment opportunity is estimated from the wrapped angular
difference and relative drift rate. The resulting RAAN waiting time is
capped at a predefined maximum of 120 days.

In addition to nodal alignment, the framework models in-plane phasing
through the relative evolution of mean anomaly. The corresponding
phasing time is obtained from the relative mean motion and capped at
30 days.

The total passive waiting time is then

\begin{equation}
t_{w,ij}
=
\alpha t_{\mathrm{RAAN},ij}
+
(1-\alpha)t_{\mathrm{phase},ij},
\end{equation}

where the implementation uses

\begin{equation}
\alpha=0.7.
\end{equation}

Thus, the dynamic leg evaluator combines secular orbital-plane evolution
and in-plane synchronization before determining whether passive or
active transfer is appropriate.

\subsection{Construction of the Orbital Transfer Network}

The objective of this stage is to transform pairwise astrodynamic
quantities into a weighted graph suitable for heuristic comparison and
GBS encoding.

For a feasible pair $(i,j)$, the implementation defines the composite
graph cost as

\begin{equation}
C_{ij}
=
\Delta v^{\mathrm{passive}}_{ij}
+
0.2\,\Delta i_{ij}^{\circ}
+
0.05\,\Delta\Omega_{ij}^{\circ}
+
0.01\,t_{w,ij},
\label{eq:graph_cost}
\end{equation}

where $\Delta i_{ij}^{\circ}$ and $\Delta\Omega_{ij}^{\circ}$ denote the
corresponding angular differences expressed in degrees.

Let

\begin{equation}
\sigma
=
\operatorname{median}
\left\{
\Delta v^{\mathrm{passive}}_{ij}:
\Delta v^{\mathrm{passive}}_{ij}<\infty
\right\}.
\end{equation}

Transfer cost is converted into an affinity through the exponential
kernel

\begin{equation}
K_{ij}
=
\exp\left(
-\frac{C_{ij}}{\sigma}
\right).
\end{equation}

\subsubsection{Heuristic-Weighted Adjacency Matrix}

Let

\begin{equation}
P_i^{(h)}\in[0,1]
\end{equation}

denote the normalized priority of object $i$ under heuristic $h$.

The heuristic-specific adjacency weight is constructed as

\begin{equation}
A_{ij}^{(h)}
=
\exp\left(
-\frac{C_{ij}}{\sigma}
\right)
\left(
P_i^{(h)}P_j^{(h)}
\right)^2
\epsilon_{ij},
\label{eq:heuristic_adjacency}
\end{equation}

where $\epsilon_{ij}$ is a multiplicative stochastic perturbation sampled
from the implementation's prescribed Gaussian jitter model.

For the \texttt{SOURCE} heuristic, an additional family-level factor is applied.
If two objects share the same non-unknown inferred source family, their
pre-normalized edge weight is multiplied by five.

The adjacency matrix is explicitly constructed symmetrically:

\begin{equation}
A_{ij}^{(h)}
=
A_{ji}^{(h)}.
\end{equation}

Finally, the matrix is spectrally normalized according to

\begin{equation}
\widetilde A^{(h)}
=
0.9
\frac{A^{(h)}}
{\|A^{(h)}\|_2},
\label{eq:spectral_normalization}
\end{equation}

whenever $\|A^{(h)}\|_2>0$.

This normalization bounds the spectral norm of the matrix below unity
before its use in the Gaussian photonic model.

\subsection{Quantum-Guided Subset Discovery}

For each heuristic that uses GBS, the normalized adjacency matrix
$\widetilde A^{(h)}$ is supplied to the GBS stage.

\subsubsection{GBS Graph Encoding}

The normalized adjacency matrix is eigendecomposed as

\begin{equation}
\widetilde A^{(h)}
=
U\Lambda U^T,
\end{equation}

where

\begin{equation}
\Lambda
=
\operatorname{diag}
(\lambda_1,\lambda_2,\ldots,\lambda_N)
\end{equation}

contains the eigenvalues and $U$ contains the corresponding eigenvectors.

The implementation uses the eigenvectors as the interferometer
transformation and maps calibrated eigenvalues to single-mode squeezing
parameters through

\begin{equation}
r_i
=
\operatorname{arctanh}(\lambda_i^{\mathrm{cal}}).
\end{equation}

The resulting Gaussian state is measured using threshold detection.
The associated sampling distribution is governed by the graph structure
and its corresponding hafnian-related Gaussian-sampling amplitudes.

\subsubsection{Photon-Number Calibration}

The target mission cardinality is $k$. To align the average photon
population with this cardinality, a global scaling factor $c$ is obtained
numerically.

Let

\begin{equation}
\mu_i=(c\lambda_i)^2.
\end{equation}

The expected photon number is

\begin{equation}
\mathbb{E}[n]
=
\sum_{i=1}^{N}
\frac{\mu_i}{1-\mu_i}
=
\sum_{i=1}^{N}
\frac{(c\lambda_i)^2}
{1-(c\lambda_i)^2}.
\end{equation}

The calibration factor is determined by solving

\begin{equation}
\sum_{i=1}^{N}
\frac{(c\lambda_i)^2}
{1-(c\lambda_i)^2}
=
k.
\label{eq:photon_calibration}
\end{equation}

The value of $c$ therefore targets an expected photon number of $k$;
it does not guarantee that every measurement outcome contains exactly
$k$ clicks.

\subsubsection{Threshold Detection and Cardinality Filtering}

For each optical mode $i$, threshold detection records

\begin{equation}
d_i=
\begin{cases}
1, & \text{if one or more photons are detected in mode }i,\\
0, & \text{otherwise}.
\end{cases}
\end{equation}

The detection pattern

\begin{equation}
\mathbf d=(d_1,d_2,\ldots,d_N)
\end{equation}

induces a target subset

\begin{equation}
S
=
\{v_i:d_i=1\}.
\end{equation}

Only outcomes satisfying

\begin{equation}
|S|=k
\end{equation}

are retained.

Duplicate subsets are removed using a set-based uniqueness test.


\subsubsection{Deterministic Fallback}

If no valid GBS subset is obtained, a deterministic fallback is invoked.

The weighted degree of vertex $i$ is

\begin{equation}
d_i
=
\sum_j \widetilde A_{ij}.
\end{equation}

The vertices with the highest weighted degrees are selected, using the
top $\max(2k,8)$ objects as the fallback candidate pool. All feasible
$k$-element combinations encountered from this pool are then appended
until the maximum candidate-subset count is reached.

This fallback is activated only when the GBS stage produces no valid
cardinality-$k$ subset. It is therefore a continuity mechanism for cases
in which the quantum sampling stage fails to produce a usable candidate.




\subsection{Multi-Objective Mission Optimization}

The candidate subsets generated by the preceding stage form the input to
the mission-planning optimizer.

\subsubsection{Route Representation}

For a subset

\begin{equation}
S=\{v_1,v_2,\ldots,v_k\},
\end{equation}

the visitation order is encoded using a continuous random-key vector

\begin{equation}
x=(x_1,x_2,\ldots,x_k),
\qquad
x_i\in[0,1].
\end{equation}

The route permutation is obtained by sorting the random keys:

\begin{equation}
\pi
=
\operatorname{argsort}(x).
\end{equation}

Thus, the evolutionary optimizer operates on a continuous encoding while
the induced mission route remains a discrete permutation of the selected
targets.

The implementation assumes that the spacecraft is initially co-located
with the first target in the induced permutation. Mission cost is
therefore accumulated from the first inter-target transfer onward.

\subsubsection{Dynamic Route Evaluation}

For each route leg $(a,b)$, the dynamic leg evaluator computes

\begin{equation}
\left(
\Delta v^{\mathrm{passive}}_{ab},
t_{w,ab},
\Delta v^{\mathrm{active}}_{ab}
\right).
\end{equation}

The passive transfer is initially preferred. The arrival time is estimated
as

\begin{equation}
t_{\mathrm{arr}}
=
t_{\mathrm{current}}
+
t_w
+
TOF
+
t_{\mathrm{service}}.
\end{equation}

For targets associated with a conjunction deadline, the corresponding
TCA is treated as a timing condition. When passive transfer results in an
arrival after the TCA, the evaluator replaces the passive maneuver with an
active-transfer alternative and sets the passive waiting time to zero.

If the active alternative still violates the TCA, the fitness function
introduces a lateness penalty. TCA handling consists of
passive evaluation, active fallback, and penalty-based treatment of
remaining deadline violations.

\subsubsection{Spacecraft Mass Feasibility}

The spacecraft mass is propagated using the Tsiolkovsky rocket equation,

\begin{equation}
m_{i+1}
=
m_i
\exp
\left(
-\frac{\Delta v_i\,1000}
{I_{sp}g_0}
\right),
\end{equation}

where $\Delta v_i$ is expressed in km/s in the implementation and the
factor $1000$ converts it to m/s.

A candidate route is feasible only if

\begin{equation}
m_i\geq m_{\mathrm{dry}}
\end{equation}

throughout mission evaluation.

Individual route legs are also restricted by the prescribed per-leg
maneuver bound. The resulting mass-depletion calculation provides the
primary propulsion-feasibility check.

\subsubsection{Bi-Objective Evolutionary Optimization}

The route evaluator exposes two objectives to the evolutionary optimizer:

\begin{equation}
f_1=\Delta v_{\mathrm{total}},
\qquad
f_2=t_{\mathrm{total}}.
\end{equation}

Two heterogeneous evolutionary islands are employed in a PyGMO
archipelago.

The first island uses NSGA-II with non-dominated sorting and
crowding-distance preservation. The second uses MOEA/D, which decomposes
the multi-objective problem into scalar subproblems.

Each island is initialized with a population of 20 candidate solutions and
evolved for 30 generations. The islands are connected using a ring
topology, allowing candidate information to migrate between the
optimization populations.

After evolution, candidate solutions with large penalty objective values
are discarded. The remaining candidates are converted from the continuous
random-key representation into explicit target permutations and are
reevaluated using the detailed mission-scoring procedure.









\subsubsection{Pareto-Front Construction}

For a set of feasible mission solutions $X$, a solution $x_1$ dominates
$x_2$ when
\begin{equation}
f_1(x_1)\leq f_1(x_2),
\qquad
f_2(x_1)\leq f_2(x_2),
\end{equation}
with at least one strict inequality.

The Pareto-optimal set is defined as
\begin{equation}
\mathcal{P}
=
\left\{
x\in X:
\forall y\in X,\; y\not\prec x
\right\},
\end{equation}
where $\prec$ denotes Pareto dominance. The corresponding Pareto front
in objective space is
\begin{equation}
\mathcal{F}
=
\left\{
\left(f_1(x),f_2(x)\right):
x\in\mathcal{P}
\right\}.
\end{equation}
\subsection{Launch-Epoch Sensitivity}

Because relative orbital geometry evolves with time, the route optimizer
is evaluated over multiple discrete launch epochs.

The mission-design reference epoch is

\begin{equation}
t_0=\text{2026-01-01T00:00:00 UTC},
\end{equation}

and launch epochs are sampled throughout the specified mission-design
window at 45-day intervals.

For each selected subset and heuristic, the evolutionary optimizer is
therefore executed independently at each sampled launch epoch. The
resulting mission solutions are subsequently pooled for heuristic-level
Pareto analysis.

This procedure represents a mission-design sensitivity analysis over
launch timing rather than an operational flight schedule.

\subsection{Integrated Quantum-Classical Optimization Workflow}

The complete framework combines orbital-state reconstruction, transfer
graph construction, heuristic weighting, Gaussian Boson Sampling, and
multi-objective evolutionary route optimization.\\

For each benchmark instance, the orbital catalog is first converted into
SGP4-derived orbital states. Pairwise transfer quantities, passive phasing
opportunities, and transfer times are then calculated. These quantities
are transformed into a weighted orbital-accessibility graph.\\

For each heuristic, object priorities are computed and incorporated into
the graph according to Equation~\eqref{eq:heuristic_adjacency}. The resulting
matrix is spectrally normalized and supplied to the GBS stage, except for
the \texttt{RANDOM\_SUBSET }baseline, which directly generates random
$k$-element subsets.\\

GBS measurement outcomes are filtered by cardinality and duplicate
removal. When no valid GBS subset is produced, the weighted-degree
fallback mechanism generates deterministic candidate subsets.
Each resulting subset is evaluated at multiple launch epochs. The
PyGMO archipelago then performs permutation-based multi-objective route
optimization for each subset and epoch. Feasible solutions are retained,
converted into explicit mission routes, and added to the heuristic-level
solution pool.\\

Finally, the solution pools from all evaluated configurations are used to
construct heuristic-specific Pareto fronts and a pooled global
non-dominated set. The resulting solutions are subsequently characterized
using the quantum-sampling, graph-theoretic, mission-level, astrodynamic,
and multi-objective metrics defined in the evaluation framework.\\

The architecture therefore performs a hierarchical reduction of the ADR
planning problem:

\begin{equation}
\begin{aligned}
\text{Orbital Catalog} 
&\rightarrow \text{Transfer Graph} 
\rightarrow \text{Heuristic Weighting} \\
&\rightarrow \text{GBS Subset Discovery} 
\rightarrow \text{Route Optimization} 
\rightarrow \text{Pareto Mission Set}.
\end{aligned}
\end{equation}
Algorithm~\ref{alg:framework} summarizes the overall execution flow of the proposed quantum-enhanced mission optimization framework.

\begin{algorithm}[h!]
\caption{Quantum-Enhanced ADR Mission Optimization Framework}
\label{alg:framework}
\footnotesize 
\linespread{0.88}\selectfont 

\KwIn{
Debris catalog $\mathcal{D}$,
conjunction data $\mathcal{C}$,
subset size $k$,
heuristic set $\mathcal{H}$
}

\KwOut{
Pooled non-dominated ADR mission architectures $\mathcal{P}_{global}$
}

Initialize global solution archive
$\mathcal{P}_{global}\leftarrow\emptyset$\;

\ForEach{benchmark seed $s$}{

    Construct benchmark debris pool $\mathcal{D}_s$\;

    Reconstruct orbital states from TLEs using SGP4\;

    Compute pairwise quantities:
    $\Delta v^{\mathrm{passive}}$,
    $\Delta i$,
    $\Delta\Omega$,
    $TOF$,
    and dynamic phasing quantities\;

    Compute inferred source-family labels\;

    Compute heuristic priority vectors $P^{(h)}$ for all $h\in\mathcal{H}$\;

    Construct Louvain communities for graph-based priorities\;

    Estimate the base-epoch waiting-time matrix $WAIT\_EST$\;


    Execute GBS sampling on the aggregate graph\;

    Compute aggregate GBS and graph-structural statistics\;

    \ForEach{heuristic $h\in\mathcal{H}$}{

        \eIf{$h=\texttt{RANDOM\_SUBSET}$}{

            Generate up to 20 random $k$-element subsets from $\mathcal{D}_s$\;

        }{

            Construct heuristic-specific graph cost:
            \[
            C_{ij} = \Delta v^{\mathrm{passive}}_{ij} + 0.2\Delta i_{ij}^{\circ} + 0.05\Delta\Omega_{ij}^{\circ} + 0.01t_{w,ij}
            \]

            Apply heuristic priority weighting\;

            \If{$h=\texttt{SOURCE}$}{
                Multiply same-family non-unknown edges by five\;
            }

            Add stochastic edge-weight perturbation\;

            Spectrally normalize the adjacency matrix\;

            Calibrate GBS squeezing to target expected photon number $k$\;

            Execute GBS threshold sampling\;

            Retain only outcomes satisfying $|S|=k$\;

            Remove duplicate subsets\;

            \If{no valid GBS subset was obtained}{
                Compute weighted vertex strengths $d_i=\sum_j\widetilde A_{ij}$\;

                Select top $\max(2k,8)$ vertices and generate $k$-element combinations\;

                Retain up to 20 fallback subsets\;
            }
        }

        \ForEach{candidate subset $S$}{

            \ForEach{launch epoch $t_0,t_0+45,t_0+90,\ldots$ in window}{

                Construct ADR route optimization problem for $S$\;

                Represent route ordering via random-key vector $x\in[0,1]^k$\;

                Decode route by sorting $x$ and evaluate leg passive phasing\;

                \If{passive arrival violates corresponding TCA}{

                    Replace passive transfer with active transfer ($t_{wait}=0$)\;

                    \If{active arrival still violates TCA}{
                        Add lateness penalty to fitness\;
                    }
                }

                Propagate mass via Tsiolkovsky rocket equation\;

                Reject solutions violating dry-mass or maneuver constraints\;

                Minimize $f_1=\Delta v_{\mathrm{total}}$ and $f_2=t_{\mathrm{total}}$ using PyGMO archipelago\;

                Initialize NSGA-II (pop 20) \& MOEA/D (pop 20) in ring topology\;

                Evolve islands for 30 generations with periodic migration\;

                Remove solutions with large penalty objectives\;

                Decode surviving vectors into explicit visitation permutations\;

                Reevaluate feasible routes and extract non-dominated set $\mathcal{P}_S$\;

                Append $\mathcal{P}_S$ to $\mathcal{P}_{global}$\;
            }
        }

        Construct the heuristic-specific Pareto set\;
    }
}

Perform final non-dominated sorting on $\mathcal{P}_{global}$\;

Construct pooled global non-dominated solution set\;

Compute required quantum, graph, astrodynamic, and Pareto metrics\;

\Return{$\mathcal{P}_{global}$}\;

\end{algorithm}
\section{Metrics}

The final stage of the framework performs a comprehensive evaluation of the
generated ADR missions. Since the proposed architecture combines
graph-theoretic modeling, GBS, astrodynamic mission
planning, and multi-objective optimization, no single metric can adequately
characterize overall performance. The evaluation therefore employs four
complementary classes of criteria: (i) quantum-sampling metrics,
(ii) graph-theoretic metrics, (iii) mission-level and astrodynamic metrics,
and (iv) multi-objective optimization metrics.

The objective of this phase is twofold. First, it quantifies the statistical
properties of the debris subsets generated by the GBS module. Second, it
evaluates the mission-level characteristics and feasibility of the ADR
architectures obtained through evolutionary optimization.

\subsection{Quantum-Sampling Evaluation}

The first category characterizes the statistical behavior of the GBS
process. GBS is a photonic quantum sampling procedure
in which a weighted adjacency matrix is encoded into a Gaussian optical
state and measurement outcomes correspond to subsets of graph vertices.
The sampled subsets therefore provide candidate debris collections for
downstream mission optimization.

Let
\[
\mathcal{Q} = \{S_1,S_2,\ldots,S_m\}
\]
denote the collection of sampled subsets, where each $S_i$ is a subset of
debris targets generated by the GBS process, and let $p_i$ denote the
empirical probability of observing subset $S_i$, estimated from the
observed sampling frequencies.
For a selected subset $S_i$, let $\mathbf{A}_{S_i}$ denote the adjacency
submatrix induced by the vertices contained in $S_i$. The \textit{Hafnian}
of a matrix is a combinatorial matrix function associated with perfect
matchings of the corresponding graph. In GBS, Hafnian amplitudes determine
the probability weights associated with photon-detection patterns for
appropriate classes of zero-displacement Gaussian states.

\subsubsection{Hafnian Concentration Ratio ($\text{HCR}$)}

The Hafnian Concentration Ratio measures the fraction of empirical sampling
probability concentrated in the five most frequently observed subsets:
\begin{equation}
\text{HCR}
=
\frac{\sum_{i=1}^{5}p_{(i)}}{\sum_j p_j},
\end{equation}
where $p_{(i)}$ denotes the $i$-th largest empirical probability.

\textit{Interpretation:} HCR measures the concentration of the GBS output
distribution. A high HCR indicates that a relatively small number of
candidate subsets account for a large fraction of the observed sampling
probability, corresponding to stronger exploitation of particular regions
of the encoded search space.

\subsubsection{Sampling Diversity ($\text{Div}$) and Catalog Coverage ($\text{Cov}$)}

Sampling Diversity measures the fraction of unique subsets among all
observed samples. Catalog Coverage measures the fraction of the complete
debris catalog represented by at least one sampled target:
\begin{equation}
\text{Div}
=
\frac{|\mathcal{Q}_{\text{unique}}|}
     {|\mathcal{Q}_{\text{total}}|},
\qquad
\text{Cov}
=
\frac{\left|\bigcup_{S_i\in\mathcal{Q}}S_i\right|}{N}.
\end{equation}

Here, $\mathcal{Q}_{\text{unique}}$ denotes the set of distinct sampled
subsets, $\mathcal{Q}_{\text{total}}$ denotes the complete collection of
observed samples, and $N$ is the total number of debris objects in the
evaluated catalog.

\textit{Interpretation:} Diversity characterizes the breadth of the sampled
subset space, whereas Coverage characterizes how broadly the sampled
targets span the available debris catalog.

\subsubsection{Effective Sample Size ($\text{ESS}$)}

The Effective Sample Size measures the effective number of equally weighted
samples represented by the empirical probability distribution. It is
computed from the inverse Simpson index:
\begin{equation}
\text{ESS}
=
\frac{1}{\sum_i p_i^2}.
\end{equation}

For a highly concentrated distribution, ESS is small because a small number
of subsets dominate the probability mass. Conversely, a more distributed
sampling probability yields a larger ESS.

\textit{Interpretation:} ESS therefore quantifies the effective breadth of
the sampled candidate distribution rather than simply counting the total
number of generated samples.

\subsubsection{Jaccard Similarity ($J_{\text{avg}}$)}

The Jaccard Similarity between two subsets $S_i$ and $S_j$ is defined as
the ratio of their intersection cardinality to the cardinality of their
union:
\begin{equation}
J(S_i,S_j)
=
\frac{|S_i\cap S_j|}
     {|S_i\cup S_j|}.
\end{equation}

The average Jaccard similarity is then
\begin{equation}
J_{\text{avg}}
=
\frac{1}{\binom{m}{2}}
\sum_{i<j}
\frac{|S_i\cap S_j|}
     {|S_i\cup S_j|}.
\end{equation}

Thus, $J_{\text{avg}}\in[0,1]$, with larger values indicating greater
overlap between sampled target subsets and smaller values indicating
greater subset-level diversity.

\textit{Interpretation:} When subsets are generated under different
selection heuristics, Jaccard similarity can additionally be used to
characterize the degree of agreement or divergence between the resulting
target selections.

\subsubsection{Log-Unadjusted Amplitude ($\log\text{UAmp}$)}

For a sampled subset $S_i$, the corresponding adjacency submatrix
$\mathbf{A}_{S_i}$ captures the weighted connectivity among the selected
targets. The internal connectivity of the subset is quantified by the sum of
its pairwise edge weights,
\begin{equation}
w(S_i)
=
\sum_{\substack{u,v\in S_i\\u<v}}
A_{uv}.
\end{equation}

To provide a reference for the connectivity of the sampled subsets, a random
baseline is computed from randomly selected subsets of the same cardinality.
Let $u_{\mathrm{rand}}$ denote the mean internal edge weight over $R$
randomly selected subsets. The log-unadjusted amplitude is then defined as
\begin{equation}
\log\text{UAmp}
=
\log_{10}
\left(
1+
\frac{
\frac{1}{m}\sum_{i=1}^{m} w(S_i)
}{
\max(u_{\mathrm{rand}},10^{-3})
}
\right).
\end{equation}

The logarithm is used to compress the numerical range of the normalized
connectivity ratio, while the random-subset reference provides a baseline
for comparison across sampled subsets.

\textit{Interpretation:} This metric characterizes the average internal
weighted connectivity of the sampled subgraphs relative to a random-subset
baseline. It therefore provides a graph-structural indicator of the degree
to which the sampled configurations exhibit stronger internal connectivity
than randomly selected subsets.

\subsubsection{Subgraph Entropy ($H_v$)}

For each debris object $k$, let $\nu_k$ denote its vertex marginal
probability, defined as the empirical frequency with which object $k$
appears across the sampled subsets. The corresponding Shannon entropy is
\begin{equation}
H_v
=
-\sum_{k=1}^{N}
\nu_k\log_2(\nu_k).
\end{equation}

Here, Shannon entropy quantifies the uncertainty or dispersion of the
vertex-selection distribution.

\textit{Interpretation:} A low $H_v$ indicates that sampling probability is
concentrated on a small number of debris objects, whereas a higher $H_v$
indicates that the sampled probability is distributed more broadly across
the catalog.

\subsection{Graph-Theoretic Evaluation}

The second category evaluates structural properties of the orbital transfer
subgraphs associated with the sampled target sets.

Let $A$ denote the weighted adjacency matrix of the orbital transfer graph.
For a subset $S$, $\bar S$ denotes its complement with respect to the
catalog, and $L=D-A$ denotes the graph Laplacian, where $D$ is the diagonal
degree matrix.

\subsubsection{Community Purity ($\text{Pur}$)}

The orbital transfer graph is partitioned into communities using the
Louvain community-detection method, which identifies groups of vertices
with comparatively dense internal connectivity. Let $\mathcal{C}$ denote
the set of identified communities. Community Purity is defined as
\begin{equation}
\text{Pur}
=
\max_{c\in\mathcal{C}}
\frac{|S\cap c|}{|S|}.
\end{equation}

\textit{Interpretation:} A high purity indicates that most selected targets
belong to the same graph community, whereas a lower value indicates that
the selected subset spans multiple communities.

\subsubsection{Graph Conductance ($\text{Cond}$)}

Graph Conductance measures the strength of the boundary separating a
subset from the remainder of the graph. It is defined as
\begin{equation}
\text{Cond}
=
\Phi(S)
=
\frac{\text{cut}(S,\bar S)}
     {\text{vol}(S)},
\end{equation}
where
\begin{equation}
\text{cut}(S,\bar S)
=
\sum_{i\in S}\sum_{j\notin S}A_{ij}
\end{equation}
is the total weight of edges connecting $S$ to its complement, and
\begin{equation}
\text{vol}(S)
=
\sum_{i\in S}\sum_j A_{ij}
\end{equation}
is the weighted volume of the subset.

\textit{Interpretation:} Lower conductance indicates that relatively less
weighted connectivity crosses the boundary of the selected subset,
indicating stronger structural separation from the remainder of the
network.

\subsubsection{Algebraic Connectivity ($\text{AlgC}$)}

Algebraic Connectivity is the second-smallest eigenvalue of the graph
Laplacian:
\begin{equation}
\text{AlgC}
=
\lambda_2(L).
\end{equation}

The quantity $\lambda_2(L)$ is commonly called the \textit{Fiedler
value}. It provides a spectral measure of graph connectivity. For a
connected graph, a larger value generally corresponds to stronger
connectivity.

\textit{Interpretation:} Within the sampled orbital subgraph, AlgC
characterizes the degree of structural connectivity among the selected
targets.

\subsubsection{Spectral Gap ($\text{SpGap}$)}

Let $\lambda_1(A)$ and $\lambda_2(A)$ denote the largest and second-largest
eigenvalues, respectively, of the adjacency matrix $A$. The spectral gap
is defined as
\begin{equation}
\text{SpGap}
=
\lambda_1(A)-\lambda_2(A).
\end{equation}

\textit{Interpretation:} The spectral gap quantifies the separation between
the dominant adjacency eigenmode and the next largest mode, providing an
indicator of the spectral organization and concentration of connectivity
within the sampled graph.

\subsection{Mission-Level and Astrodynamic Evaluation}

The third category evaluates the mission-level characteristics and
astrodynamic properties of the optimized ADR architectures.

\subsubsection{Hazard Removal Efficiency ($\text{HRE}$)}

Hazard Removal Efficiency measures the aggregate debris risk associated
with the selected targets per unit of active mission time:
\begin{equation}
\text{HRE}
=
\frac{\sum_{i\in S}\text{Risk}_i}
     {t_{\text{active}}}.
\end{equation}

Here, $\text{Risk}_i$ denotes the risk metric assigned to target $i$, and
$t_{\text{active}}$ denotes the portion of mission duration associated
with active mission operations rather than passive waiting.

\textit{Interpretation:} HRE characterizes the rate at which the selected
risk quantity is addressed relative to active operational time.

\subsubsection{Risk Capture Ratio ($\text{RCR}$)}

The Time of Closest Approach (TCA) denotes the predicted epoch associated
with the closest approach event used to define the corresponding
conjunction deadline. Risk Capture Ratio measures the fraction of the
selected risk whose targets are reached no later than their associated
TCA:
\begin{equation}
\text{RCR}
=
\frac{
\sum_{i:t_{\text{arr},i}\leq\text{TCA}_i}
\text{Risk}_i
}{
\sum_{j\in S}\text{Risk}_j
}.
\end{equation}

Here, $t_{\text{arr},i}$ is the modeled arrival time at target $i$.

\textit{Interpretation:} RCR quantifies the fraction of the selected risk
associated with targets that satisfy the specified TCA-based temporal
condition.

\subsubsection{Mission Robustness ($R$)}

Mission Robustness is evaluated through Monte Carlo perturbation analysis.
Monte Carlo analysis repeatedly evaluates a mission under randomly
perturbed conditions in order to characterize the resulting distribution
of mission costs. Let $\Delta v_{\text{MC}}$ denote the resulting
distribution of total $\Delta v$ values across the perturbation trials.
Then
\begin{equation}
R
=
1-
\frac{\sigma(\Delta v_{\text{MC}})}
     {\mu(\Delta v_{\text{MC}})}.
\end{equation}

Here, $\mu(\cdot)$ and $\sigma(\cdot)$ denote the sample mean and sample
standard deviation, respectively.

\textit{Interpretation:} The metric quantifies the relative dispersion of
the perturbed mission-cost distribution. A larger value corresponds to
lower relative variability under the perturbation model.

\subsubsection{Accessibility Index ($\text{AI}$)}

The Accessibility Index combines normalized transfer cost and waiting
time:
\begin{equation}
\text{AI}
=
\frac{1}{2}
\left(
1-\frac{\Delta v}{\Delta v_{\max}}
\right)
+
\frac{1}{2}
\left(
1-\frac{t_{\text{wait}}}{T_{\max}}
\right).
\end{equation}

Here, $\Delta v$ is the modeled mission transfer cost,
$\Delta v_{\max}$ is the corresponding normalization limit,
$t_{\text{wait}}$ is the cumulative passive waiting time, and
$T_{\max}$ is the waiting-time normalization limit.

\textit{Interpretation:} AI provides a normalized measure of accessibility
that jointly accounts for active maneuvering cost and passive phasing
requirements.

\subsubsection{Delta-V Entropy ($H_{\Delta v}$)}

For a mission containing $k-1$ inter-target transfers, let
\[
\tilde v_i
=
\frac{\Delta v_i}{\Delta v_{\text{total}}}
\]
denote the fraction of total $\Delta v$ assigned to leg $i$. The
corresponding Shannon entropy is
\begin{equation}
H_{\Delta v}
=
-\sum_{i=1}^{k-1}
\tilde v_i\log_2(\tilde v_i).
\end{equation}

\textit{Interpretation:} Lower entropy indicates a more concentrated
distribution of maneuver cost across the mission legs, whereas higher
entropy indicates a more evenly distributed allocation of $\Delta v$.

\subsubsection{Orbital Diversity ($\text{OD}$)}

Orbital Diversity measures the dispersion of the selected targets in
orbital-element space using the variance of inclination $i$ and RAAN, denoted by $\Omega$:
\begin{equation}
\text{OD}
=
\sigma_i^2+\sigma_\Omega^2.
\end{equation}

Here, $\sigma_i^2$ and $\sigma_\Omega^2$ are the sample variances of
inclination and RAAN among the selected targets.

\textit{Interpretation:} Larger OD indicates greater dispersion in these
two orbital elements, whereas smaller OD indicates greater concentration
in the corresponding orbital-element dimensions.

\subsubsection{Plane Change Burden ($\text{PCB}$)}

Plane Change Burden measures the fraction of total mission $\Delta v$
associated with out-of-plane maneuvering:
\begin{equation}
\text{PCB}
=
\frac{
\Delta v_{\text{out-of-plane}}
}{
\Delta v_{\text{total}}
}.
\end{equation}

Out-of-plane $\Delta v$ refers to the component of maneuvering associated
with correcting orbital-plane differences rather than changing orbital
energy alone.

\textit{Interpretation:} A larger PCB indicates that a greater fraction of
the mission transfer cost is associated with orbital-plane adjustment.

\subsubsection{$J_2$ Drift Utilization ($\eta_{\text{drift}}$)}

The second zonal harmonic coefficient $J_2$ represents the dominant
first-order effect of Earth's equatorial oblateness on the secular
evolution of orbital elements. In the present framework, differential
$J_2$-induced nodal precession can be exploited to passively modify the
relative RAAN between two orbital planes.

The fraction of total mission duration spent in passive waiting is defined
as
\begin{equation}
\eta_{\text{drift}}
=
\frac{t_{\text{wait}}}
     {t_{\text{total}}}.
\end{equation}

\textit{Interpretation:} Higher $\eta_{\text{drift}}$ indicates a larger
fraction of mission duration devoted to passive phasing rather than active
maneuvering.

\subsubsection{Family Diversity ($\text{FD}$)}

Each debris object is associated with a source-family label $F_i$, where
a family represents the inferred common source or fragmentation grouping
used by the benchmark construction. Family Diversity measures the number
of distinct source-family labels represented in a selected subset,
normalized by the subset size:
\begin{equation}
\text{FD}
=
\frac{
\left|
\bigcup_{i\in S}F_i
\right|
}{
|S|
}.
\end{equation}

\textit{Interpretation:} Lower FD indicates greater concentration of the
selected targets within a small number of source families, whereas higher
FD indicates representation from a larger number of distinct families.

\subsection{Multi-Objective Optimization Evaluation}

The final category evaluates the quality of the mission architectures
generated by the PyGMO evolutionary optimization framework. The missions
are evaluated simultaneously with respect to multiple objectives, rather
than through a single scalar objective.

\subsubsection{Number of Valid Missions ($N_{\text{valid}}$)}

Let $X$ denote the complete set of generated candidate mission
architectures and let $g_j(x)\leq0$ denote the $j$-th feasibility constraint
for candidate $x$. The number of valid missions is
\begin{equation}
N_{\text{valid}}
=
\left|
\left\{
x\in X
\mid
g_j(x)\leq0,\;
\forall j
\right\}
\right|.
\end{equation}

A valid mission is therefore a candidate satisfying all constraints
implemented by the mission-evaluation model.

\textit{Interpretation:} $N_{\text{valid}}$ measures how many generated
candidate architectures satisfy the imposed astrodynamic, temporal, mass,
and operational feasibility conditions.

\subsubsection{Hypervolume ($\text{HV}$)}

For a Pareto set $P$, hypervolume measures the volume of objective space
dominated by the set relative to a specified reference point
$z^{\text{ref}}$. The reference point is chosen so that it bounds the
objective region used for comparison. The hypervolume is defined as
\begin{equation}
\text{HV}(P)
=
\Lambda
\left(
\bigcup_{x\in P}
[f(x),z^{\text{ref}}]
\right),
\end{equation}
where $\Lambda$ denotes the Lebesgue measure, that is, the generalized
notion of volume used to measure the size of a set in objective space.

A \textit{Pareto-optimal} solution is a feasible solution for which no
other feasible solution is at least as good in every objective and
strictly better in at least one objective, under the adopted minimization
or maximization convention.

\textit{Interpretation:} Hypervolume jointly reflects the extent of the
objective-space region represented by the approximated Pareto front and
therefore captures both convergence and spread relative to the chosen
reference point.

\subsubsection{Global Pareto Front Hits ($\text{PF}_{\text{hits}}$)}

For heuristic $h$, let $P_h$ denote its Pareto front and let
$P_{\text{global}}^*$ denote the global non-dominated set obtained by
pooling the evaluated solutions across all strategies. Global Pareto
Front Hits are defined as
\begin{equation}
\text{PF}_{\text{hits}}
=
\left|
P_h
\cap
P_{\text{global}}^*
\right|.
\end{equation}

Here, the \textit{global Pareto front} is the set of solutions that remain
non-dominated after all candidate solutions from the evaluated strategies
are considered jointly.

\textit{Interpretation:} PF$_{\text{hits}}$ measures how many solutions
from a given heuristic also belong to the pooled non-dominated solution
set, thereby characterizing its participation in the global Pareto
solution space.\\
Table \ref{tab:comprehensive_metrics}
 summarizes these metrics alongwith their operational significance and desirable direction.

\begin{table}[h]
\centering
\caption{Comprehensive Evaluation Metrics Framework.}
\label{tab:comprehensive_metrics}
\renewcommand{\arraystretch}{1.5}
\setlength{\tabcolsep}{2pt}

\resizebox{\columnwidth}{!}{
\begin{tabular}{@{} l c l l c l @{}}
\toprule
\textbf{Evaluation Metric} &
\textbf{Symbol} &
\textbf{Mathematical Derivation} &
\textbf{Operational Significance} &
\textbf{Preferred Trend} &
\textbf{Interpretation} \\
\midrule

\multicolumn{6}{@{}l}{\textit{\textbf{Phase I: Quantum-Sampling Evaluation (Statistical Dynamics)}}} \\
\cmidrule(lr){1-6}

\hspace{0.2cm}Hafnian Concentration &
$\text{HCR}$ &
$\sum_{i=1}^{5}p_{(i)} / \sum p_j$ &
Measures concentration of probability mass in dominant sampled subsets. &
$\uparrow$ &
Greater sampling concentration. \\

\hspace{0.2cm}Sampling Diversity &
$\text{Div}$ &
$|\mathcal{Q}_{\text{unique}}| / |\mathcal{Q}_{\text{total}}|$ &
Measures the fraction of distinct subsets among observed samples. &
$\uparrow$ &
Broader subset exploration. \\

\hspace{0.2cm}Catalog Coverage &
$\text{Cov}$ &
$\left|\bigcup_{S_i}S_i\right|/N$ &
Measures the fraction of catalog objects represented in the sampled pool. &
$\uparrow$ &
Broader catalog representation. \\

\hspace{0.2cm}Effective Sample Size &
$\text{ESS}$ &
$1/\sum p_i^2$ &
Measures the effective breadth of the empirical sampling distribution. &
$\uparrow$ &
Lower probability concentration. \\

\hspace{0.2cm}Jaccard Similarity &
$J_{\text{avg}}$ &
Mean $\left(|S_i\cap S_j|/|S_i\cup S_j|\right)$ &
Measures overlap and agreement among generated target subsets. &
$\rightarrow$ &
Context-dependent. \\

\hspace{0.2cm}Log-Unadjusted Amp. &
$\log\text{UAmp}$ &
$\log_{10}\left(1+\overline{w}_S/\max(u_{\mathrm{rand}},10^{-3})\right)$ &
Characterizes the average internal weighted connectivity of sampled subgraphs relative to a random-subset baseline. &
$\uparrow$ &
Greater relative internal connectivity. \\

\hspace{0.2cm}Subgraph Entropy &
$H_v$ &
$-\sum\nu_k\log_2(\nu_k)$ &
Measures dispersion of vertex-selection probability across the catalog. &
$\rightarrow$ &
Context-dependent. \\

\addlinespace[0.2cm]

\multicolumn{6}{@{}l}{\textit{\textbf{Phase II: Graph-Theoretic Evaluation (Structural Cohesion)}}} \\
\cmidrule(lr){1-6}

\hspace{0.2cm}Community Purity &
$\text{Pur}$ &
$\max_c |S\cap c|/|S|$ &
Measures concentration of selected targets within a common graph community. &
$\rightarrow$ &
Higher concentration, but context-dependent. \\

\hspace{0.2cm}Graph Conductance &
$\text{Cond}$ &
$\text{cut}(S,\bar S)/\text{vol}(S)$ &
Measures the weighted boundary connectivity between the subset and its complement. &
$\downarrow$ &
Greater structural separation. \\

\hspace{0.2cm}Algebraic Connectivity &
$\text{AlgC}$ &
$\lambda_2(L)$ &
Quantifies structural connectivity within the selected subgraph. &
$\uparrow$ &
Stronger connectivity. \\

\hspace{0.2cm}Spectral Gap &
$\text{SpGap}$ &
$\lambda_1(A)-\lambda_2(A)$ &
Measures separation of the dominant spectral mode from the next-largest mode. &
$\rightarrow$ &
Context-dependent structural organization. \\

\addlinespace[0.2cm]

\multicolumn{6}{@{}l}{\textit{\textbf{Phase III: Mission-Level \& Astrodynamic Evaluation (Feasibility)}}} \\
\cmidrule(lr){1-6}

\hspace{0.2cm}Hazard Removal Eff. &
$\text{HRE}$ &
$\left(\sum \text{Risk}_i\right)/t_{\text{active}}$ &
Measures the rate of addressed risk relative to active mission time. &
$\uparrow$ &
Higher risk-address rate. \\

\hspace{0.2cm}Risk Capture Ratio &
$\text{RCR}$ &
$\text{Risk}_{\text{int}}/\text{Risk}_{\text{tot}}$ &
Measures the fraction of selected risk satisfying the TCA-based arrival condition. &
$\uparrow$ &
Greater timely risk capture. \\

\hspace{0.2cm}Mission Robustness &
$R$ &
$1-(\sigma_{\Delta v}/\mu_{\Delta v})$ &
Measures relative variability of mission $\Delta v$ under modeled perturbations. &
$\uparrow$ &
Lower relative sensitivity. \\

\hspace{0.2cm}Accessibility Index &
$\text{AI}$ &
Normalized $f(\Delta v,t_{\text{wait}})$ &
Combines normalized maneuver cost and passive waiting requirements. &
$\uparrow$ &
Greater modeled accessibility. \\

\hspace{0.2cm}Delta-V Entropy &
$H_{\Delta v}$ &
$-\sum\tilde v_i\log_2(\tilde v_i)$ &
Measures the distribution of $\Delta v$ across individual mission legs. &
$\rightarrow$ &
Context-dependent cost distribution. \\

\hspace{0.2cm}Orbital Diversity &
$\text{OD}$ &
$\sigma_i^2+\sigma_\Omega^2$ &
Measures dispersion of selected targets in inclination and RAAN. &
$\downarrow$ &
Lower orbital dispersion. \\

\hspace{0.2cm}Plane Change Burden &
$\text{PCB}$ &
$\Delta v_{\text{oop}}/\Delta v_{\text{total}}$ &
Measures the fraction of mission $\Delta v$ associated with out-of-plane maneuvering. &
$\downarrow$ &
Lower plane-change expenditure. \\

\hspace{0.2cm}$J_2$ Drift Utilization &
$\eta_{\text{drift}}$ &
$t_{\text{wait}}/t_{\text{total}}$ &
Measures the fraction of mission duration spent in passive phasing. &
$\rightarrow$ &
Context-dependent time--fuel trade-off. \\

\hspace{0.2cm}Family Diversity &
$\text{FD}$ &
$|\bigcup F_i|/|S|$ &
Measures dispersion of selected targets across inferred source families. &
$\rightarrow$ &
Context-dependent family concentration. \\

\addlinespace[0.3cm]

\multicolumn{6}{@{}l}{\textit{\textbf{Phase IV: Multi-Objective Optimization Evaluation (Pareto Quality \& Viability)}}} \\
\cmidrule(lr){1-6}

\hspace{0.2cm}Valid Missions &
$N_{\text{valid}}$ &
$\left|\{x\in X\mid g_j(x)\leq0,\ \forall j\}\right|$ &
Measures the number of generated architectures satisfying all imposed feasibility constraints. &
$\uparrow$ &
Greater feasible-solution yield. \\

\hspace{0.2cm}Hypervolume &
$\text{HV}$ &
$\Lambda\left(\bigcup[f(x),z^{\text{ref}}]\right)$ &
Measures the objective-space region dominated by the approximated Pareto front. &
$\uparrow$ &
Greater Pareto-front extent. \\

\hspace{0.2cm}Global PF Hits &
$\text{PF}_{\text{hits}}$ &
$\left|P_h\cap P_{\text{global}}^*\right|$ &
Measures participation of a heuristic's solutions in the pooled global non-dominated set. &
$\uparrow$ &
Greater global Pareto participation. \\

\bottomrule
\end{tabular}
}

\vspace{1mm}
\begin{flushleft}
\scriptsize
\textbf{Legend:}
$\uparrow$ Higher values indicate greater magnitude of the measured property;
$\downarrow$ Lower values indicate greater magnitude of the preferred property;
$\rightarrow$ No universally preferred direction; interpretation is context-dependent.
\end{flushleft}
\end{table}
\section{Experimental Details}
\label{sec:experimental_details}

\subsection{Physical Constants and Mission Parameters}

All experiments were performed using a common set of physical,
spacecraft, and mission parameters to maintain consistency across the
three benchmark environments, seven target-selection strategies, and
all $(N,k)$ scaling configurations. Orbital-state generation and
propagation employed the standard Earth gravitational parameter,
\[
\mu = 398600.4418~\mathrm{km^3\,s^{-2}},
\]
Earth's equatorial radius,
\[
R_E = 6378.137~\mathrm{km},
\]
the second zonal harmonic coefficient,
\[
J_2 = 1.08263\times10^{-3},
\]
and the standard gravitational acceleration,
\[
g_0 = 9.80665~\mathrm{m\,s^{-2}}.
\]
These constants were used throughout the orbital-state construction,
secular nodal-drift calculations, transfer estimation, and
propellant-mass propagation. TLE records were converted to orbital
states using the SGP4 propagator and subsequently represented using
the orbital elements required by the transfer and graph-construction
models.\\

The servicing spacecraft was initialized with a wet mass of
$9200~\mathrm{kg}$, a dry-mass floor of $500~\mathrm{kg}$, and a
constant specific impulse of $I_{sp}=300~\mathrm{s}$. The selected mass
values are treated as conceptual mission-design parameters rather than
as a direct representation of a particular flight vehicle. They were
chosen to provide a relatively high-capacity servicer appropriate for
a multi-target ADR campaign while maintaining a substantial propulsive
margin.\\


Spacecraft mass depletion after each transfer maneuver was computed
using the classical Tsiolkovsky rocket equation. Rather than imposing
an independent cumulative $\Delta v$ limit in the reported mission
model, cumulative propulsive expenditure was bounded implicitly through
the requirement that the propagated spacecraft mass remain above the
specified dry-mass floor. The resulting admissible cumulative
propulsive expenditure is therefore determined by the initial mass,
dry mass, and specific impulse through the rocket equation. This
formulation directly couples route-dependent maneuver expenditure to
remaining spacecraft mass and prevents solutions that exhaust the
available propellant reserve.\\

A fixed servicing interval of $t_{\mathrm{service}}=50$ days was added
to each inter-target leg. This interval is intentionally conservative
and represents a bundled mission-operation allowance rather than a
single maneuver duration. It is intended to accommodate the sequence
of proximity and rendezvous operations, target observation and
inspection, capture or berthing, servicing or removal operations, and
the possibility of repeated or contingency attempts before departure toward the next target. 
Importantly, it should
not be interpreted as a validated operational servicing timeline for
all ADR missions.\\

Mission feasibility was further constrained by the propagated
spacecraft dry-mass condition and by an inclination-difference
admissibility condition. Candidate transfer pairs for which the
inclination difference exceeded $100^\circ$ were rejected before
downstream optimization. Additional route feasibility was enforced
through finite-transfer checks and spacecraft-mass feasibility during
the evolutionary evaluation. For conjunction-sensitive targets, TCA
information from the SOCRATES catalogue was also incorporated into the
arrival-time evaluation when such information was available.\\

Table~\ref{tab:experimental_parameters} summarizes the principal
physical, spacecraft, mission, and optimization parameters.

\begin{table*}[!h]
\centering
\scriptsize
\caption{Experimental constants and implementation parameters used
throughout the benchmark simulations.}
\label{tab:experimental_parameters}
\renewcommand{\arraystretch}{1.25}
\resizebox{\textwidth}{!}{
\begin{tabular}{lll}
\toprule
\textbf{Category} & \textbf{Parameter} & \textbf{Value} \\
\midrule

\multirow{4}{*}{\textbf{Physical Constants}}
 & Earth's gravitational parameter ($\mu$)
 & $398600.4418~\mathrm{km^3\,s^{-2}}$ \\
 & Earth's equatorial radius ($R_E$)
 & $6378.137~\mathrm{km}$ \\
 & Earth's second zonal harmonic ($J_2$)
 & $1.08263\times10^{-3}$ \\
 & Standard gravity ($g_0$)
 & $9.80665~\mathrm{m\,s^{-2}}$ \\
\midrule

\multirow{4}{*}{\textbf{Spacecraft Parameters}}
 & Initial spacecraft wet mass
 & $9200~\mathrm{kg}$ \\
 & Dry-mass floor
 & $500~\mathrm{kg}$ \\
 & Specific impulse ($I_{sp}$)
 & $300~\mathrm{s}$ \\
 & Service interval per inter-target leg
 & $50~\mathrm{days}$ \\
\midrule

\multirow{3}{*}{\textbf{Mission Constraints}}
 & Mass-feasibility condition
 & $m_f \geq m_{\mathrm{dry}}$ \\
 & Maximum inclination difference
 & $100^\circ$ \\
 & Launch search window
 & $800~\mathrm{days}$ \\
\midrule

\multirow{5}{*}{\textbf{Experimental Design}}
 & Candidate-pool sizes ($N$)
 & $\{15,18,22,25\}$ \\
 & Target subset sizes ($k$)
 & $\{3,5\}$ \\
 & Benchmark environments
 & SOCRATES, COLLISION, RANDOM \\
 & Target-selection strategies
 & 7 strategies \\
 & Random seeds
 & $\{15,287,323,192,67\}$ \\
\midrule

\multirow{5}{*}{\textbf{Optimization Settings}}
 & GBS samples per execution
 & $400$ requested samples \\
 & Maximum unique subsets retained
 & $20$ \\
 & Evolutionary framework
 & PyGMO Archipelago \\
 & Evolutionary algorithms
 & NSGA-II + MOEA/D \\
 & Population size per island
 & $20$ \\
\bottomrule
\end{tabular}}
\end{table*}

\subsection{Experimental Benchmarks}

The proposed framework was evaluated under a systematic benchmark
protocol in which the candidate-pool size, target cardinality,
environmental structure, target-selection strategy, and stochastic
seed were varied while the spacecraft and physical parameters were
held fixed. This design isolates the effects of combinatorial scaling
and environmental topology from changes in the underlying mission
model.\\

The candidate-pool size was varied as
\[
N\in\{15,18,22,25\},
\]
while the number of targets selected for a mission was varied as
\[
k\in\{3,5\}.
\]
The Cartesian product therefore produced eight distinct $(N,k)$
scaling configurations. Each configuration was evaluated in three
benchmark environments using seven target-selection strategies. This
gives
\[
4\times2\times3\times7=168
\]
suite-scale-strategy configurations. Each configuration was
repeated using five predetermined stochastic seeds,
\[
\{15,287,323,192,67\},
\]
giving $840$ seeded experimental instances in total.\\

The three benchmark environments were generated using distinct
catalogue-selection procedures. For the Random suite, $N$
objects were sampled randomly from the available catalogue. For the
SOCRATES Minefield suite, objects were assigned a
conjunction-risk score from the SOCRATES data, the highest-risk tier
of 200 objects was retained, and $N$ objects were sampled from this
tier. For the Collision Cloud suite, objects were grouped
according to an inferred source-family label obtained from their
catalogue names, the most populated identifiable family was selected,
and $N$ objects were sampled from that family. The family label is
therefore an implementation-level provenance proxy derived from the
catalogue naming convention rather than an independently determined
genealogical reconstruction. The corresponding sampling procedures
were repeated independently for each experimental seed.\\

Within each benchmark environment, seven target-selection strategies
were evaluated. The four object-centric strategies were
\texttt{SIZE}, \texttt{BSTAR}, \texttt{SOCRATES\_RISK}, and
\texttt{SOURCE}. These were compared with the graph-theoretic
strategies \texttt{CENTRALITY} and \texttt{COMMUNITY}, together with
the \texttt{RANDOM\_SUBSET} baseline. The graph-theoretic methods
operate on the weighted orbital-transfer network, whereas
\texttt{RANDOM\_SUBSET} directly generates uniformly sampled target
subsets and therefore bypasses the GBS stage.\\

This distinction is important for interpreting the experimental
results. For the six graph- or object-informed strategies, the
corresponding priority values were incorporated into the graph used
for candidate-subset generation. The \texttt{RANDOM\_SUBSET} strategy
instead generated 20 random target subsets directly, providing a
non-GBS baseline against which the structured candidate-generation
mechanisms could be compared.

\subsection{Gaussian Boson Sampling Configuration}
\label{sec:gbs_config}

The GBS stage was implemented using the weighted orbital-transfer
graph generated for each benchmark configuration. For every
suite-scale-seed combination and every graph-informed prioritization
strategy, the weighted adjacency matrix was constructed and
spectrally normalized before Gaussian-state preparation. Because the
physical GBS model requires a symmetric matrix representation, the
quantum sampling graph is represented as an undirected weighted graph,
while the directional orbital-transfer quantities remain available
in the downstream mission-evaluation layer.\\

The weighted graph was eigendecomposed as
\[
A=U\Lambda U^{T}.
\]
The adjacency matrix was normalized using
\[
A_{\mathrm{norm}}
=
0.9\frac{A}{\|A\|_2},
\]
after which a scalar eigenvalue-scaling factor was obtained numerically
using Brent's root-finding method. The scaling was selected to satisfy
the prescribed mean-photon-number condition associated with the target
cardinality $k$. The resulting eigenvalues determine the single-mode
squeezing parameters, while the eigenvectors define the interferometer
transformation implemented in the Strawberry Fields Gaussian circuit \cite{killoran_strawberry_2019}.\\

A total of 400 GBS samples were requested for each execution using
threshold detectors. A sampled detection pattern was converted into a
candidate target subset by identifying the modes with non-zero
threshold detections and retaining only samples containing exactly $k$
distinct active modes. Duplicate subsets were removed. Candidate
collection terminated when 20 unique valid subsets had been obtained
or when the requested samples had been exhausted. The
number of valid raw samples available for the sampling-statistics
metrics can be smaller than 400 because the implementation may
terminate candidate collection early after reaching the 20-subset
limit. Sampling frequencies were retained from the valid raw samples for the
calculation of HCR, ESS, entropy, and related GBS statistics. The
retained candidate subsets were then forwarded to the evolutionary
mission-planning stage.\\

If a GBS execution failed to return a valid collection of target
subsets, the implementation invoked a deterministic graph-based
fallback. In this case, nodes were ranked according to their weighted
graph strength and candidate combinations were generated from the
highest-scoring nodes. This fallback was included to prevent numerical
or sampling failures from terminating the downstream optimization
pipeline. Table \ref{tab:gbs_parameters} summarizes the GBS execution model. 

\begin{table}[!h]
\centering
\caption{GBS implementation parameters used in the benchmark
experiments.}
\label{tab:gbs_parameters}
\renewcommand{\arraystretch}{1.3}
\begin{tabular}{ll}
\toprule
\textbf{Parameter} & \textbf{Value} \\
\midrule
Graph representation & Weighted symmetric adjacency matrix \\
Quantum model & Gaussian Boson Sampling \\
Detector model & Threshold detectors \\
Requested samples per execution & $400$ \\
Maximum unique subsets retained & $20$ \\
Valid subset criterion & Exactly $k$ distinct detected modes \\
Duplicate handling & Duplicate subsets discarded \\
Spectral normalization & $0.9\,\|A\|_2$ \\
Matrix decomposition & $A=U\Lambda U^T$ \\
State calibration & Brent's root-finding algorithm \\
Candidate ranking statistics & Sampling frequency \\
Fallback & Weighted-node graph combination search \\
\bottomrule
\end{tabular}
\end{table}

\subsection{Implementation of the Orbital Transfer Graph}
\label{sec:graph_implementation}

For each benchmark instance, the propagated debris catalogue was
transformed into a weighted orbital-transfer graph. Each vertex
represents a debris object, while each edge represents the estimated
accessibility of transferring between two objects. In the
implementation, the transfer graph is symmetric at the GBS stage and
therefore treated as an undirected weighted graph. The underlying
orbital quantities used to construct each transfer estimate include
transfer $\Delta v$, inclination difference, RAAN separation, and
passive-phasing wait time.\\

The graph construction uses the following composite transfer cost:
\[
C_{ij}
=
\Delta v^{\mathrm{passive}}_{ij}
+
0.2\,\Delta i_{ij}
+
0.05\,\Delta\Omega_{ij}
+
0.01\,t^{\mathrm{wait}}_{ij},
\]
where angular quantities are evaluated in degrees within the
implementation and $t^{\mathrm{wait}}_{ij}$ is the estimated passive
phasing time. The corresponding graph weight is generated using an
exponential kernel,
\[
w_{ij}
=
\exp\!\left(-\frac{C_{ij}}{\sigma}\right)
\left(p_i p_j\right)^2\xi_{ij},
\]
where $\sigma$ is the median of the finite passive-transfer-cost
values and $p_i,p_j$ are the normalized heuristic priorities.\\

The implementation additionally applies a source-family enhancement.
When a heuristic-specific graph is constructed with family labels,
two objects sharing the same non-\texttt{UNKNOWN} inferred
source-family label receive an explicit
\[
\times 5
\]
multiplicative edge-weight boost. This factor is applied within
the heuristic-specific graph construction, including the
\texttt{SOURCE} strategy. It is not applied to the separate aggregate
GBS summary graph, which is constructed using the
\texttt{CENTRALITY}-priority vector without source-family labels.\\

To introduce controlled stochastic variation while preserving the
dominant graph structure, multiplicative Gaussian jitter was applied
to the constructed edge weights:
\[
\xi_{ij}
=
\max\left(0.5,\mathcal{N}(1,0.05)\right).
\]
The lower bound prevents an individual perturbation from becoming
excessively small while maintaining a nominal unit-mean perturbation
with only $5\%$ standard deviation.\\

Three network descriptors were used to characterize the graph
structure and construct graph-based heuristics. PageRank, eigenvector
centrality, and betweenness centrality were computed to characterize
node importance, while Louvain community detection was used to
identify densely connected graph communities. These graph properties
were used both to generate target-selection priorities and to quantify
the structural properties of the sampled subgraphs.


\subsection{Dynamic Orbital Transfer and Phasing Model}
\label{sec:dynamic_transfer_model}

The mission-evaluation layer propagates the orbital state from the
reference epoch and incorporates secular $J_2$ nodal drift into
inter-target transfer evaluation. The secular RAAN drift rate is
computed from the semimajor axis, eccentricity, inclination, and
Earth's $J_2$ coefficient. At a given mission time, the model updates
the predicted RAAN of each object using its corresponding secular
drift rate.\\

For each transfer leg, passive waiting is determined by combining
relative RAAN drift and mean-anomaly phasing. In the implementation,
the waiting time is represented as
\[
t_{\mathrm{wait}}
=
0.7\,t_{\mathrm{RAAN}}
+
0.3\,t_{\mathrm{phase}},
\]
where the RAAN-alignment contribution is capped at 120 days and the
mean-anomaly phasing contribution is capped at 30 days. The passive
transfer cost and transfer time are subsequently propagated into the
mission chronology.\\

When a passive-phasing solution would result in an arrival after an
available TCA deadline, the implementation replaces the passive
solution with an active-maneuver transfer and sets the passive waiting
time to zero. The resulting arrival time is then reevaluated against
the deadline. This mechanism allows the mission planner to trade
propellant expenditure against temporal feasibility when servicing
deadlines are present.\\

The duration assigned to each inter-target leg is
\[
t_{\mathrm{leg}}
=
t_{\mathrm{wait}}
+
t_{\mathrm{tof}}
+
t_{\mathrm{service}},
\]
so that the fixed $50$-day servicing allocation is applied at every
inter-target stage. For a five-target mission, for example, four
inter-target legs are evaluated and therefore four servicing
intervals are included in the computed campaign duration.

\subsection{Evolutionary Optimization Configuration}
\label{sec:evolutionary_configuration}

The GBS-generated candidate target subsets were evaluated using the
PyGMO Archipelago framework (summarized in Table \ref{tab:evolutionary_parameters}). A two-island architecture was employed,
with one island executing NSGA-II and the second executing MOEA/D.
The two islands were connected through a ring topology and evolved
candidate solutions under identical spacecraft and mission
constraints. This architecture introduces algorithmic heterogeneity
while allowing the two search processes to exchange candidate
solutions through the archipelago mechanism.\\

Each island used a population of 20 individuals. NSGA-II was evolved
for 30 generations, while MOEA/D was likewise evolved for 30
generations. The optimization objectives were total mission
$\Delta v$ and total mission duration. Candidate solutions were
represented through continuous decision variables whose ordering
induces the corresponding target permutation. For each target subset
and launch epoch, the two optimization islands were evolved
independently before their populations were evaluated through the
full mission-scoring function.\\

The mission-scoring procedure propagated spacecraft mass after every
transfer using the rocket equation and rejected solutions that
violated the dry-mass feasibility condition. For selected targets
having available SOCRATES TCA information, arrival-time conditions
were evaluated during route scoring. The resulting feasible
solutions were retained for Pareto-front construction with respect to
the two optimization objectives.

\begin{table}[!h]
\centering
\caption{Evolutionary optimization parameters used throughout the
experiments.}
\label{tab:evolutionary_parameters}
\renewcommand{\arraystretch}{1.3}
\begin{tabular}{ll}
\toprule
\textbf{Parameter} & \textbf{Value} \\
\midrule
Optimization framework & PyGMO Archipelago \\
Archipelago topology & Ring \\
Number of islands & $2$ \\
Island 1 algorithm & NSGA-II \\
Island 2 algorithm & MOEA/D \\
Population size per island & $20$ \\
NSGA-II generations & $30$ \\
MOEA/D generations & $30$ \\
Optimization objectives & Minimize total $\Delta v$ and mission duration \\
Route representation & Continuous ordering variables \\
Mass feasibility & $m_f\geq m_{\mathrm{dry}}$ \\
Output & Feasible Pareto solutions \\
\bottomrule
\end{tabular}
\end{table}

\subsection{Launch Epoch Sampling}
\label{sec:launch_sampling}

To account for temporal variation in orbital geometry, each GBS
candidate subset was evaluated over multiple launch epochs rather than
a single departure date. The mission-design horizon began on
1 January 2026 (UTC) and extended over 800 days. Launch opportunities
were evaluated at uniform 45-day intervals. With the implementation
\[
\texttt{range}(0,800,45),
\]
the sampled offsets were
\[
0,45,90,\ldots,765~\mathrm{days},
\]
corresponding to 18 evaluated launch epochs.\\

At each launch epoch, the selected target subset was evaluated through
the dynamic transfer model and the PyGMO evolutionary optimizer. The
mission chronology was propagated from the selected launch epoch,
including passive-phasing delays, transfer times, servicing intervals,
and any active-maneuver fallback required by TCA constraints. The
resulting solutions therefore reflect the sensitivity of the mission
objectives to changes in orbital configuration over the modeled
mission-design horizon.\\

Because the same launch-epoch grid was applied across the benchmark
configurations, the comparison isolates the effects of candidate-pool
size, target cardinality, environmental topology, and
target-selection strategy while maintaining a common temporal
sampling protocol.

\subsection{Stochastic Repetition and Statistical Treatment}
\label{sec:stochastic_protocol}

The complete benchmark sweep was repeated using five predetermined
random seeds,
\[
\{15,287,323,192,67\}.
\]
The seeds control stochastic components of catalogue sampling, graph
perturbation, GBS sampling, Louvain community detection, and
evolutionary optimization. Performance statistics were subsequently
aggregated across the repeated runs for each suite, $(N,k)$
configuration, and selection strategy.\\

For the evaluation of GBS behavior, sampling-distribution statistics
including HCR, ESS, sampling diversity, coverage, entropy, and
Jaccard similarity were computed from the valid raw detection samples.
Mission-level statistics were obtained from the feasible solutions
returned by the evolutionary optimization stage. Pareto-front
performance was evaluated through hypervolume and related
multi-objective indicators.\\

The stochastic repetition therefore provides repeated realizations of
the same experimental configuration rather than independent physical
orbital environments. Reported averages should consequently be
interpreted as seed-aggregated computational performance measures
within the specified benchmark-generation procedure.

\subsection{Orbital Dynamics and Mission Execution Model}
\label{sec:dynamics_model}

Mission feasibility was evaluated using a high-fidelity orbital propagation and transfer model that combines analytical astrodynamics with realistic spacecraft operational constraints. Rather than assuming static orbital elements or simplified impulsive transfers, each candidate servicing mission was propagated dynamically throughout the optimization process to account for the time-varying geometry of the orbital environment.\\

All resident space objects were propagated using the SGP4 orbit propagator initialized from publicly available TLE data. Orbital evolution was evaluated at each candidate launch epoch, providing updated orbital states for subsequent transfer calculations. Secular perturbations due to the Earth's oblateness ($J_2$) were explicitly incorporated through analytical RAAN drift models, helping the optimizer to exploit naturally occurring orbital alignment opportunities instead of relying exclusively on propulsive maneuvers.\\

For each servicing leg, the optimizer first evaluated the feasibility of passive phasing through secular $J_2$ drift. When acceptable orbital alignment could be achieved within the allowable waiting horizon, the spacecraft remained in a drift phase until the required RAAN alignment was obtained, reducing the propellant required for the subsequent rendezvous maneuver. If passive alignment was insufficient to satisfy mission constraints, an active maneuver was performed to complete the transfer. This adaptive strategy enabled the optimizer to trade mission duration against propellant expenditure according to the selected Pareto objective.\\

Mission execution was additionally constrained by conjunction timing requirements and transfer deadlines to ensure that each servicing sequence remained operationally feasible throughout the optimization window. Following every transfer, the spacecraft mass was updated using the classical Tsiolkovsky rocket equation, allowing propellant consumption to propagate throughout the complete servicing campaign. All subsequent transfer calculations accounted for the continuously changing spacecraft mass, preserving physical consistency across the entire mission sequence.

\section{Results}
This section presents the empirical evaluation of the proposed framework, analyzed systematically across environmental suites, candidate pool sizes ($N$), target cardinalities ($k$), and selection heuristics.

\subsection{Suite-Wise Environmental Topology and Analysis}

The three benchmark environments were designed to represent distinct classes of orbital organization and therefore provide complementary tests of the proposed GBS-PyGMO framework. The SOCRATES Minefield represents a conjunction-risk-driven environment with structured hazard corridors, the Collision Cloud represents a fragmentation-driven environment with strong local orbital coherence, and the Random suite provides a topology-neutral stress test in which no explicit organizational structure is imposed. The following analysis examines how these environmental characteristics propagate through the graph-theoretic, quantum-sampling, astrodynamic, and mission-planning layers of the framework.

\subsubsection{The SOCRATES Minefield Environment Topology}

The SOCRATES Minefield environment represents a conjunction-risk-driven ADR scenario in which high-risk debris objects are distributed within structured orbital corridors. This environment therefore tests the ability of the framework to exploit risk-associated transfer structure while simultaneously satisfying the physical constraints of multi-target mission construction. Table~\ref{tab:socrates_all_metrics} summarizes the changes in graph topology, quantum-sampling behavior, astrodynamic characteristics, and mission performance when the problem scales from $N=15,\;k=3$ to $N=25,\;k=5$.\\

\begin{table}[h!]
\centering
\caption{Comprehensive Performance Profile for the SOCRATES Minefield Suite. Values represent global averages across all valid heuristic runs for the two bounding configurations.}
\label{tab:socrates_all_metrics}
\renewcommand{\arraystretch}{1.2}
\resizebox{\textwidth}{!}{
\begin{tabular}{llccc}
\toprule
\textbf{Category} & \textbf{Metric} & \textbf{Baseline ($N=15,k=3$)} & \textbf{Extreme Scale ($N=25,k=5$)} & \textbf{Change} \\
\midrule
\multirow{4}{*}{\textbf{Mission Output}}
 & Mean $\Delta v$ (km/s) & 3.53 & 6.67 & +89.0\% \\
 & Mean Duration (days) & 136.96 & 326.56 & +138.4\% \\
 & Mean Wait (\%) & 19.9\% & 35.4\% & +77.9\% \\
 & Mean Fuel Consumed (kg) & 5559.71 & 8056.42 & +44.9\% \\
\midrule
\multirow{11}{*}{\textbf{GBS / Graph Topology}}
 & Hafnian-Concentration Ratio (HCR) & 0.99 & 0.54 & -45.5\% \\
 & Effective Sample Size (ESS) & 1.7 & 10.9 & +541.2\% \\
 & Sampling Diversity (Div) & 0.24 & 0.71 & +195.8\% \\
 & Target Coverage (Cov) & 0.35 & 0.53 & +51.4\% \\
 & Community Purity (Pur) & 0.93 & 0.86 & -7.5\% \\
 & Utility Amplification ($\log U_{amp}$) & 1.04 & 0.91 & -12.5\% \\
 & Conductance (Cond) & 0.29 & 0.40 & +37.9\% \\
 & Algebraic Connectivity (AlgC) & 0.46 & 0.16 & -65.2\% \\
 & Spectral Gap (SpGap) & 0.70 & 0.45 & -35.7\% \\
 & Jaccard Similarity ($J_{avg}$) & 0.73 & 0.43 & -41.1\% \\
 & Subgraph Entropy ($H_v$) & 0.53 & 0.71 & +34.0\% \\
\midrule
\multirow{9}{*}{\textbf{Astrodynamic \& Risk}}
 & Orbital Dispersion (OD) & 0.20 & 0.20 & 0.0\% \\
 & Family Diversity (FD) & 0.21 & 0.28 & +33.3\% \\
 & Drift Efficiency ($\eta_{drift}$) & 0.33 & 0.17 & -48.5\% \\
 & Accessibility Index (AI) & 0.972 & 0.974 & +0.2\% \\
 & Plane Change Burden (PCB) & 0.33 & 0.32 & -3.0\% \\
 & Maneuver Entropy ($H_{\Delta v}$) & 0.61 & 0.80 & +31.1\% \\
 & Hazard Reduction Efficiency (HRE) & 0.13 & 0.17 & +30.8\% \\
 & Route Conjunction Reliability (RCR) & 0.55 & 0.26 & -52.7\% \\
 & Kinematic Robustness ($R$) & 0.96 & 0.97 & +1.0\% \\
\bottomrule
\end{tabular}}
\end{table}

A defining characteristic of the SOCRATES Minefield environment is the presence of organized risk corridors rather than a uniformly dispersed debris population. Despite the increase in problem scale, the aggregate orbital-dispersion and transfer-accessibility measures remain relatively stable. OD remains at $0.20$, AI changes only marginally ($0.972 \rightarrow 0.974$), and PCB remains essentially unchanged ($0.33 \rightarrow 0.32$). These observations indicate that the risk-associated targets retain relatively coherent orbital relationships across the two bounding configurations. From the perspective of mission construction, this provides a structured accessibility environment in which favorable transfer relationships remain available even as the campaign becomes more demanding.\\

Despite this geometric stability, mission-level costs increase substantially as the framework transitions from the baseline configuration $(N=15,k=3)$ to the extreme-scale configuration $(N=25,k=5)$. Mean mission $\Delta v$ rises from $3.53$ km/s to $6.67$ km/s, representing an $89\%$ increase, while mission duration expands from $136.96$ to $326.56$ days. Fuel consumption increases by approximately $45\%$. These changes occur despite the relatively stable OD, AI, and PCB values, indicating that the dominant scaling challenge is associated with constructing longer servicing sequences rather than with a major change in the aggregate accessibility of the environment. As the mission expands from three to five removals, favorable local transfer opportunities become progressively exhausted, requiring the optimizer to incorporate a larger fraction of less favorable links into the servicing sequence.\\

The graph metrics show that the SOCRATES environment produces a highly concentrated quantum-sampling landscape at small scale. For the baseline configuration $(N=15,k=3)$, HCR reaches $0.99$, while ESS remains only $1.7$. Together, these metrics indicate that probability mass is concentrated within a small number of dominant target subsets. This interpretation is also consistent with the high Pur ($0.93$), large SpGap ($0.70$), and relatively high AlgC ($0.46$), which characterize the strong local organization present in the transfer network. In this regime, the GBS sampler therefore concentrates on a limited number of highly coherent and high-accessibility candidate subgraphs.\\

As the problem scale increases, the sampling distribution undergoes a marked transition. HCR decreases by approximately $46\%$ ($0.99 \rightarrow 0.54$), while ESS increases from $1.7$ to $10.9$. Div increases from $0.24$ to $0.71$, Cov rises by $51\%$, and $H_v$ increases from $0.53$ to $0.71$. Collectively, these changes indicate a transition from concentrated exploitation toward broader exploration. The enlarged problem no longer concentrates sampling on a very small number of dominant target subsets. Instead, probability mass is distributed across a broader collection of competing candidate subgraphs.\\

The transition is also reflected in the structural metrics of the selected subgraphs. AlgC declines by approximately $65\%$, SpGap decreases by $36\%$, and $J_{avg}$ decreases from $0.73$ to $0.43$, while Cond increases from $0.29$ to $0.40$. These changes are consistent with a reduction in internal structural cohesion as increasingly diverse target combinations are incorporated into the candidate pool. Nevertheless, Pur remains high at $0.86$, indicating that the GBS distribution continues to retain a substantial association with organized regions of the transfer network rather than becoming fully diffuse.\\

The astrodynamic metrics reveal an additional scaling trade-off. Mission duration increases substantially while $\eta_{drift}$ decreases from $0.33$ to $0.17$. At first sight, this may appear counterintuitive because longer missions can provide additional opportunities for passive phasing. However, the simultaneous increase in HRE ($0.13 \rightarrow 0.17$), decline in RCR ($0.55 \rightarrow 0.26$), and increase in maneuver entropy ($0.61 \rightarrow 0.80$) are consistent with a shift toward more active and varied transfer strategies under increased sequence complexity. Within the evaluated configurations, the optimizer appears to accept greater transfer complexity in order to preserve servicing progress rather than relying exclusively on prolonged passive alignment.\\

The increase in $H_{\Delta v}$ further indicates that route construction becomes more heterogeneous as the servicing sequence expands. Instead of repeatedly using a small set of highly similar transfer patterns, the optimizer must accommodate a broader range of maneuver requirements to connect the selected targets. Nevertheless, $R$ remains nearly unchanged ($0.96 \rightarrow 0.97$), indicating that the generated solutions remain within the implemented kinematic feasibility criteria despite the increase in mission complexity.\\

Overall, the SOCRATES Minefield environment exhibits a structured risk-driven topology that supports concentrated quantum sampling at small scale but transitions toward broader exploration as the candidate population and servicing sequence are expanded. The principal scaling effects are observed at the mission level, where longer sequences lead to increased $\Delta v$, duration, maneuver diversity, and reduced conjunction reliability. Within the evaluated benchmark, the key challenge therefore shifts from identifying structured risk corridors to constructing longer feasible servicing chains while preserving accessibility and route reliability.

\subsubsection{The Collision Cloud Environment Topology}

The Collision Cloud suite represents a fragmentation-driven environment in which candidate objects are generated from a common breakup event and therefore exhibit strong local orbital coherence. Unlike the conjunction-driven SOCRATES Minefield, the dominant structure in this environment arises from shared fragmentation ancestry and the resulting similarity in orbital characteristics. The transfer network consequently contains dense local connectivity and favorable accessibility relationships. Table~\ref{tab:collision_all_metrics} summarizes the corresponding changes between the two bounding configurations.\\

\begin{table}[h!]
\centering
\caption{Comprehensive Performance Profile for the Collision Cloud Suite. Values represent global averages across all valid heuristic runs for the two bounding configurations.}
\label{tab:collision_all_metrics}
\renewcommand{\arraystretch}{1.2}
\resizebox{\textwidth}{!}{
\begin{tabular}{llccc}
\toprule
\textbf{Category} & \textbf{Metric} & \textbf{Baseline ($N=15, k=3$)} & \textbf{Extreme Scale ($N=25, k=5$)} & \textbf{Change} \\
\midrule
\multirow{4}{*}{\textbf{Mission Output}} 
 & Mean $\Delta v$ (km/s) & 3.30 & 5.58 & +69.1\% \\
 & Mean Duration (days) & 134.3 & 233.6 & +73.9\% \\
 & Mean Wait (\%) & 18.0\% & 11.7\% & -35.0\% \\
 & Mean Fuel Consumed (kg) & 5129.86 & 7383 & +43.9\% \\
\midrule
\multirow{10}{*}{\textbf{GBS / Graph Topology}} 
 & Hafnian-Concentration Ratio (HCR) & 0.94 & 0.51 & -45.7\% \\
 & Effective Sample Size (ESS) & 2.9 & 11.6 & +300.0\% \\
 & Sampling Diversity (Div) & 0.15 & 0.64 & +326.7\% \\
 & Target Coverage (Cov) & 0.35 & 0.61 & +74.3\% \\
 & Community Purity (Pur) & 1.00 & 0.89 & -11.0\% \\
 & Utility Amplification ($\log U_{amp}$) & 1.24 & 0.92 & -25.8\% \\
 & Conductance (Cond) & 0.32 & 0.43 & +34.4\% \\
 & Algebraic Connectivity (AlgC) & 0.56 & 0.15 & -73.2\% \\
 & Spectral Gap (SpGap) & 0.78 & 0.38 & -51.3\% \\
 & Jaccard Similarity ($J_{avg}$) & 0.70 & 0.43 & -38.6\% \\
 & Subgraph Entropy ($H_v$) & 0.53 & 0.74 & +39.6\% \\
\midrule
\multirow{9}{*}{\textbf{Astrodynamic \& Risk}} 
 & Orbital Dispersion (OD) & 0.10 & 0.20 & +100.0\% \\
 & Family Diversity (FD) & 0.00 & 0.00 & 0.0\% \\
 & Drift Efficiency ($\eta_{drift}$) & 0.32 & 0.21 & -34.4\% \\
 & Accessibility Index (AI) & 0.974 & 0.971 & -0.3\% \\
 & Plane Change Burden (PCB) & 0.20 & 0.20 & 0.0\% \\
 & Maneuver Entropy ($H_{\Delta v}$) & 0.59 & 0.79 & +33.9\% \\
 & Hazard Reduction Efficiency (HRE) & 0.11 & 0.16 & +45.5\% \\
 & Route Conjunction Reliability (RCR) & 0.70 & 0.24 & -65.7\% \\
 & Kinematic Robustness ($R$) & 0.96 & 0.97 & +1.0\% \\
\bottomrule
\end{tabular}}
\end{table}

A defining characteristic of the Collision Cloud environment is its strong orbital coherence. FD remains identically zero ($FD=0.00$), consistent with the single-fragmentation-family construction of the benchmark. OD remains low, increasing from $0.10$ to $0.20$, while PCB remains fixed at $0.20$ and AI changes only marginally ($0.974 \rightarrow 0.971$). Together, these metrics indicate that the target population remains concentrated within a relatively compact orbital neighborhood with favorable transfer geometry. In contrast to the SOCRATES Minefield, accessibility in the Collision Cloud is primarily inherited from the physical organization of the fragmentation event.\\

This structure produces a strongly concentrated quantum-sampling distribution at the baseline scale. At $(N=15,k=3)$, Pur reaches $1.00$, while $\log U_{amp}$ reaches $1.24$. AlgC ($0.56$) and SpGap ($0.78$) are also high, consistent with strong structural separation and connectivity within the transfer graph. The HCR of $0.94$, together with an ESS of only $2.9$ and Div of $0.15$, further indicates that the GBS sampler concentrates probability on a small number of highly coherent target subsets.\\

As the candidate pool and mission cardinality increase, the sampling landscape undergoes a gradual dilution. HCR decreases from $0.94$ to $0.51$, while ESS increases from $2.9$ to $11.6$. Div rises from $0.15$ to $0.64$, Cov increases from $0.35$ to $0.61$, and $H_v$ rises from $0.53$ to $0.74$. These changes indicate a transition from concentrated exploitation toward broader exploration. In this environment, the transition is consistent with the progressive incorporation of more peripheral fragments as the servicing sequence expands beyond the most locally coherent portion of the fragmentation cloud. Nevertheless, Pur remains high at $0.89$, indicating that fragmentation-derived structure continues to influence the selected subgraphs at the larger scale.\\

The reduction in topological cohesion is also reflected in AlgC, SpGap, and $J_{avg}$. AlgC decreases by approximately $73\%$ ($0.56 \rightarrow 0.15$), SpGap decreases by more than $50\%$ ($0.78 \rightarrow 0.38$), and $J_{avg}$ decreases from $0.70$ to $0.43$, while Cond increases from $0.32$ to $0.43$. These changes indicate that the selected target subsets become more structurally diverse as additional objects are incorporated into the servicing sequence.\\

The favorable transfer geometry is also reflected in mission-level performance. Mean mission $\Delta v$ increases from $3.30$ km/s to $5.58$ km/s, corresponding to a $69.1\%$ increase, while mission duration increases from $134.3$ to $233.6$ days. Although these changes are substantial, the average waiting fraction decreases from $18.0\%$ to $11.7\%$, while $\eta_{drift}$ decreases from $0.32$ to $0.21$. Within the evaluated configurations, this combination is consistent with greater use of direct transfers between nearby fragments rather than increased dependence on prolonged passive phasing.\\

The increase in $H_{\Delta v}$ ($0.59 \rightarrow 0.79$) indicates greater diversity in maneuver requirements as the servicing chain expands. At the same time, HRE increases from $0.11$ to $0.16$, whereas RCR decreases from $0.70$ to $0.24$. Thus, increasing mission scale produces greater environmental benefit but also introduces additional temporal and conjunction-related constraints.\\

Kinematic Robustness remains nearly unchanged ($0.96 \rightarrow 0.97$), indicating that the feasible missions generated under the implemented constraints remain kinematically admissible despite the increase in search-space complexity. The combination of low dispersion, low plane-change burden, and stable accessibility therefore provides a favorable structural basis for multi-target routing, although the mission burden still increases as more targets are incorporated into a single campaign.\\

Overall, the Collision Cloud environment exhibits strong fragmentation-driven organization, characterized by low orbital dispersion, zero family diversity, high purity, and favorable accessibility. As $N$ and $k$ increase, the GBS distribution broadens and the selected subgraphs become less cohesive, but the underlying fragmentation structure remains sufficiently pronounced to support direct transfer-based routing across a substantial portion of the candidate population.

\subsubsection{The Random Environment Topology}

The Random suite serves as a topology-neutral stress-test environment. Unlike the SOCRATES Minefield, which contains conjunction-risk structure, and the Collision Cloud, which exhibits fragmentation-driven orbital coherence, the Random catalog does not impose an explicit organizational structure based on collision risk, fragmentation ancestry, or transfer accessibility. This provides a complementary test of whether the proposed graph-based pipeline can identify locally favorable transfer structures within a heterogeneous candidate population. Table~\ref{tab:random_all_metrics} summarizes the behavior across the two bounding configurations.\\

\begin{table}[h!]
\centering
\caption{Comprehensive Performance Profile for the Random Suite. Values represent global averages across all valid heuristic runs for the two bounding configurations.}
\label{tab:random_all_metrics}
\renewcommand{\arraystretch}{1.2}
\resizebox{\textwidth}{!}{
\begin{tabular}{llccc}
\toprule
\textbf{Category} & \textbf{Metric} & \textbf{Baseline ($N=15, k=3$)} & \textbf{Extreme Scale ($N=25, k=5$)} & \textbf{Change} \\
\midrule
\multirow{4}{*}{\textbf{Mission Output}} 
 & Mean $\Delta v$ (km/s) & 4.24 & 6.67 & +57.3\% \\
 & Mean Duration (days) & 153.7 & 326.6 & +112.5\% \\
 & Mean Wait (\%) & 28.1\% & 35.4\% & +26.0\% \\
 & Mean Fuel Consumed (kg) & 5961 & 8056 & +35.1\% \\
\midrule
\multirow{10}{*}{\textbf{GBS / Graph Topology}} 
 & Hafnian-Concentration Ratio (HCR) & 0.85 & 0.48 & -43.5\% \\
 & Effective Sample Size (ESS) & 4.1 & 13.1 & +219.5\% \\
 & Sampling Diversity (Div) & 0.31 & 0.71 & +129.0\% \\
 & Target Coverage (Cov) & 0.51 & 0.50 & -2.0\% \\
 & Community Purity (Pur) & 0.99 & 0.89 & -10.1\% \\
 & Utility Amplification ($\log U_{amp}$) & 0.99 & 0.96 & -3.0\% \\
 & Conductance (Cond) & 0.51 & 0.46 & -9.8\% \\
 & Algebraic Connectivity (AlgC) & 0.37 & 0.25 & -32.4\% \\
 & Spectral Gap (SpGap) & 0.59 & 0.51 & -13.6\% \\
 & Jaccard Similarity ($J_{avg}$) & 0.52 & 0.47 & -9.6\% \\
 & Subgraph Entropy ($H_v$) & 0.63 & 0.69 & +9.5\% \\
\midrule
\multirow{9}{*}{\textbf{Astrodynamic \& Risk}} 
 & Orbital Dispersion (OD) & 1.10 & 1.40 & +27.3\% \\
 & Family Diversity (FD) & 0.73 & 0.79 & +8.2\% \\
 & Drift Efficiency ($\eta_{drift}$) & 0.43 & 0.51 & +18.6\% \\
 & Accessibility Index (AI) & 0.961 & 0.915 & -4.8\% \\
 & Plane Change Burden (PCB) & 0.45 & 0.47 & +4.4\% \\
 & Maneuver Entropy ($H_{\Delta v}$) & 0.64 & 0.74 & +15.6\% \\
 & Hazard Reduction Efficiency (HRE) & 0.19 & 0.17 & -10.5\% \\
 & Route Conjunction Reliability (RCR) & 0.62 & 0.35 & -43.5\% \\
 & Kinematic Robustness ($R$) & 0.96 & 0.97 & +1.0\% \\
\bottomrule
\end{tabular}}
\end{table}

The Random suite represents the least explicitly organized benchmark environment considered in the study. Targets are sampled from heterogeneous orbital populations without regard to conjunction risk, fragmentation lineage, or transfer accessibility. The resulting environment therefore provides a stress test of the ability of the weighted transfer graph and downstream optimization framework to identify locally coherent servicing opportunities without relying on predefined environmental structure.\\

The astrodynamic metrics show that the Random environment contains the largest orbital dispersion among the three suites. OD increases from $1.10$ to $1.40$, while FD increases from $0.73$ to $0.79$. PCB also remains relatively high ($0.45 \rightarrow 0.47$), indicating larger geometric separation and more demanding plane-change requirements than those observed in the more organized Collision Cloud environment. AI decreases from $0.961$ to $0.915$ as the problem scale increases, consistent with progressively less favorable direct accessibility as additional targets are incorporated into the servicing sequence.\\

The GBS metrics further characterize the diffuse nature of the sampling landscape. At $(N=15,k=3)$, HCR is $0.85$, considerably lower than the baseline values obtained for the SOCRATES Minefield and Collision Cloud suites. ESS is already $4.1$, while Div reaches $0.31$, Cov is $0.51$, and $H_v$ is $0.63$. These values indicate that probability mass is distributed over a broader range of candidate subsets rather than concentrated within a small number of overwhelmingly dominant structures.\\

 Pur remains high at $0.99$, even in the baseline Random environment. This should not be interpreted as evidence for physical fragmentation communities. Rather, it indicates that the weighted transfer graph can still contain locally coherent connectivity structures even when the catalog itself has no explicitly imposed organizational pattern. In this case, the identified local structure arises from transfer accessibility encoded in the graph rather than from common physical ancestry or risk concentration.\\

At the extreme-scale configuration $(N=25,k=5)$, HCR decreases further from $0.85$ to $0.48$, while ESS increases from $4.1$ to $13.1$. Div rises from $0.31$ to $0.71$, and $H_v$ increases from $0.63$ to $0.69$. Cov remains effectively unchanged ($0.51 \rightarrow 0.50$), indicating that the increase in sampling diversity does not result from a comparable increase in coverage. Instead, the GBS distribution explores a broader set of competing target combinations within a similarly sized portion of the candidate population.\\

The spectral metrics reinforce this interpretation. AlgC decreases from $0.37$ to $0.25$, while SpGap decreases from $0.59$ to $0.51$. These changes are more modest than the corresponding reductions observed in the structured environments, consistent with the fact that the Random environment begins from a comparatively weakly organized topology rather than from a single dominant community. The modest decline in $J_{avg}$ from $0.52$ to $0.47$ likewise indicates continued differentiation among the target subsets generated by the different heuristics.\\

The mission-level consequences of this topological disorder are substantial. Mean mission $\Delta v$ increases from $4.24$ km/s to $6.67$ km/s, while mission duration increases from $153.7$ to $326.6$ days. Fuel consumption increases by $35.1\%$. At the same time, the waiting fraction increases from $28.1\%$ to $35.4\%$, and $\eta_{drift}$ rises from $0.43$ to $0.51$. The simultaneous increase in waiting and drift efficiency is consistent with greater use of passive $J_2$-assisted phasing to overcome unfavorable direct-transfer geometry.\\

The increase in $H_{\Delta v}$ from $0.64$ to $0.74$ indicates that the optimizer employs a wider range of maneuver requirements as the mission becomes more complex. HRE decreases from $0.19$ to $0.17$, while RCR decreases from $0.62$ to $0.35$. Thus, within this benchmark, increasing mission complexity does not produce a proportional increase in hazard-reduction efficiency and is accompanied by lower route-conjunction reliability.\\

Despite these changes, Kinematic Robustness remains effectively constant at approximately $0.96$-$0.97$. This indicates that the optimization framework continues to generate solutions satisfying the implemented kinematic constraints even as the environment becomes more heterogeneous and mission duration increases. The result is consistent with a gradual degradation in mission efficiency rather than a complete loss of feasible routing.\\

Overall, the Random suite exhibits a comparatively diffuse sampling landscape, elevated orbital dispersion, reduced direct accessibility, and greater reliance on passive phasing. The framework nevertheless continues to identify locally coherent transfer neighborhoods through the weighted orbital graph, indicating that graph-based candidate generation can retain structured information even when the underlying catalog lacks explicit physical organization.

\subsubsection{Concluding Quantitative Synthesis of Environmental Topologies}

The aggregated results reveal a consistent relationship between environmental topology, graph structure, quantum-sampling behavior, and downstream mission characteristics. Table~\ref{tab:topology_synthesis_uniform} presents the minimum and maximum observed values across the full set of candidate-pool sizes, target cardinalities, and heuristic configurations for the three benchmark environments.

\begin{table}[!h]
\centering
\caption{Uniform Quantitative Synthesis of Orbital Topologies. Values denote the minimum and maximum observed average values across all pool sizes ($N$), subset cardinalities ($k$), and heuristic configurations.}
\label{tab:topology_synthesis_uniform}
\renewcommand{\arraystretch}{1.4}
\resizebox{\textwidth}{!}{
\begin{tabular}{l | l | l | l | p{5.8cm}}
\toprule
\textbf{Environment} &
\textbf{GBS / Graph Topology} &
\textbf{Astrodynamic Characteristics} &
\textbf{Mission Performance Bounds} &
\textbf{Operational Interpretation} \\
\midrule

\multirow{4}{*}{\textbf{Collision Cloud}} 
& $\mathbf{HCR:}\;0.51-0.94$
& $\mathbf{OD:}\;0.10-0.20$
& $\mathbf{Mean~\Delta v:}\;1.29-6.81~\mathrm{km/s}$
& \multirow{4}{5.8cm}{%
Dense fragmentation communities support localized transfers with favorable accessibility and comparatively low reliance on passive orbital drift.
} \\

& $\mathbf{ESS:}\;2.9-11.6$
& $\mathbf{FD:}\;0.00$
& $\mathbf{\eta_{drift}:}\;0.21-0.36$
& \\

& $\mathbf{Pur:}\;0.89-1.00$
& $\mathbf{PCB:}\;0.12-0.28$
& $\mathbf{RCR:}\;0.21-0.70$
& \\

& $\mathbf{J_{avg}:}\;0.35-0.74$
& $\mathbf{AI:}\;0.963-0.981$
& $\mathbf{Mean~Duration:}\;112.4-265.1~\mathrm{days}$
& \\

\midrule

\multirow{4}{*}{\textbf{SOCRATES Minefield}} 
& $\mathbf{HCR:}\;0.53-0.99$
& $\mathbf{OD:}\;0.10-0.30$
& $\mathbf{Mean~\Delta v:}\;1.74-7.59~\mathrm{km/s}$
& \multirow{4}{5.8cm}{%
Risk-concentrated hazard corridors create strong sampling concentration while requiring transfer and phasing across multiple structured target regions.
} \\

& $\mathbf{ESS:}\;1.7-11.6$
& $\mathbf{FD:}\;0.21-0.38$
& $\mathbf{\eta_{drift}:}\;0.16-0.36$
& \\

& $\mathbf{Pur:}\;0.86-1.00$
& $\mathbf{PCB:}\;0.29-0.40$
& $\mathbf{RCR:}\;0.25-0.68$
& \\

& $\mathbf{J_{avg}:}\;0.31-0.75$
& $\mathbf{AI:}\;0.969-0.975$
& $\mathbf{Mean~Duration:}\;118.5-289.8~\mathrm{days}$
& \\

\midrule

\multirow{4}{*}{\textbf{Random}} 
& $\mathbf{HCR:}\;0.48-0.85$
& $\mathbf{OD:}\;0.60-2.60$
& $\mathbf{Mean~\Delta v:}\;1.35-7.75~\mathrm{km/s}$
& \multirow{4}{5.8cm}{%
Topologically heterogeneous populations require greater exploration, greater transfer dispersion, and stronger reliance on passive orbital phasing.
} \\

& $\mathbf{ESS:}\;4.1-13.1$
& $\mathbf{FD:}\;0.63-0.95$
& $\mathbf{\eta_{drift}:}\;0.38-0.52$
& \\

& $\mathbf{Pur:}\;0.89-0.99$
& $\mathbf{PCB:}\;0.39-0.50$
& $\mathbf{RCR:}\;0.33-0.68$
& \\

& $\mathbf{J_{avg}:}\;0.36-0.64$
& $\mathbf{AI:}\;0.895-0.961$
& $\mathbf{Mean~Duration:}\;124.1-462.1~\mathrm{days}$
& \\

\bottomrule
\end{tabular}
}
\end{table}

A clear structural differentiation emerges across the three environments. The Collision Cloud exhibits the lowest orbital dispersion ($OD=0.10-0.20$), zero family diversity ($FD=0$), high Purity ($0.89-1.00$), and favorable accessibility ($AI=0.963-0.981$). These properties are consistent with a fragmentation-driven transfer network containing dense local communities and comparatively favorable inter-target geometry. Across the evaluated design space, the corresponding mission ranges include mean $\Delta v$ values of $1.29-6.81$ km/s and mean durations of $112.4-265.1$ days.\\

The SOCRATES Minefield occupies a structurally different regime. Its OD range remains relatively low ($0.10-0.30$), while the HCR range of $0.53-0.99$ and Purity range of $0.86-1.00$ indicate persistent concentration around structured hazard-related regions. At the same time, its PCB range of $0.29-0.40$ and AI range of $0.969-0.975$ indicate that favorable transfer accessibility coexists with greater geometric separation between some of the selected targets. The resulting mission behavior therefore reflects a balance between concentrated hazard structure and the increasing routing burden associated with larger servicing campaigns.\\

The Random environment represents the comparatively least organized regime. Its OD range of $0.60-2.60$, FD range of $0.63-0.95$, PCB range of $0.39-0.50$, and AI range of $0.895-0.961$ indicate greater orbital heterogeneity and less uniform accessibility. Its HCR range of $0.48-0.85$, ESS range of $4.1-13.1$, and broader entropy range are consistent with a more diffuse GBS sampling distribution. The higher drift-efficiency range ($0.38-0.52$) is likewise consistent with greater reliance on passive $J_2$-assisted phasing in the presence of less favorable direct-transfer geometry.\\

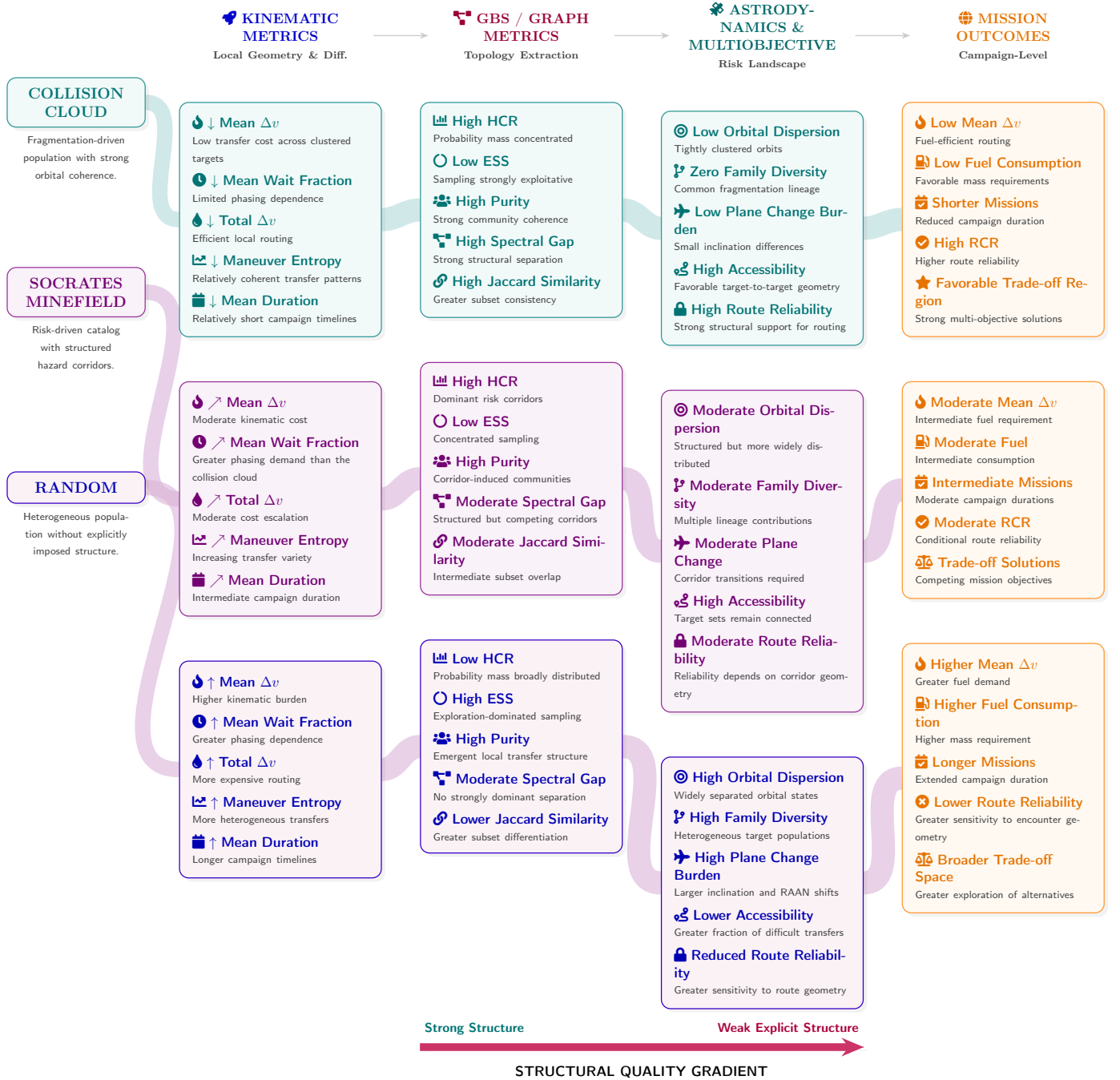
\begin{figure}[!h]
\centering

\newcommand{\meta}[1]{{\scriptsize\color{darkgray}#1}}

\resizebox{\textwidth}{!}{%
\begin{tikzpicture}[
    font=\sffamily,
    colTeal/.style={draw=teal!80, fill=teal!5, text=teal!90!black},
    colSoc/.style={draw=violet!80, fill=violet!5, text=violet!90!black},
    colRan/.style={draw=blue!70!purple, fill=blue!5!purple!5, text=blue!70!black},
    colOut/.style={draw=orange!80, fill=orange!5, text=orange!90!black},
    uiCard/.style={
        rounded corners=1.5ex,
        thick,
        text width=4.8cm,
        align=left,
        inner sep=10pt,
        drop shadow={shadow xshift=3pt, shadow yshift=-3pt, opacity=0.1}
    },
    envBadge/.style={
        rounded corners=1.5ex,
        thick,
        align=center,
        text width=3.2cm,
        inner sep=8pt,
        font=\bfseries\large,
        drop shadow={shadow xshift=3pt, shadow yshift=-3pt, opacity=0.1}
    },
    headerBox/.style={
        text width=4.8cm,
        align=center,
        font=\bfseries\normalsize
    },
    ribbonTeal/.style={
        teal!30,
        line width=18pt,
        opacity=0.6
    },
    ribbonSoc/.style={
        violet!30,
        line width=18pt,
        opacity=0.6
    },
    ribbonRan/.style={
        blue!30!purple!30,
        line width=18pt,
        opacity=0.6
    }
]


\node[headerBox] (H1) {};

\node[
    headerBox,
    right=0.5cm of H1,
    text=blue!80!black
] (H2) {
    \faRocket\ KINEMATIC METRICS\\
    \meta{Local Geometry \& Diff.}
};

\node[
    headerBox,
    right=1.5cm of H2,
    text=purple!80!black
] (H3) {
    \faProjectDiagram\ GBS / GRAPH METRICS\\
    \meta{Topology Extraction}
};

\node[
    headerBox,
    right=1.5cm of H3,
    text=teal!80!black
] (H4) {
    \faSatellite\ ASTRODYNAMICS \&\\
    MULTIOBJECTIVE\\
    \meta{Risk Landscape}
};

\node[
    headerBox,
    right=1.5cm of H4,
    text=orange!90!black
] (H5) {
    \faGlobe\ MISSION OUTCOMES\\
    \meta{Campaign-Level}
};

\draw[-{Stealth}, thick, gray!40] (H2.east) -- (H3.west);
\draw[-{Stealth}, thick, gray!40] (H3.east) -- (H4.west);
\draw[-{Stealth}, thick, gray!40] (H4.east) -- (H5.west);


\node[
    envBadge,
    colTeal,
    below=1cm of H1
] (Env1) {
    COLLISION\\CLOUD
};

\node[
    below=0.1cm of Env1,
    text width=3.2cm,
    align=center
] {
    \meta{Fragmentation-driven population with strong orbital coherence.}
};

\node[
    uiCard,
    colTeal,
    below=1cm of H2
] (K1) {
    \faFire\ \textbf{$\downarrow$ Mean $\Delta v$}\\[-0.2ex]
    \meta{Low transfer cost across clustered targets}\\[1ex]
    \faClock\ \textbf{$\downarrow$ Mean Wait Fraction}\\[-0.2ex]
    \meta{Limited phasing dependence}\\[1ex]
    \faTint\ \textbf{$\downarrow$ Total $\Delta v$}\\[-0.2ex]
    \meta{Efficient local routing}\\[1ex]
    \faChartLine\ \textbf{$\downarrow$ Maneuver Entropy}\\[-0.2ex]
    \meta{Relatively coherent transfer patterns}\\[1ex]
    \faCalendar\ \textbf{$\downarrow$ Mean Duration}\\[-0.2ex]
    \meta{Relatively short campaign timelines}
};

\node[
    uiCard,
    colTeal,
    below=1cm of H3
] (G1) {
    \faChartBar\ \textbf{High HCR}\\[-0.2ex]
    \meta{Probability mass concentrated}\\[1ex]
    \faCircleNotch\ \textbf{Low ESS}\\[-0.2ex]
    \meta{Sampling strongly exploitative}\\[1ex]
    \faUsers\ \textbf{High Purity}\\[-0.2ex]
    \meta{Strong community coherence}\\[1ex]
    \faProjectDiagram\ \textbf{High Spectral Gap}\\[-0.2ex]
    \meta{Strong structural separation}\\[1ex]
    \faLink\ \textbf{High Jaccard Similarity}\\[-0.2ex]
    \meta{Greater subset consistency}
};

\node[
    uiCard,
    colTeal,
    below=1cm of H4
] (A1) {
    \faBullseye\ \textbf{Low Orbital Dispersion}\\[-0.2ex]
    \meta{Tightly clustered orbits}\\[1ex]
    \faCodeBranch\ \textbf{Zero Family Diversity}\\[-0.2ex]
    \meta{Common fragmentation lineage}\\[1ex]
    \faPlane\ \textbf{Low Plane Change Burden}\\[-0.2ex]
    \meta{Small inclination differences}\\[1ex]
    \faRoute\ \textbf{High Accessibility}\\[-0.2ex]
    \meta{Favorable target-to-target geometry}\\[1ex]
    \faLock\ \textbf{High Route Reliability}\\[-0.2ex]
    \meta{Strong structural support for routing}
};

\node[
    uiCard,
    colOut,
    below=1cm of H5
] (O1) {
    \faFire\ \textbf{Low Mean $\Delta v$}\\[-0.2ex]
    \meta{Fuel-efficient routing}\\[1ex]
    \faGasPump\ \textbf{Low Fuel Consumption}\\[-0.2ex]
    \meta{Favorable mass requirements}\\[1ex]
    \faCalendarCheck\ \textbf{Shorter Missions}\\[-0.2ex]
    \meta{Reduced campaign duration}\\[1ex]
    \faCheckCircle\ \textbf{High RCR}\\[-0.2ex]
    \meta{Higher route reliability}\\[1ex]
    \faStar\ \textbf{Favorable Trade-off Region}\\[-0.2ex]
    \meta{Strong multi-objective solutions}
};


\node[
    envBadge,
    colSoc,
    below=3.8cm of Env1
] (Env2) {
    SOCRATES\\MINEFIELD
};

\node[
    below=0.1cm of Env2,
    text width=3.2cm,
    align=center
] {
    \meta{Risk-driven catalog with structured hazard corridors.}
};

\node[
    uiCard,
    colSoc,
    below=1.2cm of K1
] (K2) {
    \faFire\ \textbf{$\nearrow$ Mean $\Delta v$}\\[-0.2ex]
    \meta{Moderate kinematic cost}\\[1ex]
    \faClock\ \textbf{$\nearrow$ Mean Wait Fraction}\\[-0.2ex]
    \meta{Greater phasing demand than the collision cloud}\\[1ex]
    \faTint\ \textbf{$\nearrow$ Total $\Delta v$}\\[-0.2ex]
    \meta{Moderate cost escalation}\\[1ex]
    \faChartLine\ \textbf{$\nearrow$ Maneuver Entropy}\\[-0.2ex]
    \meta{Increasing transfer variety}\\[1ex]
    \faCalendar\ \textbf{$\nearrow$ Mean Duration}\\[-0.2ex]
    \meta{Intermediate campaign duration}
};

\node[
    uiCard,
    colSoc,
    below=1.2cm of G1
] (G2) {
    \faChartBar\ \textbf{High HCR}\\[-0.2ex]
    \meta{Dominant risk corridors}\\[1ex]
    \faCircleNotch\ \textbf{Low ESS}\\[-0.2ex]
    \meta{Concentrated sampling}\\[1ex]
    \faUsers\ \textbf{High Purity}\\[-0.2ex]
    \meta{Corridor-induced communities}\\[1ex]
    \faProjectDiagram\ \textbf{Moderate Spectral Gap}\\[-0.2ex]
    \meta{Structured but competing corridors}\\[1ex]
    \faLink\ \textbf{Moderate Jaccard Similarity}\\[-0.2ex]
    \meta{Intermediate subset overlap}
};

\node[
    uiCard,
    colSoc,
    below=1.2cm of A1
] (A2) {
    \faBullseye\ \textbf{Moderate Orbital Dispersion}\\[-0.2ex]
    \meta{Structured but more widely distributed}\\[1ex]
    \faCodeBranch\ \textbf{Moderate Family Diversity}\\[-0.2ex]
    \meta{Multiple lineage contributions}\\[1ex]
    \faPlane\ \textbf{Moderate Plane Change}\\[-0.2ex]
    \meta{Corridor transitions required}\\[1ex]
    \faRoute\ \textbf{High Accessibility}\\[-0.2ex]
    \meta{Target sets remain connected}\\[1ex]
    \faLock\ \textbf{Moderate Route Reliability}\\[-0.2ex]
    \meta{Reliability depends on corridor geometry}
};

\node[
    uiCard,
    colOut,
    below=1.2cm of O1
] (O2) {
    \faFire\ \textbf{Moderate Mean $\Delta v$}\\[-0.2ex]
    \meta{Intermediate fuel requirement}\\[1ex]
    \faGasPump\ \textbf{Moderate Fuel}\\[-0.2ex]
    \meta{Intermediate consumption}\\[1ex]
    \faCalendarCheck\ \textbf{Intermediate Missions}\\[-0.2ex]
    \meta{Moderate campaign durations}\\[1ex]
    \faCheckCircle\ \textbf{Moderate RCR}\\[-0.2ex]
    \meta{Conditional route reliability}\\[1ex]
    \faBalanceScale\ \textbf{Trade-off Solutions}\\[-0.2ex]
    \meta{Competing mission objectives}
};


\node[
    envBadge,
    colRan,
    below=4.2cm of Env2
] (Env3) {
    RANDOM
};

\node[
    below=0.1cm of Env3,
    text width=3.2cm,
    align=center
] {
    \meta{Heterogeneous population without explicitly imposed structure.}
};

\node[
    uiCard,
    colRan,
    below=1.2cm of K2
] (K3) {
    \faFire\ \textbf{$\uparrow$ Mean $\Delta v$}\\[-0.2ex]
    \meta{Higher kinematic burden}\\[1ex]
    \faClock\ \textbf{$\uparrow$ Mean Wait Fraction}\\[-0.2ex]
    \meta{Greater phasing dependence}\\[1ex]
    \faTint\ \textbf{$\uparrow$ Total $\Delta v$}\\[-0.2ex]
    \meta{More expensive routing}\\[1ex]
    \faChartLine\ \textbf{$\uparrow$ Maneuver Entropy}\\[-0.2ex]
    \meta{More heterogeneous transfers}\\[1ex]
    \faCalendar\ \textbf{$\uparrow$ Mean Duration}\\[-0.2ex]
    \meta{Longer campaign timelines}
};

\node[
    uiCard,
    colRan,
    below=1.2cm of G2
] (G3) {
    \faChartBar\ \textbf{Low HCR}\\[-0.2ex]
    \meta{Probability mass broadly distributed}\\[1ex]
    \faCircleNotch\ \textbf{High ESS}\\[-0.2ex]
    \meta{Exploration-dominated sampling}\\[1ex]
    \faUsers\ \textbf{High Purity}\\[-0.2ex]
    \meta{Emergent local transfer structure}\\[1ex]
    \faProjectDiagram\ \textbf{Moderate Spectral Gap}\\[-0.2ex]
    \meta{No strongly dominant separation}\\[1ex]
    \faLink\ \textbf{Lower Jaccard Similarity}\\[-0.2ex]
    \meta{Greater subset differentiation}
};

\node[
    uiCard,
    colRan,
    below=1.2cm of A2
] (A3) {
    \faBullseye\ \textbf{High Orbital Dispersion}\\[-0.2ex]
    \meta{Widely separated orbital states}\\[1ex]
    \faCodeBranch\ \textbf{High Family Diversity}\\[-0.2ex]
    \meta{Heterogeneous target populations}\\[1ex]
    \faPlane\ \textbf{High Plane Change Burden}\\[-0.2ex]
    \meta{Larger inclination and RAAN shifts}\\[1ex]
    \faRoute\ \textbf{Lower Accessibility}\\[-0.2ex]
    \meta{Greater fraction of difficult transfers}\\[1ex]
    \faLock\ \textbf{Reduced Route Reliability}\\[-0.2ex]
    \meta{Greater sensitivity to route geometry}
};

\node[
    uiCard,
    colOut,
    below=1.2cm of O2
] (O3) {
    \faFire\ \textbf{Higher Mean $\Delta v$}\\[-0.2ex]
    \meta{Greater fuel demand}\\[1ex]
    \faGasPump\ \textbf{Higher Fuel Consumption}\\[-0.2ex]
    \meta{Higher mass requirement}\\[1ex]
    \faCalendarCheck\ \textbf{Longer Missions}\\[-0.2ex]
    \meta{Extended campaign duration}\\[1ex]
    \faTimesCircle\ \textbf{Lower Route Reliability}\\[-0.2ex]
    \meta{Greater sensitivity to encounter geometry}\\[1ex]
    \faBalanceScale\ \textbf{Broader Trade-off Space}\\[-0.2ex]
    \meta{Greater exploration of alternatives}
};


\begin{scope}[on background layer]

    \draw[ribbonTeal]
        (K1.east) to[out=0, in=180] (G1.west);

    \draw[ribbonTeal]
        (G1.east) to[out=0, in=180] (A1.west);

    \draw[ribbonTeal]
        (A1.east) to[out=0, in=180] (O1.west);

    \draw[ribbonSoc]
        (K2.east) to[out=0, in=180] (G2.west);

    \draw[ribbonSoc]
        (G2.east) to[out=0, in=180] (A2.west);

    \draw[ribbonSoc]
        (A2.east) to[out=0, in=180] (O2.west);

    \draw[ribbonRan]
        (K3.east) to[out=0, in=180] (G3.west);

    \draw[ribbonRan]
        (G3.east) to[out=0, in=180] (A3.west);

    \draw[ribbonRan]
        (A3.east) to[out=0, in=180] (O3.west);

    \draw[ribbonTeal, line width=12pt]
        (Env1.east) to[out=0, in=180] (K1.west);

    \draw[ribbonSoc, line width=12pt]
        (Env2.east) to[out=0, in=180] (K2.west);

    \draw[ribbonRan, line width=12pt]
        (Env3.east) to[out=0, in=180] (K3.west);

\end{scope}


\coordinate (ArrowYRef) at ($(K3.south) + (0, -4.6)$);

\coordinate (ArrowStart) at (G3.west |- ArrowYRef);
\coordinate (ArrowEnd)   at (A3.east |- ArrowYRef);

\draw[
    left color=teal!80,
    right color=purple!80,
    draw=purple!80,
    fill=purple!80,
    -{Stealth[length=6mm, width=5mm]},
    line width=6pt,
    drop shadow={opacity=0.15}
]
    (ArrowStart) -- (ArrowEnd)
    node[
        midway,
        below=8pt,
        font=\bfseries\sffamily
    ] {STRUCTURAL QUALITY GRADIENT};

\node[
    above=6pt of ArrowStart,
    anchor=south west,
    font=\sffamily\bfseries\small,
    text=teal!90!black
] {Strong Structure};

\node[
    above=6pt of ArrowEnd,
    anchor=south east,
    font=\sffamily\bfseries\small,
    text=purple!90!black
] {Weak Explicit Structure};




\end{tikzpicture}
}

\caption{Topology-to-mission relationship across the three benchmark environments. The diagram summarizes how environmental structure propagates through kinematic, GBS/graph, and astrodynamic layers to produce distinct mission-level characteristics across the evaluated ADR scenarios.}
\label{fig:full_flow_diagram}

\end{figure}

Across the three suites, the results indicate that environmental topology propagates through each stage of the optimization pipeline. Strongly organized transfer structure tends to concentrate GBS probability mass on a smaller collection of coherent candidate subgraphs, whereas heterogeneous topology produces broader sampling and greater exploration. These differences subsequently influence the physical routing layer through variations in transfer cost, waiting requirements, maneuver diversity, accessibility, and mission duration. The observed relationship therefore reflects the interaction between object-level properties and the emergent structure of the orbital-transfer network rather than any single environmental metric (Figure \ref{fig:full_flow_diagram}).

\subsection{Synthesis of Combinatorial Scaling Dynamics: The Interplay between $N$ and $k$}

The complexity of multi-target ADR is fundamentally governed by the combinatorial size of the target-selection space,

\begin{equation}
\binom{N}{k},
\end{equation}

where $N$ denotes the number of candidate debris objects and $k$ denotes the number of targets selected for remediation. Although both parameters increase the size of the combinatorial search space, they influence the optimization process through distinct mechanisms.\\

Increasing $N$ expands the orbital-transfer graph by introducing additional debris objects and, under the evaluated graph-construction procedure, additional transfer corridors and intermediate rendezvous opportunities. This broadens the set of accessible routing alternatives and increases the diversity of candidate subgraphs explored by the GBS stage. In contrast, increasing $k$ extends the servicing sequence itself. As additional targets are incorporated into a single campaign, locally accessible orbital neighborhoods can become progressively exhausted, requiring transfers across more distant orbital regions and increasing the physical difficulty of mission execution.\\

The scaling behavior of the framework therefore emerges from the interaction between graph enrichment and route extension. Larger catalogs primarily increase combinatorial opportunity and routing flexibility, whereas larger target sets primarily increase astrodynamic and operational burden through greater transfer requirements, longer phasing intervals, and increased dependence on drift-assisted alignment. To characterize these effects, the principal graph-theoretic, quantum-sampling, and astrodynamic performance metrics are first classified according to their empirical sensitivity to variations in $N$ and $k$.

\subsubsection{Parametric Sensitivity Analysis}

Table~\ref{tab:sensitivity_ranking} summarizes the relative sensitivity of the principal graph-theoretic, quantum-sampling, and astrodynamic performance metrics to variations in candidate-pool size ($N$) and target cardinality ($k$). The classifications represent the dominant empirical trends observed across the evaluated $(N,k)$ configurations rather than universal monotonic laws.

\begin{table}[h]
\centering
\caption{Scaling Sensitivity Ranking. Empirical classification of the relative sensitivity of the principal performance metrics to changes in candidate-pool size ($N$) and target cardinality ($k$) across the evaluated $(N,k)$ configurations. The dominant driver column identifies the parameter associated with the stronger observed effect.}
\label{tab:sensitivity_ranking}
\renewcommand{\arraystretch}{1.3}
\setlength{\tabcolsep}{5pt}
\small

\begin{tabularx}{\textwidth}{
    >{\raggedright\arraybackslash}p{3.7cm}
    >{\centering\arraybackslash}p{2.0cm}
    >{\centering\arraybackslash}p{2.0cm}
    >{\raggedright\arraybackslash}X
}
\toprule
\textbf{Performance Metric} &
\textbf{$N \uparrow$} &
\textbf{$k \uparrow$} &
\textbf{Dominant Driver and Observed Implication} \\
\midrule

\textbf{Hafnian-Concentration (HCR)}
&
Moderate
&
High
&
\textbf{Cardinality ($k$):} Probability concentration on dominant pathways decreases as the target-selection requirement becomes more demanding.
\\

\textbf{Effective Sample Size (ESS)}
&
High
&
High
&
\textbf{Combined ($N,k$):} Both parameters enlarge the accessible combinatorial search space.
\\

\textbf{Sampling Diversity}
&
High
&
High
&
\textbf{Combined ($N,k$):} Probability mass is distributed across a broader set of candidate subsets.
\\

\textbf{Mean and Worst $\Delta v$}
&
Moderate
&
Very High
&
\textbf{Cardinality ($k$):} Increasing sequence length progressively exhausts local low-cost transfer opportunities.
\\

\textbf{Mission Duration}
&
Moderate
&
Very High
&
\textbf{Cardinality ($k$):} Longer servicing sequences increase phasing requirements and dependence on drift-assisted routing.
\\

\textbf{Drift Efficiency ($\eta_{drift}$)}
&
Low
&
High
&
\textbf{Cardinality ($k$):} Increasing sequence length increases the role of time as a substitute for impulsive propellant expenditure.
\\

\textbf{Target Coverage}
&
High
&
Low
&
\textbf{Pool Size ($N$):} Larger catalogs provide a broader population from which candidate targets can be sampled.
\\

\textbf{Jaccard Similarity}
&
High
&
Moderate
&
\textbf{Pool Size ($N$):} Increased catalog size creates additional competing target combinations and reduces overlap among independently generated selections.
\\

\bottomrule
\end{tabularx}
\end{table}

The sensitivity analysis reveals a separation between combinatorial and astrodynamic scaling effects. Metrics associated with graph exploration, target coverage, and sampling diversity are influenced strongly by catalog expansion, whereas propellant expenditure, mission duration, and drift utilization exhibit greater sensitivity to target cardinality. The resulting behavior can therefore be interpreted through two dominant scaling modes: an exploration mode controlled primarily by $N$, and a mission-complexity mode controlled primarily by $k$. The interaction between these modes produces the characteristic operating regimes examined below.

\subsubsection{Regime Transition Across the $(N,k)$ Design Space}

The effects of catalog size ($N$) and target cardinality ($k$) cannot be interpreted independently because the proposed architecture consists of two coupled stages. The GBS layer first explores the combinatorial structure of the orbital-transfer graph and generates candidate target subsets. These subsets are subsequently refined by the PyGMO optimization layer, which solves the associated trajectory-planning problem. Cchanges in $(N,k)$ propagate through both the graph-theoretic and astrodynamic components of the framework.\\

The two parameters primarily affect different stages of the optimization pipeline. Increasing the catalog size $N$ modifies the topology of the orbital-transfer graph by adding nodes and potentially additional transfer corridors and community structures. This broadens the distribution of candidate subsets explored by GBS. Increasing the target cardinality $k$, in contrast, primarily affects the physical routing problem by extending the servicing sequence, increasing transfer complexity, and increasing the need to exploit favorable phasing opportunities. The combined influence of these parameters is summarized in Table~\ref{tab:causal_cascade}, which traces the transition from graph-level sampling behavior to routing-level mission characteristics.\\

\begin{table}[h]
\centering
\caption{Causal Cascade Matrix of the GBS-PyGMO Pipeline. The table traces how boundary parameter configurations propagate through the topological sampling layer and the trajectory-routing layer to produce distinct optimization regimes.}
\label{tab:causal_cascade}
\renewcommand{\arraystretch}{1.5}
\small
\begin{tabular}{@{} p{1.8cm} c p{4.5cm} c p{4.2cm} c p{3.2cm} @{}}
\toprule

\textbf{Boundary State}
&
&
\textbf{Stage 1: GBS Topology}
&
&
\textbf{Stage 2: PyGMO Routing}
&
&
\textbf{Emergent Regime}
\\

\cmidrule{1-1} \cmidrule{3-3} \cmidrule{5-5} \cmidrule{7-7}

$\mathbf{N=15}$ \newline $\mathbf{k=3}$
&
$\rightarrow$
&
\textbf{Concentrated Sampling:}\newline
High HCR, low ESS, and strong community coherence. Probability mass remains concentrated within a small number of highly probable subgraphs.
&
$\rightarrow$
&
\textbf{Local Routing:}\newline
Short transfers exploiting dense, locally accessible orbital structures.
&
$\rightarrow$
&
\textbf{Exploitation-Dominated Regime}\newline
Concentrated exploration of a small number of high-probability routing alternatives.
\\

\midrule

$\mathbf{N=25}$ \newline $\mathbf{k=3}$
&
$\rightarrow$
&
\textbf{Distributed Sampling:}\newline
Reduced HCR and increased ESS. Probability mass is distributed across a broader collection of feasible subgraphs, increasing subset diversity.
&
$\rightarrow$
&
\textbf{Accessible Routing:}\newline
Broader physical accessibility with multiple low-cost transfer alternatives.
&
$\rightarrow$
&
\textbf{Exploration-Dominated Regime}\newline
Broader search across competing routing alternatives.
\\

\midrule

$\mathbf{N=15}$ \newline $\mathbf{k=5}$
&
$\rightarrow$
&
\textbf{Constrained Sampling:}\newline
High HCR and increased subset overlap. The limited candidate pool promotes repeated selection of structurally similar target sets.
&
$\rightarrow$
&
\textbf{Extended Routing:}\newline
Longer servicing sequences requiring greater phasing and higher $\Delta v$ expenditure.
&
$\rightarrow$
&
\textbf{Resource-Constrained Regime}\newline
Concentrated exploitation of the limited set of physically feasible corridors.
\\

\midrule

$\mathbf{N=25}$ \newline $\mathbf{k=5}$
&
$\rightarrow$
&
\textbf{Diffuse Sampling:}\newline
Low HCR, high ESS, and reduced community coherence. Probability mass is distributed across numerous competing subgraphs, producing the highest observed sampling diversity.
&
$\rightarrow$
&
\textbf{Drift-Assisted Routing:}\newline
Greater reliance on long-duration, $J_2$-assisted phasing across dispersed orbital states.
&
$\rightarrow$
&
\textbf{High-Complexity Regime}\newline
High combinatorial diversity combined with elevated astrodynamic complexity.
\\

\bottomrule
\end{tabular}
\end{table}

The four boundary configurations illustrate how changes in $N$ and $k$ shift the balance between combinatorial exploration and astrodynamic execution. Increasing $N$ primarily broadens the set of candidate structures available to the sampling layer, whereas increasing $k$ extends the physical routing problem and increases the demand for phasing, propellant, and feasibility margins. The resulting regimes provide a compact interpretation of the observed behavior across the evaluated design space.

\subsubsection{Operational Scaling Laws}

Table~\ref{tab:master_scaling_matrix} integrates the empirical trends across the principal graph, sampling, astrodynamic, temporal, and mission-level metrics. The directional symbols summarize the observed behavior as $N$ and $k$ increase, while the final columns connect these trends to the corresponding algorithmic and mission-level effects. The entries should be interpreted as empirical scaling tendencies within the evaluated design space rather than as formal analytical scaling laws.\\

\begin{table}[!h]
\centering
\caption{Master Operational Scaling Matrix. The table synthesizes the observed trends associated with increasing candidate-pool size ($N$) and subset cardinality ($k$), together with their dominant drivers, algorithmic effects, and mission-level consequences within the GBS-PyGMO framework.}
\label{tab:master_scaling_matrix}
\renewcommand{\arraystretch}{1.35}
\resizebox{\textwidth}{!}{
\begin{tabular}{@{} p{4.2cm} c c c c p{4.8cm} p{4.8cm} @{}}
\toprule
\textbf{Performance Metric}
&
\textbf{$N\uparrow$}
&
\textbf{$k\uparrow$}
&
\textbf{Environment of Strongest Observed Effect}
&
\textbf{Dominant Driver}
&
\textbf{Primary Algorithmic Effect}
&
\textbf{Mission-Level Consequence}
\\
\midrule

\multicolumn{7}{c}{\textbf{Graph Exploration, Entropy and Target Coverage}}
\\
\midrule

Effective Sample Size (ESS)
& $\uparrow\uparrow$
& $\uparrow\uparrow$
& RANDOM
& \multirow{7}{*}{$N$}
& \multirow{7}{4.8cm}{
Expands accessible subgraphs, increases sampling diversity, and distributes probability mass across a larger candidate population.
}
& \multirow{7}{4.8cm}{
Broadens exploration of the combinatorial search space and increases the range of candidate environmental configurations available for remediation.
}
\\

Sampling Diversity
& $\uparrow$
& $\uparrow\uparrow$
& RANDOM
& & & \\

Target Coverage
& $\uparrow\uparrow$
& $\uparrow$
& RANDOM
& & & \\

Subset Entropy
& $\uparrow$
& $\uparrow$
& RANDOM
& & & \\

Probability Concentration
& $\downarrow$
& $\downarrow$
& SOCRATES
& & & \\

Hafnian-Concentration Ratio (HCR)
& $\downarrow$
& $\downarrow\downarrow$
& SOCRATES
& & & \\

Amplification Uniformity (logUAmp)
& $\uparrow$
& $\downarrow$
& COLLISION
& & & \\

\midrule

\multicolumn{7}{c}{\textbf{Topological Accessibility and Structure}}
\\
\midrule

Transfer Accessibility
& $\uparrow$
& $\downarrow$
& COLLISION
& \multirow{5}{*}{$N$}
& \multirow{5}{4.8cm}{
Introduces additional transfer corridors and intermediate rendezvous opportunities within the active orbital regime.
}
& \multirow{5}{4.8cm}{
Broadens routing flexibility and increases the availability of low-cost transfer alternatives within clustered debris populations.
}
\\

Graph Connectivity
& $\uparrow$
& $\approx$
& COLLISION
& & & \\

Community Purity
& $\approx$
& $\downarrow$
& COLLISION
& & & \\

Jaccard Similarity
& $\downarrow$
& $\uparrow$
& SOCRATES
& & & \\

Orbital Dispersion (OD)
& $\downarrow$
& $\uparrow\uparrow$
& RANDOM
& & & \\

\midrule

\multicolumn{7}{c}{\textbf{Propellant Expenditure and Astrodynamic Cost}}
\\
\midrule

Mean $\Delta v$
& $\downarrow$
& $\uparrow\uparrow$
& SOCRATES
& \multirow{3}{*}{$k$}
& \multirow{3}{4.8cm}{
Increasing sequence length exhausts locally favorable transfer clusters more rapidly and increases the complexity of graph-based routing.
}
& \multirow{3}{4.8cm}{
Higher maneuver costs and reduced propellant margins for the servicing vehicle.
}
\\[1.5ex]

Worst-Case $\Delta v$
& $\downarrow$
& $\uparrow\uparrow$
& SOCRATES
& & & \\[1.5ex]

Plane Change Burden
& $\downarrow$
& $\uparrow$
& RANDOM
& & & \\[1.5ex]

\midrule

\multicolumn{7}{c}{\textbf{Temporal Constraints and Drift Utilization}}
\\
\midrule

Total Mission Duration
& $\approx$
& $\uparrow\uparrow$
& RANDOM
& \multirow{4}{*}{$k$}
& \multirow{4}{4.8cm}{
Requires longer phasing intervals and greater dependence on secular orbital-alignment opportunities associated with $J_2$ dynamics.
}
& \multirow{4}{4.8cm}{
Longer campaign timelines with increasing reliance on passive, drift-assisted routing.
}
\\[1ex]

Mission Wait Time (Days)
& $\downarrow$
& $\uparrow\uparrow$
& RANDOM
& & & \\[1ex]

Wait Percentage (\%)
& $\approx$
& $\downarrow\downarrow$
& RANDOM
& & & \\[1ex]

Drift Utilization
& $\approx$
& $\uparrow\uparrow$
& RANDOM
& & & \\[1ex]

\midrule

\multicolumn{7}{c}{\textbf{Reliability and Objective-Space Diversity}}
\\
\midrule

Route Conjunction Reliability
& $\uparrow$
& $\downarrow$
& SOCRATES
& \multirow{4}{*}{$N+k$}
& \multirow{4}{4.8cm}{
Longer mission chains increase accumulated operational uncertainty while also increasing the complexity of the objective-space routing problem.
}
& \multirow{4}{4.8cm}{
Reduced campaign robustness as sequence length and routing complexity increase, together with a broader set of accessible Pareto trade-offs.
}
\\[1ex]

Hypervolume ($H_v$)
& $\uparrow$
& $\uparrow$
& RANDOM
& & & \\[1ex]

Pareto Diversity
& $\uparrow$
& $\uparrow$
& RANDOM
& & & \\[1ex]

Optimization Difficulty
& $\uparrow$
& $\uparrow\uparrow$
& RANDOM
& & & \\[1ex]

\bottomrule
\end{tabular}}
\end{table}

The trends summarized in Table~\ref{tab:master_scaling_matrix} reveal a separation between the principal effects of catalog expansion and target-sequence extension. Increasing $N$ primarily modifies the combinatorial and topological structure of the problem. A larger candidate catalog introduces additional nodes and potential transfer corridors, broadening the space of accessible candidate subsets and increasing the scope of graph exploration. The strongest consequences of $N$ are therefore observed within the graph-theoretic and quantum-sampling components of the framework, where the richness of the candidate population influences sampling diversity, target coverage, and connectivity-related measures.\\

Increasing $k$, in contrast, primarily modifies the physical complexity of the resulting mission. As the number of required targets increases, locally coherent transfer opportunities can become progressively exhausted, requiring transfers across more dispersed orbital regions. This manifests in higher propellant expenditure, larger plane-change requirements, increased use of passive $J_2$ drift, and longer mission durations. The temporal table should therefore be interpreted together with both absolute waiting time and Wait Percentage, since the latter is a normalized quantity whose scaling also depends on the total mission-time composition.\\

The benchmark environments modulate these scaling mechanisms through their underlying orbital organization. The Collision Cloud environment retains pronounced local fragmentation structure and associated transfer relationships, whereas the SOCRATES Minefield exhibits stronger probability concentration around structured hazard corridors. The Random environment is characterized by weaker persistent organization and comparatively greater dispersion in the graph and sampling metrics. These differences explain why the same change in $(N,k)$ can propagate differently through the GBS and trajectory-optimization stages.\\

Taken together, the results provide an empirical scaling picture for the proposed GBS-PyGMO architecture. Catalog expansion ($N\uparrow$) primarily increases combinatorial opportunity through graph enrichment and broader candidate exploration, whereas target expansion ($k\uparrow$) primarily increases the astrodynamic and temporal burden associated with executing longer servicing sequences. Overall mission behavior therefore emerges from the interaction between these effects: increasing $N$ broadens the set of potentially accessible routing structures, while increasing $k$ raises the physical requirements for realizing them.
\subsection{Detailed Comparative Evaluation of the Selection Heuristics}

This section provides a comprehensive comparison of the seven target-selection heuristics across the evaluated orbital environments and scaling configurations. The analysis examines their behavior from complementary perspectives, including objective-space quality, kinematic efficiency, temporal performance, mission feasibility, search-space diversity, and the topology of the GBS sampling distribution.

\subsubsection{Objective-Space Performance and Robustness}

Table~\ref{tab:heuristic_dominance_hv} summarizes the objective-space performance of all seven evaluated heuristics across the three environmental suites. To quantify cross-topology robustness, three derived metrics are considered alongside the raw Hypervolume (HV) values. Let
$\mathcal{E}=\{\text{SOCRATES Minefield},\text{Collision Cloud},\text{Random}\}$
denote the evaluated environments. For each heuristic, the mean Hypervolume is

\[
\bar{HV}=\frac{1}{|\mathcal{E}|}\sum_{e\in\mathcal{E}}HV_e.
\]

Environmental sensitivity is quantified through

\[
\sigma_{HV}=
\sqrt{\frac{1}{|\mathcal{E}|}\sum_{e\in\mathcal{E}}
(HV_e-\bar{HV})^2},
\]

with relative sensitivity measured using

\[
CV=\frac{\sigma_{HV}}{\bar{HV}}\times100\%.
\]

Finally, Randomization Loss is defined as

\[
L_{rand}=
\frac{\max(HV)-HV_{Random}}
{\max(HV)}
\times100\%.
\]

\begin{table}[H]
\centering
\caption{Global Heuristic Objective-Space Performance Matrix. Hypervolume (HV) values are macro-averaged across the evaluated environmental suites and scaling configurations ($N \in \{15,18,22,25\}$, $k \in \{3,5\}$). Environmental Sensitivity ($\sigma_{HV}$) and the Coefficient of Variation (CV) quantify cross-topology variability, while Randomization Loss ($L_{rand}$) measures the reduction in Hypervolume in the Random suite relative to the highest HV attained by the corresponding heuristic.}
\label{tab:heuristic_dominance_hv}
\renewcommand{\arraystretch}{1.3}
\resizebox{\textwidth}{!}{
\begin{tabular}{ll|ccc|cccc}
\toprule
& & \multicolumn{3}{c|}{\textbf{Mean Hypervolume (HV) by Suite}} & \multicolumn{4}{c}{\textbf{Derived Robustness Metrics}} \\
\textbf{Category} & \textbf{Heuristic} & \textbf{\texttt{SOCRATES}} & \textbf{\texttt{COLLISION}} & \textbf{Random} & \textbf{Mean HV ($\bar{HV}$)} & \textbf{Env. Sens. ($\sigma_{HV}$)} & \textbf{CV (\%)} & \textbf{Rand. Loss ($L_{rand}$)} \\
\midrule
\multirow{2}{*}{\textbf{Topological}} 
 & \texttt{CENTRALITY} & 0.990 & 0.988 & 0.952 & 0.977 & 0.017 & \textbf{1.7\%} & \textbf{3.84\%} \\
 & \texttt{COMMUNITY}  & 0.987 & 0.989 & 0.955 & 0.977 & \textbf{0.016} & \textbf{1.6\%} & \textbf{3.44\%} \\
\midrule
\multirow{2}{*}{\textbf{Environmental}} 
 & \texttt{SOURCE} & \textbf{0.991} & \textbf{0.997} & 0.958$^\dagger$ & \textbf{0.982} & 0.017 & 1.7\% & 3.91\% \\
 & \texttt{SOCRATES\_RISK} & 0.950 & 0.975 & 0.960$^*$ & 0.962 & 0.010 & 1.0\% & 1.54\% \\
\midrule
\multirow{2}{*}{\textbf{Kinematic}} 
 & \texttt{SIZE} & \textbf{0.991} & 0.988 & 0.870 & 0.950 & 0.056 & 5.9\% & 12.21\% \\
 & \texttt{BSTAR} & 0.971 & 0.981 & 0.832 & 0.928 & 0.068 & 7.3\% & 15.19\% \\
\midrule
\textbf{Baseline} 
 & \texttt{RANDOM\_SUBSET} & 0.955 & 0.978 & 0.872 & 0.935 & 0.046 & 4.9\% & 10.84\% \\
\bottomrule
\multicolumn{9}{l}{\footnotesize $^*$ \textit{Average excludes $N=15,18$ ($k=5$) due to 100\% mission-generation failure (0 valid sequences).}} \\
\multicolumn{9}{l}{\footnotesize $^\dagger$ \textit{Average excludes $N=15$ ($k=5$) due to 100\% mission-generation failure.}}
\end{tabular}}
\end{table}

\paragraph{Objective-Space Performance}
The Hypervolume values show consistently high objective-space quality across the evaluated suites. \texttt{SOURCE} attains the highest mean HV of $0.982$, while both \texttt{CENTRALITY} and \texttt{COMMUNITY} achieve $0.977$. The difference between \texttt{SOURCE} and either graph-theoretic heuristic is $0.005$ in absolute HV, indicating similar aggregate objective-space coverage despite the use of different selection criteria. The high HV values obtained across the structured environments further indicate that the downstream evolutionary optimizer can identify substantial Pareto-efficient regions from the candidate subsets supplied by the selection stage.

\paragraph{Cross-Topology Robustness}
The Environmental Sensitivity ($\sigma_{HV}$) and Coefficient of Variation (CV) provide a measure of how objective-space performance changes with environmental topology. \texttt{CENTRALITY} and \texttt{COMMUNITY} exhibit CV values below $2\%$, while \texttt{SIZE} and \texttt{BSTAR} show CV values of $5.9\%$ and $7.3\%$, respectively. Thus, the graph-theoretic strategies exhibit smaller relative variation in Hypervolume across the evaluated environments.

\paragraph{Resilience to Topological Randomization}
The Randomization Loss metric provides a direct comparison of the reduction in objective-space quality in the Random suite. \texttt{CENTRALITY} and \texttt{COMMUNITY} exhibit losses of $3.84\%$ and $3.44\%$, respectively, whereas \texttt{SIZE} and \texttt{BSTAR} exhibit substantially larger reductions of $12.21\%$ and $15.19\%$. This indicates that the objective-space performance of the graph-based strategies changes less when the structured environmental organization is replaced by an unstructured background.

\paragraph{The \texttt{SOCRATES\_RISK} Anomaly}
The relatively small CV and Randomization Loss obtained by \texttt{SOCRATES\_RISK} require cautious interpretation. Several high-complexity configurations produced no valid mission sequences and were therefore excluded from the corresponding averages. Therefore, the reported HV statistics describe the surviving feasible instances rather than the complete set of attempted configurations. This distinction is important because objective-space quality and mission-generation feasibility capture different aspects of heuristic behavior.

\subsubsection{Global Kinematic Performance and Algorithmic Reliability}

While Hypervolume characterizes objective-space coverage, the kinematic performance of the resulting missions is reflected directly in propellant expenditure and feasibility. Table~\ref{tab:heuristic_kinematics} therefore aggregates the Mean $\Delta v$, Median $\Delta v$, 90th percentile, Interquartile Range (IQR), and Convergence Rate across the evaluated environmental configurations.

\begin{table}[h]
\centering
\caption{Global Kinematic Performance Profile. Metrics are macro-aggregated across all topological configurations ($N \in \{15,18,22,25\}$, $k \in \{3,5\}$). The Convergence Rate denotes the percentage of evaluated configurations for which at least one feasible mission satisfying the implemented astrodynamic and mass constraints was generated.}
\label{tab:heuristic_kinematics}
\renewcommand{\arraystretch}{1.3}
\resizebox{\textwidth}{!}{
\begin{tabular}{l | c c c c | c}
\toprule
\textbf{Heuristic} & \textbf{Mean $\Delta v$ (km/s)} & \textbf{Median $\Delta v$ (km/s)} & \textbf{90th Pctl (km/s)} & \textbf{IQR (km/s)} & \textbf{Conv. Rate} \\
\midrule
\texttt{COMMUNITY}       & \textbf{3.21} & \textbf{0.63} & 7.37 & 5.75 & \textbf{100\%} \\
\texttt{CENTRALITY}      & 3.32 & 1.44 & 7.38 & 5.92 & \textbf{100\%} \\
\texttt{SOURCE}          & 3.58 & 2.12 & 7.35 & 5.43 & 95.8\% \\
\texttt{SIZE}            & 3.89 & 3.36 & 7.35 & \textbf{4.45} & \textbf{100\%} \\
\texttt{BSTAR}           & 4.12 & 2.87 & 7.91 & 5.48 & \textbf{100\%} \\
\texttt{SOCRATES\_RISK}  & 4.35 & 3.74 & \textbf{8.07} & 4.33 & 91.7\% \\
\midrule
\texttt{RANDOM\_SUBSET}  & 4.95 & 4.30 & 7.69 & 3.86 & \textbf{100\%} \\
\bottomrule
\end{tabular}}
\end{table}

\paragraph{Propellant Efficiency and Central Tendency}
The Mean and Median $\Delta v$ values show substantial differences among the selection strategies. \texttt{COMMUNITY} and \texttt{CENTRALITY} have the lowest mean mission costs, at $3.21$ km/s and $3.32$ km/s, respectively, compared with $4.95$ km/s for the \texttt{RANDOM\_SUBSET} baseline. These values correspond to reductions of $35.2\%$ and $32.9\%$, respectively, relative to the baseline. The large difference between mean and median values, particularly for \texttt{COMMUNITY}, indicates a strongly right-skewed cost distribution in which relatively low-cost missions occur frequently despite the presence of substantially more expensive cases.

\paragraph{Distributional Variability}
The IQR provides complementary information on the spread of mission costs. \texttt{COMMUNITY} and \texttt{CENTRALITY} exhibit IQR values of $5.75$ km/s and $5.92$ km/s, respectively, indicating broad variation in attainable mission costs across the evaluated configurations. In contrast, the \texttt{RANDOM\_SUBSET} baseline has a narrower IQR of $3.86$ km/s, but this narrower distribution is accompanied by a substantially higher median cost of $4.30$ km/s. Hence, a narrower cost distribution does not necessarily imply lower-cost mission architectures.

\paragraph{Upper-Tail Kinematic Performance}
The 90th-percentile values characterize the high-cost portion of the mission distribution. \texttt{COMMUNITY} and \texttt{CENTRALITY} remain near $7.37$-$7.38$ km/s, whereas \texttt{SOCRATES\_RISK} reaches $8.07$ km/s and \texttt{BSTAR} reaches $7.91$ km/s. These higher upper-tail values indicate greater exposure to expensive transfer configurations under unfavorable orbital geometries.

\paragraph{Algorithmic Reliability}
The Convergence Rate shows that \texttt{SOURCE} and \texttt{SOCRATES\_RISK} do not maintain complete feasibility coverage, with rates of $95.8\%$ and $91.7\%$, respectively. The remaining heuristics achieve $100\%$ convergence across the evaluated configurations. The reduced coverage of the specialized strategies is consistent with target selections that can become difficult to service under the transfer and mass constraints when object-level priorities are not aligned with orbital accessibility.

\subsubsection{Temporal Dynamics and Phasing Efficiency}

The temporal behavior of the generated missions provides a complementary perspective on transfer accessibility. In the implemented trajectory model, mission duration includes transfer time, passive waiting, and the fixed servicing allocation. The present analysis therefore considers the resulting mission duration together with the fraction of time assigned to passive waiting and the corresponding sensitivity to environmental topology. Temporal Degradation, $L_T$, is defined as

\[
L_T =
\frac{T_{Random}-T_{\text{structured}}}
{T_{\text{structured}}}
\times 100\%,
\]

where $T_{\text{structured}}$ denotes the average mission duration across the SOCRATES Minefield and Collision Cloud suites, and $T_{Random}$ denotes the corresponding value for the Random suite. Table~\ref{tab:temporal_dynamics} summarizes these quantities for each heuristic.

\begin{table}[h]
\centering
\caption{Temporal Dynamics and Phasing Efficiency. Mean mission duration and average Wait Fraction across all scaling configurations ($N \in \{15,18,22,25\}$, $k \in \{3,5\}$). Temporal Degradation ($L_T$) quantifies the increase in mission duration under topological randomization.}
\label{tab:temporal_dynamics}
\renewcommand{\arraystretch}{1.3}
\resizebox{\textwidth}{!}{
\begin{tabular}{l | c c | c c | c c | c}
\toprule
& \multicolumn{2}{c|}{\textbf{SOCRATES Minefield}} & \multicolumn{2}{c|}{\textbf{Collision Cloud}} & \multicolumn{2}{c|}{\textbf{Random}} & \textbf{Temporal} \\
\textbf{Heuristic} & \textbf{Mean Mission Duration (d)} & \textbf{Wait \%} & \textbf{Mean Mission Duration (d)} & \textbf{Wait \%} & \textbf{Mean Mission Duration (d)} & \textbf{Wait \%} & \textbf{Degradation ($L_T$)} \\
\midrule
\texttt{CENTRALITY}      & 175.5 & 12.6\% & 181.4 & 13.2\% & 232.7 & 30.2\% & \textbf{30.4\%} \\
\texttt{COMMUNITY}       & 179.7 & 13.2\% & 183.2 & 14.2\% & 230.5 & 29.1\% & \textbf{27.0\%} \\
\midrule
\texttt{SOURCE}          & 179.2 & 13.3\% & 179.5 & 12.3\% & 211.7$^*$ & 28.6\%$^*$ & 18.0\%$^*$ \\
\texttt{SOCRATES\_RISK}  & 197.7 & 20.7\% & 176.9 & 12.9\% & 190.5$^*$ & 26.4\%$^*$ & 1.7\%$^*$ \\
\midrule
\texttt{SIZE}            & 171.5 & 11.3\% & 186.0 & 15.6\% & 282.2 & 39.3\% & \textbf{57.9\%} \\
\texttt{BSTAR}           & 191.2 & 19.6\% & 190.6 & 17.8\% & 287.1 & 44.9\% & \textbf{50.4\%} \\
\midrule
\texttt{RANDOM\_SUBSET}  & 189.0 & 16.7\% & 194.0 & 18.7\% & 256.4 & 33.2\% & 33.9\% \\
\bottomrule
\multicolumn{8}{l}{\footnotesize $^*$ \textit{Averages calculated over feasible mission instances; excludes high-complexity configurations with 100\% mission-generation failure.}}
\end{tabular}}
\end{table}

\paragraph{Mission Duration and Temporal Efficiency}
The mean mission durations in the Random suite range from $190.5$ to $287.1$ days among the evaluated heuristics. \texttt{COMMUNITY} and \texttt{CENTRALITY} yield durations of $230.5$ and $232.7$ days, respectively, compared with $282.2$ days for \texttt{SIZE} and $287.1$ days for \texttt{BSTAR}. Relative to \texttt{BSTAR}, these correspond to reductions of approximately $19.7\%$ and $19.0\%$.

\paragraph{Passive Phasing and $J_2$ Utilization}
The Wait Fraction increases systematically in the Random suite. For \texttt{CENTRALITY} and \texttt{COMMUNITY}, the waiting contribution rises from approximately $13\%$ in the structured environments to $30.2\%$ and $29.1\%$, respectively, in the Random suite. The corresponding increases for \texttt{SIZE} and \texttt{BSTAR} reach $39.3\%$ and $44.9\%$. This behavior is consistent with increased use of passive orbital phasing when favorable transfer corridors are less readily available.

\paragraph{Temporal Robustness Across Orbital Topologies}
The Temporal Degradation values are $27.0\%$ for \texttt{COMMUNITY} and $30.4\%$ for \texttt{CENTRALITY}, compared with $57.9\%$ for \texttt{SIZE} and $50.4\%$ for \texttt{BSTAR}. The apparently small degradation of $1.7\%$ for \texttt{SOCRATES\_RISK} should be interpreted cautiously because infeasible high-complexity configurations are excluded from the corresponding averages. The reported value therefore characterizes the surviving feasible missions and should not be interpreted independently of the associated convergence rate.

\subsubsection{Mission Feasibility and Pareto Dominance}

\begin{figure}[!h]
    \centering
    \includegraphics[width=1\linewidth]{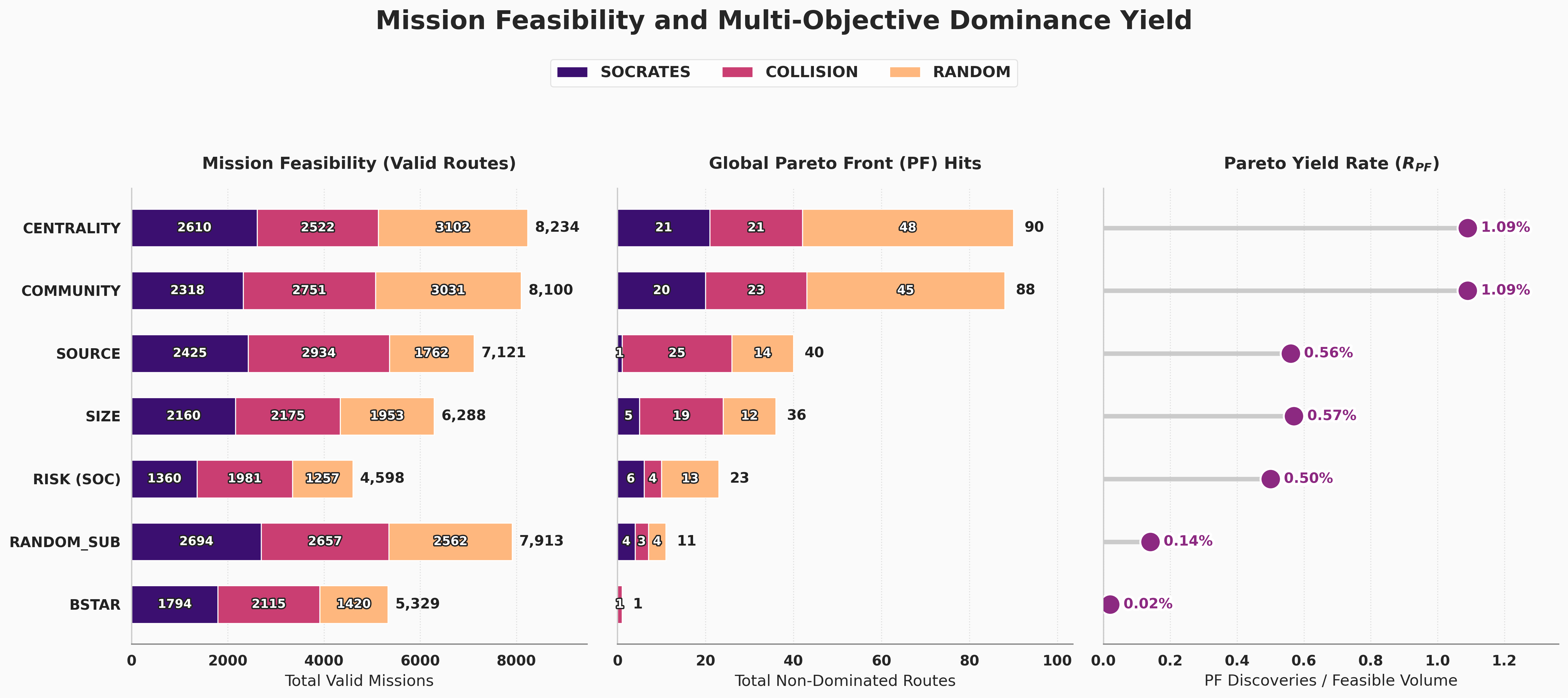}
    \caption{Feasible mission generation and global Pareto-front participation across selection heuristics. The figure compares the total number of valid mission sequences with the number of sequences identified on the global Pareto front across the evaluated environmental suites and scaling configurations.}
    \label{fig:pareto}
\end{figure}

Hypervolume characterizes continuous objective-space coverage, whereas the number of Valid Missions and Global Pareto Front (PF) Hits provide discrete measures of feasibility and global objective-space participation. Figure~\ref{fig:pareto} compares these quantities across the evaluated heuristics. To normalize the number of Pareto-front hits by the available feasible search volume, the Pareto Yield Rate is defined as

\[
R_{PF} =
\frac{\text{PF Hits}}
{\text{Valid Missions}}
\times100\%.
\]

\paragraph{Pareto Dominance and Network Awareness}
The discrete Pareto-front statistics show that \texttt{CENTRALITY} and \texttt{COMMUNITY} account collectively for $178$ of the $289$ identified global Pareto-front hits, corresponding to more than $61\%$ of the total. In the Random suite, the two heuristics account for $93$ of the $136$ available PF hits, or approximately $68\%$. \texttt{SOURCE}, in contrast, records a concentration of Pareto-front hits in the structured Collision Cloud environment, where it obtains $25$ hits.

\paragraph{Algorithmic Yield and Feasibility}
The number of valid missions provides a complementary measure of the feasible search volume. \texttt{SOCRATES\_RISK} produces the lowest global validity count, with $4{,}598$ feasible mission sequences. The lower feasible volume is consistent with the possibility that prioritization of conjunction-risk targets can produce target combinations that are difficult to service under the orbital-transfer and mass constraints. This demonstrates that a favorable objective-space statistic does not necessarily correspond to broad mission-generation coverage.

\paragraph{Pareto Yield Efficiency}
Normalizing Pareto-front hits by feasible mission volume provides an additional measure of search efficiency. \texttt{RANDOM\_SUBSET} generates $7{,}913$ valid missions, compared with $8{,}234$ for \texttt{CENTRALITY}, yet its Pareto Yield Rate is approximately eight times lower. Both \texttt{CENTRALITY} and \texttt{COMMUNITY} convert approximately $1.1\%$ of feasible mission architectures into globally Pareto-optimal solutions, compared with approximately $0.14\%$ for the \texttt{RANDOM\_SUBSET} baseline. Thus, the proportion of feasible solutions reaching the global Pareto front differs substantially even when the overall feasible search volumes are of similar magnitude.

\subsubsection{Inter-Heuristic Diversity and Search-Space Exploration}

The preceding analyses characterize objective-space quality and mission feasibility, while the Jaccard analysis examines whether the different selection heuristics actually explore similar regions of the combinatorial search space. Inter-Algorithmic Overlap is quantified using the Jaccard Similarity between the target sequences generated by each pair of heuristics across the evaluated environmental suites.

\begin{figure}[!htp]
    \centering
    \includegraphics[width=1\linewidth]{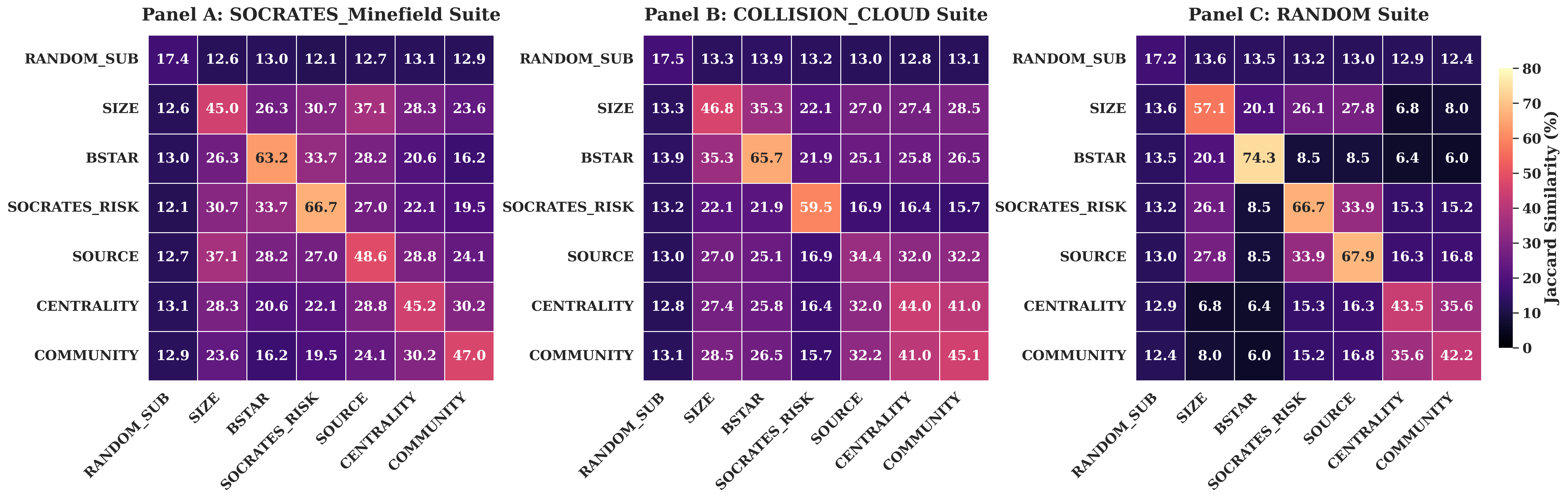}
    \caption{Inter-Algorithmic Overlap (Jaccard Similarity) by orbital topology. Values represent the macro-averaged intersection-over-union of target sequences selected by each heuristic across all scaling configurations ($N \in \{15,18,22,25\}$, $k \in \{3,5\}$). Panels A, B, and C correspond to the SOCRATES Minefield, Collision Cloud, and Random suites, respectively.}
    \label{fig:jaccard_matrices}
\end{figure}

\paragraph{Inter-Algorithmic Target Divergence}
The suite-wise Jaccard matrices show that target-set overlap depends strongly on the underlying orbital topology. In the structured \texttt{SOCRATES} and \texttt{COLLISION} environments, graph-theoretic heuristics exhibit moderate overlap with conventional strategies, indicating partial agreement on structurally prominent targets. In the Random suite, however, overlap with object-centric heuristics decreases substantially. \texttt{COMMUNITY} shares only $6.0\%$ overlap with \texttt{BSTAR} and $8.0\%$ with \texttt{SIZE}, while \texttt{CENTRALITY} exhibits overlaps of $6.4\%$ and $6.8\%$, respectively. In contrast, the overlap between \texttt{CENTRALITY} and \texttt{COMMUNITY} remains comparatively high across all three suites, at $30.2\%$, $41.0\%$, and $35.6\%$. These values indicate that the two graph-theoretic strategies retain related selection behavior even when their overlap with object-centric heuristics becomes small.

\paragraph{Implications for Search-Space Exploration}
The low overlap between graph-theoretic and object-centric heuristics indicates that the graph-based selection stage is not simply reproducing conventional object rankings. Instead, it identifies structurally distinct target subsets based on relationships encoded in the transfer network. This provides evidence that network topology contributes an additional source of information to candidate generation that is not explicitly represented by isolated object-level ranking criteria.

\paragraph{Implications for Quantum-Guided Selection}
The Jaccard results also clarify the role of the GBS stage within the overall framework. Because candidate subsets are sampled from graph embeddings that encode transfer accessibility, the downstream optimizer receives target collections reflecting network-level relationships before detailed trajectory optimization is performed. The distinct target selections observed across heuristics therefore arise not only from different object-prioritization rules, but also from how those rules interact with the topology-aware candidate-generation process.

\subsubsection{Quantum Sampling Topology and Graph-Theoretic Profile}

GBS operates as the sampling stage preceding evolutionary trajectory optimization, and its aggregate metrics are therefore primarily determined by the structure of the environmental orbital graphs rather than by the individual target-selection heuristics. Table~\ref{tab:gbs_metrics} summarizes the HCR, ESS, $H_v$, Pur, Cond, and SpGap across the three environmental suites and all eight $(N,k)$ scaling configurations.


\begin{table}[h]
\centering
\caption{Global GBS and Graph-Theoretic Profile by Environmental Suite. Metrics are macro-averaged across all scaling configurations. Sampling behavior is classified according to the observed relationship between probability concentration, distributional diversity, and graph connectivity.}
\label{tab:gbs_metrics}
\renewcommand{\arraystretch}{1.3}
\resizebox{\textwidth}{!}{
\begin{tabular}{l | c c c c c c | l}
\toprule
\textbf{Suite (Topology)} & \textbf{HCR} & \textbf{ESS} & \textbf{Entropy ($H_v$)} & \textbf{Purity} & \textbf{Cond.} & \textbf{Sp. Gap} & \textbf{Sampling Behavior Classification} \\
\midrule
\texttt{SOCRATES} &
\textbf{0.77} & 6.18 & 0.65 & 0.92 & \textbf{0.39} & 0.53 &
\textbf{Exploitation-Dominant} (Hazard-Biased) \\

\texttt{COLLISION} &
0.74 & \textbf{7.26} & \textbf{0.68} & 0.95 & 0.40 & 0.53 &
\textbf{Balanced} (Structure-Guided) \\

Random &
0.71 & 7.19 & 0.66 & \textbf{0.96} & \textbf{0.47} & \textbf{0.55} &
\textbf{Exploration-Dominant} (Diffuse) \\
\bottomrule
\end{tabular}}
\end{table}

\paragraph{Exploitation-Dominant Sampling in the SOCRATES Minefield}
The \texttt{SOCRATES} suite exhibits the highest HCR value ($0.77$) and the lowest Conductance ($0.39$). Together, these values are consistent with a more localized graph structure in which probability concentration is associated with a smaller set of densely connected regions. The corresponding ESS of $6.18$ is the lowest among the three suites, indicating a comparatively concentrated sampling distribution.

\paragraph{Balanced Topological Extraction in the Collision Cloud}
The \texttt{COLLISION} suite exhibits the highest ESS ($7.26$) and the highest von Neumann Entropy ($H_v=0.68$). Its Cond of $0.40$ remains close to that of the SOCRATES environment, while Pur reaches $0.95$. These characteristics correspond to a sampling distribution that remains structurally concentrated while exhibiting greater dispersion than the SOCRATES suite.

\paragraph{Exploration-Dominant Dispersion in the Random Suite}
The Random suite exhibits the lowest HCR ($0.71$) and the highest Cond ($0.47$). The increased Cond is consistent with weaker community boundaries and a less localized connectivity structure. The SpGap also reaches the largest value among the three suites ($0.55$). Together with the relatively high ESS ($7.19$), these metrics are consistent with broader dispersion of the GBS sampling distribution across the underlying candidate-subgraph space.

\paragraph{Cross-Topology Sampling Transition}
Despite the variation in HCR, ESS, and entropy, Pur remains high across all three environments, ranging from $0.92$ to $0.96$. Thus, the increase in distributional diversity in the Random suite does not correspond to a uniform sampling distribution. Instead, the GBS output retains preferential concentration on structurally favorable regions while broadening the set of candidate subsets considered. The transition from the SOCRATES environment toward the Random suite is therefore characterized by decreasing probability concentration and increasing distributional dispersion, consistent with a shift from localized exploitation toward broader exploration.

\subsubsection{Suite-Specific Heuristic Profile}

Having evaluated the selection heuristics across objective-space quality, kinematic efficiency, temporal behavior, mission feasibility, and search-space diversity, the results indicate that heuristic performance is strongly dependent on the structure of the underlying orbital environment. The SOCRATES Minefield favors strategies that can exploit concentrated risk and transfer structure, the Collision Cloud is particularly responsive to source- and topology-related organization, while the Random suite places greater emphasis on transfer-network structure and the ability to maintain feasible candidate diversity in the absence of strong environmental correlations. Accordingly, no single heuristic dominates every metric and environmental condition simultaneously (Table \ref{tab:suite_heuristic_profile}). Instead, the results reveal distinct suite-heuristic associations, with different strategies exhibiting the strongest observed behavior under different combinations of topology and mission objectives.

\begin{table}[!h]
\centering
\caption{Suite-Specific Heuristic Performance Profile. The table summarizes the heuristics associated with the strongest observed results for selected metrics within each environmental suite. The entries are metric-specific and should not be interpreted as a universal ranking of heuristics.}
\label{tab:suite_heuristic_profile}
\renewcommand{\arraystretch}{1.45}
\setlength{\tabcolsep}{7pt}

\begin{tabularx}{\textwidth}{
    >{\raggedright\arraybackslash}p{2.8cm}
    >{\raggedright\arraybackslash}p{3.6cm}
    >{\raggedright\arraybackslash}p{3.4cm}
    >{\raggedright\arraybackslash}X
}
\toprule
\textbf{Environmental Suite} &
\textbf{Characteristic Structure} &
\textbf{Strong Observed Heuristic(s)} &
\textbf{Primary Evidence} \\
\midrule

\texttt{SOCRATES Minefield} &
Concentrated conjunction-risk corridors and structured transfer relationships &
\texttt{SOURCE}, \texttt{SIZE}, \texttt{CENTRALITY} &
\texttt{SOURCE} and \texttt{SIZE} attain the highest suite-level HV ($0.991$); \texttt{SIZE} also gives the shortest mean mission duration ($171.5$ d) among the evaluated heuristics. The result indicates that both environmental and network-related prioritization can identify favorable candidates within the highly structured conjunction-risk environment. \\[0.3cm]

\texttt{COLLISION Cloud} &
Fragmentation-driven local structure and source-family organization &
\texttt{SOURCE}, \texttt{COMMUNITY}, \texttt{SOCRATES\_RISK} &
\texttt{SOURCE} attains the highest suite-level HV ($0.997$); \texttt{COMMUNITY} shows high objective-space stability, while \texttt{SOCRATES\_RISK} gives a mean mission duration of $176.9$ d. The concentration of strong results among source- and topology-related strategies is consistent with the pronounced local structure associated with fragmentation-generated debris. \\[0.3cm]

\texttt{Random} &
Weak environmental organization and reduced topological localization &
\texttt{CENTRALITY}, \texttt{COMMUNITY} &
\texttt{COMMUNITY} and \texttt{CENTRALITY} maintain high HV values of $0.955$ and $0.952$, respectively, together with complete mission-generation convergence. Their mean mission durations are $230.5$ and $232.7$ d, respectively, indicating that topology-based selection continues to provide structured candidate subsets even when strong environmental correlations are absent. \\

\bottomrule
\end{tabularx}
\end{table}

\subsection{Granular Astrodynamic Case Studies of Representative Pareto Solutions}

While aggregate performance metrics reveal the macroscopic behavior of the proposed GBS-PyGMO framework, they do not fully explain the physical mechanisms through which different mission architectures emerge. The Pareto-optimal trajectories arise from the interaction between orbital topology, target-selection strategy, secular dynamics, conjunction urgency, and spacecraft resource constraints. To expose these mechanisms, five representative solutions are extracted from the final Pareto frontiers and analyzed at the leg level. Collectively, these case studies reveal distinct routing regimes that emerge naturally from the optimization process despite not being explicitly encoded within the framework.\\

A key observation is that the optimizer does not converge toward a universal servicing strategy. Instead, qualitatively different trajectory structures appear depending on the topology of the selected subset and the operational characteristics of the surrounding environment. The observed transitions are primarily associated with three factors: (i) the degree of orbital clustering within the selected subset, (ii) the availability of passive secular alignment opportunities, and (iii) the presence of time-critical conjunction constraints.

\subsubsection{Case Study I: Deep-Drift Dominance in a Dispersed Environment}

The most pronounced fuel-conserving trajectory emerged within the Random environment at the largest combinatorial scale ($N=25$, $k=5$). The resulting mission achieved a total transfer cost of only $0.629$ km/s despite servicing a dispersed set of five candidate objects.

\begin{table}[H]
\centering
\caption{Astrodynamic Characteristics: Case Study I (Random, $N=25$, $k=5$). Selected using \texttt{CENTRALITY}. Total $\Delta v = 0.629$ km/s, total duration = $538.4$ days, and total propellant consumption = $1{,}771.6$ kg.}
\label{tab:case1_random}
\renewcommand{\arraystretch}{1.2}
\resizebox{\textwidth}{!}{
\begin{tabular}{cllcccccc}
\toprule
\textbf{Leg} & \textbf{Departure Target} & \textbf{Arrival Target} & \textbf{$\Delta v$ (km/s)} & \textbf{Wait (days)} & \textbf{TOF (days)} & \textbf{$\Delta i$ ($^\circ$)} & \textbf{$\Delta \text{Alt}$ (km)} & \textbf{$\Delta \Omega$ ($^\circ$)} \\
\midrule
1 & COSMOS 2601        & CZ-6A DEB          & 0.323 & 84.4 & 0.03 & 1.95 & 212.3 & 20.9 \\
2 & CZ-6A DEB          & FENGYUN 1C DEB     & 0.101 & 84.5 & 0.03 & 0.26 & 171.8 & 298.6 \\
3 & FENGYUN 1C DEB     & THORAD AGENA D DEB & 0.181 & 84.9 & 0.04 & 1.19 & 105.5 & 43.7 \\
4 & THORAD AGENA D DEB & THORAD AGENA D DEB & 0.024 & 84.5 & 0.04 & 0.03 & 47.5  & 198.1 \\
\bottomrule
\end{tabular}}
\end{table}

\paragraph{Astrodynamic Interpretation}
The trajectory is characterized by repeated $\sim84$ day waiting intervals and relatively small impulsive corrections. Rather than correcting large RAAN offsets through direct maneuvering, the optimizer exploits differential $J_2$-induced nodal precession associated with differences in orbital altitude. The most prominent example occurs during Leg~2, where a RAAN separation approaching $300^\circ$ is resolved through passive drift while expending only $0.101$ km/s. The resulting trajectory therefore represents a \emph{deep-drift regime}, in which increased mission duration is exchanged for reduced propellant expenditure. This case illustrates how long-term secular dynamics can provide an alternative mechanism for establishing effective connectivity between orbitally separated targets.

\subsubsection{Case Study II: Risk-Driven Interception under Conjunction Constraints}

A fundamentally different behavior emerges within the SOCRATES Minefield environment, where target urgency is influenced by conjunction risk. The representative Pareto solution identified for the $N=15$, $k=5$ configuration exhibits a substantially different control strategy.

\begin{table}[H]
\centering
\caption{Astrodynamic Characteristics: Case Study II (SOCRATES Minefield, $N=15$, $k=5$). Selected using \texttt{CENTRALITY}. Total $\Delta v = 3.250$ km/s, total duration = $200.1$ days, and total propellant consumption = $6{,}152.4$ kg.}
\label{tab:case2_socrates}
\renewcommand{\arraystretch}{1.2}
\resizebox{\textwidth}{!}{
\begin{tabular}{cllcccccc}
\toprule
\textbf{Leg} & \textbf{Departure Target} & \textbf{Arrival Target} & \textbf{$\Delta v$ (km/s)} & \textbf{Wait (days)} & \textbf{TOF (days)} & \textbf{$\Delta i$ ($^\circ$)} & \textbf{$\Delta \text{Alt}$ (km)} & \textbf{$\Delta \Omega$ ($^\circ$)} \\
\midrule
1 & STARLINK-4056  & KUIPER-00205   & 0.449 & 0.0 & 0.03 & 1.35 & 65.4  & 36.2 \\
2 & KUIPER-00205   & STARLINK-31027 & 0.213 & 0.0 & 0.03 & 1.28 & 108.5 & 24.3 \\
3 & STARLINK-31027 & STARLINK-1017  & 1.619 & 0.0 & 0.03 & 0.09 & 108.9 & 75.3 \\
4 & STARLINK-1017  & STARLINK-1281  & 0.968 & 0.0 & 0.03 & 0.11 & 10.0  & 5.6  \\
\bottomrule
\end{tabular}}
\end{table}

\paragraph{Astrodynamic Interpretation}
Unlike the previous case, all waiting times collapse to zero. The optimizer therefore favors immediate transfers between consecutive targets rather than passive phasing. The large $1.619$ km/s maneuver required in Leg~3 would be energetically unfavorable under a purely propellant-minimizing objective, but can become part of a Pareto-efficient solution when temporal and conjunction-related objectives are considered. This trajectory therefore represents a \emph{risk-driven interception regime}, in which temporal urgency can dominate the trade-off between transfer efficiency and mission duration. The case illustrates that the framework does not inherently favor minimum-$\Delta v$ solutions, but instead generates different trade-offs according to the relative importance of the modeled mission objectives and constraints.

\subsubsection{Case Study III: Kinematic Degeneracy within Fragmentation Structures}

The lowest-energy trajectory observed throughout the experimental campaign occurred within the Collision Cloud environment.

\begin{table}[H]
\centering
\caption{Astrodynamic Characteristics: Case Study III (Collision Cloud, $N=15$, $k=3$). Selected using \texttt{SIZE}. Total $\Delta v = 0.008$ km/s, total duration = $193.1$ days, and total propellant consumption = $24.3$ kg.}
\label{tab:case3_collision}
\renewcommand{\arraystretch}{1.2}
\resizebox{\textwidth}{!}{
\begin{tabular}{cllcccccc}
\toprule
\textbf{Leg} & \textbf{Departure Target} & \textbf{Arrival Target} & \textbf{$\Delta v$ (km/s)} & \textbf{Wait (days)} & \textbf{TOF (days)} & \textbf{$\Delta i$ ($^\circ$)} & \textbf{$\Delta \text{Alt}$ (km)} & \textbf{$\Delta \Omega$ ($^\circ$)} \\
\midrule
1 & STARLINK-34149 & STARLINK-6156 & 0.000 & 93.0 & 0.03 & 0.00 & 0.0 & 120.5 \\
2 & STARLINK-6156  & STARLINK-6319 & 0.008 & 0.0  & 0.03 & 0.00 & 0.0 & 0.1   \\
\bottomrule
\end{tabular}}
\end{table}

\paragraph{Astrodynamic Interpretation}
The total mission expenditure of only $0.008$ km/s indicates that the selected objects occupy a highly concentrated region of orbital state space. The significance of this case lies not primarily in the routing strategy, but in the subset-selection process that precedes trajectory optimization. The selected debris objects possess highly correlated orbital characteristics, resulting in extremely low-cost transfers between consecutive targets. This observation indicates that mission complexity is strongly influenced by the topology of the selected subset. In highly clustered environments, topology-aware target selection can reduce the difficulty of the downstream trajectory-optimization problem by presenting the optimizer with intrinsically compatible target combinations.

\subsubsection{Case Study IV: Adaptive Hybrid Control}

Between the two extremes of deep-drift exploitation and rapid interception lies a hybrid routing regime. The trajectory identified within the SOCRATES Minefield environment for $N=18$, $k=5$ illustrates this behavior.

\begin{table}[H]
\centering
\caption{Astrodynamic Characteristics: Case Study IV (SOCRATES Minefield, $N=18$, $k=5$). Selected using \texttt{COMMUNITY}. Total $\Delta v = 1.071$ km/s, total duration = $285.2$ days, and total propellant consumption = $2{,}808.0$ kg.}
\label{tab:case4_hybrid}
\renewcommand{\arraystretch}{1.2}
\resizebox{\textwidth}{!}{
\begin{tabular}{cllcccccc}
\toprule
\textbf{Leg} & \textbf{Departure Target} & \textbf{Arrival Target} & \textbf{$\Delta v$ (km/s)} & \textbf{Wait (days)} & \textbf{TOF (days)} & \textbf{$\Delta i$ ($^\circ$)} & \textbf{$\Delta \text{Alt}$ (km)} & \textbf{$\Delta \Omega$ ($^\circ$)} \\
\midrule
1 & STARLINK-1281 & STARLINK-4056  & 0.043 & 85.1 & 0.03 & 0.05 & 74.9 & 20.5 \\
2 & STARLINK-4056 & KUIPER-00275   & 0.324 & 0.0  & 0.03 & 1.36 & 75.7 & 41.2 \\
3 & KUIPER-00275 & KUIPER-00205   & 0.017 & 0.0  & 0.03 & 0.01 & 10.3 & 5.0  \\
4 & KUIPER-00205 & STARLINK-31027 & 0.688 & 0.0  & 0.03 & 1.28 & 108.5 & 24.3 \\
\bottomrule
\end{tabular}}
\end{table}

\paragraph{Astrodynamic Interpretation}
The trajectory begins with an extended passive phasing interval that resolves a moderate RAAN separation at low propellant cost. Once the servicing vehicle enters a densely connected target region, however, waiting times disappear and the trajectory transitions to immediate impulsive transfers. The resulting mission therefore exhibits two distinct phases: an initial energy-conservation stage followed by a rapid-intervention stage. Importantly, this transition is not explicitly programmed into the optimization framework. Instead, it emerges from the interaction between orbital accessibility, temporal conditions, and the competing mission objectives. The case illustrates how the optimization process can combine passive and active transfer mechanisms within a single servicing sequence.

\subsubsection{Case Study V: Secular Rephasing in Dispersed Fragmentation Families}

The final case study highlights a regime that can arise within dispersed fragmentation-derived populations. Although objects may share a common origin, long-term orbital evolution can progressively separate them within orbital-element space.

\begin{table}[h]
\centering
\caption{Astrodynamic Characteristics: Case Study V (Collision Cloud, $N=22$, $k=5$). Selected using \texttt{SIZE}. Total $\Delta v = 0.400$ km/s, total duration = $452.6$ days, and total propellant consumption = $1{,}170.2$ kg.}
\label{tab:case5_raan}
\renewcommand{\arraystretch}{1.2}
\resizebox{\textwidth}{!}{
\begin{tabular}{cllcccccc}
\toprule
\textbf{Leg} & \textbf{Departure Target} & \textbf{Arrival Target} & \textbf{$\Delta v$ (km/s)} & \textbf{Wait (days)} & \textbf{TOF (days)} & \textbf{$\Delta i$ ($^\circ$)} & \textbf{$\Delta \text{Alt}$ (km)} & \textbf{$\Delta \Omega$ ($^\circ$)} \\
\midrule
1 & STARLINK-1420  & STARLINK-3758  & 0.100 & 84.0 & 0.03 & 0.16 & 171.2 & 315.2 \\
2 & STARLINK-3758  & STARLINK-3771  & 0.000 & 0.0  & 0.03 & 0.00 & 0.0   & 0.0   \\
3 & STARLINK-3771  & STARLINK-32823 & 0.200 & 84.1 & 0.03 & 0.06 & 351.1 & 318.3 \\
4 & STARLINK-32823 & STARLINK-11266 & 0.099 & 84.3 & 0.03 & 0.01 & 171.2 & 272.2 \\
\bottomrule
\end{tabular}}
\end{table}

\paragraph{Astrodynamic Interpretation}
Several transfers involve RAAN separations exceeding $300^\circ$, yet the optimizer avoids expensive direct corrections by repeatedly exploiting differential nodal regression. The trajectory alternates between extended phasing intervals and opportunistic near-zero-cost transfers, effectively reconstructing favorable orbital connectivity through time-dependent dynamics rather than instantaneous geometry. This behavior defines a \emph{secular rephasing regime} and illustrates how common-origin debris populations can retain favorable servicing opportunities despite substantial dynamical dispersion.

\subsubsection{Origins of the Emergent Trajectory Regimes}

An important observation is that the trajectory regimes identified above originate from the interaction between the evolutionary optimizer and the topological structure established during the subset-selection stage. The GBS module biases sampling toward subsets that simultaneously exhibit strong transfer accessibility and high heuristic utility. As  a result, the evolutionary layer does not operate on arbitrary target collections, but on topologically filtered subsets possessing distinct structural characteristics.

Dense communities amplified by the Hafnian-dependent sampling distribution can produce kinematic-degeneracy regimes, whereas sparse yet accessible subsets can favor deep-drift trajectories when passive secular alignment provides a favorable trade-off. Similarly, subsets containing conjunction-critical objects can lead to interception-oriented solutions. The observed routing patterns therefore reflect the combined influence of quantum-guided subset discovery and astrodynamic trajectory optimization. This coupling is a central feature of the proposed architecture.

\subsubsection{Synthesis of Emergent Operational Regimes}

The five case studies reveal that the dominant mission strategy depends on the relationship between orbital topology and operational constraints. Table~\ref{tab:regime_synthesis} summarizes the principal routing regimes identified throughout the experimental campaign.

\begin{table}[h]
\centering
\caption{Synthesis of routing regimes identified through detailed astrodynamic trajectory analysis of PyGMO-evolved mission sequences across the benchmark debris environments.}
\label{tab:regime_synthesis}
\renewcommand{\arraystretch}{1.4}
\resizebox{\textwidth}{!}{
\begin{tabular}{>{\bfseries}c l l p{5.5cm} p{5.5cm}}
\toprule
Case & Emergent Routing Regime & Environment & Key Astrodynamic Strategy & Core Algorithmic Implication \\
\midrule
I & Deep-Drift Dominance & Random & 
Extended passive waiting to exploit differential nodal precession for RAAN realignment. & 
Temporal flexibility allows the optimizer to trade mission duration for reduced propellant expenditure in weakly structured environments. \\

II & Risk-Driven Interception & SOCRATES Minefield & 
Zero-wait, sequential impulsive transfers used to satisfy time-sensitive servicing requirements. & 
Temporal constraints can shift the Pareto-efficient solution toward more propulsive routing strategies despite increased $\Delta v$. \\

III & Kinematic Degeneracy & Collision Cloud & 
Near-zero-energy transfers between objects occupying highly correlated orbital states. & 
Topology-aware subset selection can substantially reduce downstream routing difficulty before evolutionary trajectory optimization. \\

IV & Adaptive Hybrid Control & SOCRATES Minefield & 
An initial passive phasing interval followed by rapid, immediate transfers within a localized target cluster. & 
The optimizer can combine passive and active mechanisms when fuel and temporal objectives compete within a single mission. \\

V & Secular Rephasing & Collision Cloud & 
Repeated passive phasing intervals interspersed with opportunistic low-cost transfers across a dispersed target family. & 
Long-term secular dynamics can provide favorable transfer opportunities even when instantaneous orbital separation is substantial. \\
\bottomrule
\end{tabular}}
\end{table}

The five case studies demonstrate that the Pareto solutions do not follow a single transfer philosophy. Instead, the optimization process balances passive secular dynamics, impulsive maneuvering, temporal constraints, target urgency, and orbital accessibility according to the structure of the selected subset. Deep-drift exploitation emerges when temporal flexibility permits prolonged phasing, risk-driven interception becomes prominent when conjunction urgency constrains the available timeline, and hybrid strategies arise when both propellant and temporal objectives must be satisfied simultaneously. These observations reinforce the interpretation of multi-target ADR as a coupled graph-theoretic and astrodynamic optimization problem in which both subset topology and orbital dynamics influence the resulting mission architectures.

\section{Discussion}

A central contribution of this study is the systematic evaluation of the proposed ADR framework across three fundamentally distinct orbital environments. Most ADR optimization studies assess performance on a limited debris population or under a single environmental configuration, making it difficult to determine whether observed behavior originates from the optimization architecture or from structural properties of the underlying catalogue. By constructing conjunction-driven, fragmentation-driven, and unstructured environments with substantially different graph and astrodynamic characteristics, the present study enables topology-dependent behavior to be examined within a common computational framework. The SOCRATES Minefield environment represents a conjunction-driven setting in which collision relevance is prominent while the transfer network retains substantial structure through concentrated accessibility and community organization. The Collision Cloud environment captures highly correlated fragmentation dynamics, in which objects originating from common breakup events occupy comparatively clustered orbital regions. The Random environment removes these explicit sources of structure and produces greater orbital dispersion and weaker large-scale connectivity. Together, these environments provide complementary tests of whether topology-aware selection remains effective as the structure available to the optimization process changes. The comparative results indicate that environmental topology is an important determinant of ADR mission complexity, while the benefits associated with incorporating network-level information persist across substantially different debris configurations. These observations support the use of topology-aware planning as a means of adapting target selection and routing to the structural characteristics of the debris environment.\\

The advantage of graph-theoretic selection arises from a fundamental distinction between object-centric and relational representations of the debris population. Classical heuristics such as \texttt{SIZE}, \texttt{BSTAR}, and conjunction-risk ranking assign utility primarily at the individual-object level. Such criteria can identify physically important targets, but they do not explicitly account for whether several selected targets can be serviced efficiently as a collective mission. A set of individually attractive objects may therefore contain large transfer costs between successive targets. In contrast, the graph-theoretic heuristics operate directly on the transfer network, incorporating $\Delta v$ accessibility, orbital similarity, phasing opportunities, and connectivity into the selection process. Measures such as centrality and community structure consequently evaluate each object in relation to its surrounding orbital neighborhood rather than in isolation. The resulting selection process favors targets that provide intrinsic value and also useful connectivity to other candidate objects. Across the evaluated configurations, the strong Hypervolume and Pareto-front performance associated with \texttt{CENTRALITY} and \texttt{COMMUNITY} is consistent with this interpretation. These findings indicate that network-level structural information can provide a complementary basis for target selection beyond isolated object attributes, particularly when the mission objective involves servicing multiple targets within a constrained transfer budget. The results therefore support a network-optimization interpretation of multi-target ADR in which the mission-level value of a target depends partly on its position within the transfer topology.\\

Within this architecture, GBS functions primarily as a probabilistic combinatorial reduction mechanism rather than as a standalone mission optimizer. The sampler operates on an astrodynamically informed transfer network whose edge weights encode transfer accessibility and other mission-relevant structural information. The sampled subsets are not arbitrary graph configurations but candidate target combinations shaped by the topology embedded in the GBS input matrix. The role of GBS is therefore intermediate: it generates a distribution over candidate target subsets, while the subsequent multi-objective evolutionary stage determines whether those subsets can be transformed into feasible and competitive mission architectures. This hierarchical decomposition separates combinatorial candidate generation from detailed trajectory optimization and  limits the number of subsets subjected to computationally expensive mission evaluation. The sampling statistics further show that increasing problem scale is accompanied by decreasing HCR and increasing ESS, Div, Cov, and $H_v$. These changes indicate a redistribution of sampling probability across a broader collection of structurally distinct candidate subsets rather than concentration on a small set of dominant configurations. Importantly, deterioration in graph-connectivity measures such as AlgC and SpGap does not produce a proportional deterioration in mission-level solution quality. This behavior suggests that the utility of GBS within the proposed framework is not restricted to identifying a single dominant transfer corridor. Instead, it can provide diverse candidate subsets from multiple regions of the search landscape. The resulting role is therefore best understood as topology-aware candidate generation that preserves exploration capacity while reducing the number of combinatorial alternatives passed to detailed optimization.\\

The effects of candidate-pool size $N$ and subset cardinality $k$ are also qualitatively distinct. Increasing $N$ primarily modifies the topology of the available transfer network by introducing additional candidate objects, transfer pathways, and alternative routing opportunities. The associated reductions in inter-heuristic Jaccard similarity and increases in ESS and sampling entropy indicate a broader solution landscape as the candidate pool expands. In contrast, increasing $k$ directly increases the depth of the servicing sequence and therefore the physical burden of executing the resulting mission. As additional targets are incorporated, locally favorable clusters become depleted and the optimizer may need to establish connections between less compatible orbital regions, producing greater $\Delta v$ requirements, longer mission durations, and reduced route reliability. The two parameters therefore influence different layers of the optimization process: $N$ primarily controls topological flexibility and the breadth of the combinatorial search space, whereas $k$ primarily controls the physical complexity of the resulting servicing campaign. An important interaction is that increasing $N$ can partially compensate for the additional burden introduced by larger $k$ by providing intermediate stepping-stone objects and alternative transfer pathways. The resulting behavior indicates that catalogue expansion should not be interpreted solely as an increase in computational complexity.  Under the evaluated conditions, it can also create additional opportunities for low-cost routing.\\

Relative to conventional ADR optimization formulations, the proposed framework combines several components that are commonly treated separately: a topology-aware representation of the debris environment, probabilistic candidate-subset generation, and detailed multi-objective mission optimization. The weighted orbital-transfer graph incorporates relational accessibility into target selection, while GBS provides a mechanism for sampling candidate subsets without explicitly enumerating the full $\binom{N}{k}$ target space. The subsequent evolutionary stage then operates on this reduced collection of candidates rather than directly on the complete combinatorial space. The use of multiple environmental suites further allows the framework to be evaluated under different structural conditions rather than relying on a single debris distribution. This architecture should therefore be interpreted primarily as a decomposition strategy for controlling the candidate-evaluation burden: GBS performs topology-aware candidate generation, while evolutionary optimization resolves the detailed mission-design problem. The experiments support the feasibility of this hierarchical formulation, but do not by themselves establish a general computational speedup or end-to-end scalability for operationally sized catalogues.\\

Several non-trivial behaviors also emerged from the experimental campaign. The Accessibility Index (AI) remained relatively stable across the structured SOCRATES Minefield and Collision Cloud environments despite increases in candidate-pool size and subset cardinality, whereas the Random environment exhibited a clearer decline as orbital dispersion increased. This suggests that the consequences of increasing mission scale depend strongly on the underlying transfer topology. Mission duration also did not translate monotonically into greater drift utilization. In several conjunction-driven cases, extended mission durations coincided with reduced reliance on passive $J_2$ exploitation, indicating a shift toward more impulsive transfers when temporal constraints become dominant. A further observation is that degradation in graph-connectivity metrics did not produce a proportional degradation in mission quality. Reduced AlgC, SpGap, and inter-heuristic Jaccard similarity were accompanied by continued discovery of competitive servicing sequences, indicating that the optimization process remained capable of exploiting multiple structurally distinct regions of the candidate space. These observations reinforce the interpretation that increasing combinatorial and topological complexity does not necessarily translate directly into equivalent degradation in mission quality, because additional candidate structures can provide alternative routes through the feasible design space.\\

Taken together, the findings support a topology-aware formulation of multi-target ADR in which the debris environment is treated as a collection of individual targets and also as a structured orbital-transfer network. Within this formulation, GBS provides probabilistic candidate generation, graph structure provides relational information for target selection, and multi-objective evolutionary optimization resolves the detailed trajectory and scheduling problem. The resulting decomposition provides a mechanism for exploring large combinatorial candidate spaces while retaining physically meaningful mission-level evaluation. More broadly, the study indicates that the interaction between debris-network structure and mission design variables can be exploited as an explicit component of ADR optimization rather than treated solely as a consequence of downstream trajectory planning.

\section{Limitations and Future Research Directions}

While the proposed framework demonstrates the viability of topology-driven ADR planning through the integration of graph theory, GBS, and multi-objective astrodynamic optimization, several important research directions remain open.\\

A primary limitation of the current framework is the use of static, surrogate-based transfer costs during graph construction. Although physically motivated, these approximations cannot fully capture the temporal evolution of orbital geometry or higher-order perturbative effects. Future work should therefore transition toward dynamic orbital transfer networks,
\[
G(t)=(V,E(t),W(t)),
\]
where edge weights evolve through high-fidelity propagation models incorporating $J_2$, atmospheric drag, solar radiation pressure, luni-solar perturbations, and long-term covariance growth. Such formulations would enable GBS to identify both spatially coherent target subsets and temporally optimal servicing opportunities.\\

Similarly, the current adjacency matrix assumes deterministic orbital-state knowledge and neglects tracking uncertainty, conjunction-prediction errors, maneuver-execution uncertainty, and navigation covariance. Future uncertainty-aware formulations could employ stochastic adjacency matrices,
\[
A_{ij}=\mathbb{E}[U_{ij}]-\lambda\,\mathrm{Var}(U_{ij}),
\]
allowing the quantum sampler to prioritize robust servicing opportunities rather than merely nominally optimal transfers.Also, the present framework utilizes a single weighted transfer graph. Future extensions could incorporate multiplex graph representations integrating transfer accessibility, collision risk, fragmentation ancestry, mission priority, and economic value simultaneously. Such multi-layer formulations may better capture the inherently multi-dimensional nature of operational ADR planning.\\

Furthermore, the empirical evaluations in this work rely on TLE catalogue snapshots. TLEs represent mean orbital elements fitted using the SGP4/SDP4 propagation model, and their predictive accuracy depends on factors including the quality and quantity of observations, orbital regime, and environmental conditions. Prediction uncertainty can therefore increase as the propagation interval grows, particularly when a single element set is extrapolated over an extended planning horizon. Accordingly, the trajectory solutions, candidate sequences, and schedule timelines presented here should be interpreted as computational mission-design results rather than operational flight plans. Transitioning the framework toward operational mission planning will require higher-fidelity orbit determination and ephemeris products, together with associated state-estimation and covariance information.\\

Another promising direction is the development of closed-loop quantum-evolutionary optimization. Rather than treating subset selection and trajectory optimization as sequential stages, feedback from the PyGMO optimizer could be used to adaptively reweight the sampling graph, progressively steering the GBS distribution toward high-performing regions of the solution space. From a theoretical perspective, the scalability properties of GBS also remain largely unexplored. Establishing formal relationships between graph topology, Hafnian statistics, entropy, and subset quality may provide approximation guarantees and theoretical performance bounds for quantum-guided ADR planning.\\

The current framework also assumes complete prior knowledge of the debris catalog. Future systems should support online and receding-horizon ADR planning in which conjunction alerts, catalog updates, and newly generated fragments dynamically modify the transfer graph throughout mission execution. The astrodynamic layer can likewise be expanded beyond impulsive transfers to include low-thrust propulsion, hybrid maneuver architectures, finite-burn dynamics, and autonomous rendezvous constraints. Future formulations should therefore incorporate operational factors currently omitted from the objective function, including capture complexity, communication constraints, servicing risk, and economic cost.\\

Finally, the largest experiments considered in this work remain modest compared with future remediation campaigns involving hundreds or thousands of candidate objects. Extending the framework to catalog-scale optimization, while simultaneously coupling mission planning with long-term debris evolution models such as NASA LEGEND or ESA MASTER, represents an important step toward operational deployment. Beyond ADR, the proposed graph-based GBS paradigm may also be applicable to satellite servicing, constellation maintenance, orbital logistics, in-orbit assembly, and cislunar mission planning.\\

Collectively, these directions point toward a unified quantum-assisted framework in which graph topology, uncertainty quantification, orbital dynamics, and quantum sampling are treated as tightly coupled components of large-scale space-network optimization.

\section{Conclusion}

The increasing complexity of multi-target Active Debris Removal (ADR) exposes a fundamental limitation in existing mission-planning formulations: target selection is commonly treated either as an object-level prioritization problem or as part of a monolithic combinatorial trajectory-optimization problem. Such formulations do not fully exploit the relational information contained in the orbital-transfer structure of the debris population and can require the simultaneous exploration of target-selection and trajectory-design decisions. This work addressed this gap by investigating whether the topology of the orbital environment could be incorporated directly into target selection and exploited through quantum sampling before computationally intensive trajectory optimization.\\

To this end, we developed a topology-aware quantum-classical framework in which ADR target selection is represented as a weighted orbital-transfer network and Gaussian Boson Sampling (GBS) is used to generate candidate target subsets from this network. The selected subsets are subsequently evaluated through multi-objective evolutionary trajectory optimization, decoupling the combinatorial discovery of mission candidates from the detailed optimization of their physical execution. This decomposition also reduces the effective search space presented to the trajectory optimizer, since the detailed optimization is performed only after a target subset of prescribed cardinality has been identified rather than simultaneously searching over the full debris catalogue. The framework further incorporates multiple complementary target-prioritization strategies, enabling the role of intrinsic object properties and network-level structural information to be examined within a common optimization architecture.\\

The central contribution of this work is therefore twofold. First, it introduces a network-centric perspective for multi-target ADR in which the utility of a debris object is considered in relation to its position within the broader orbital-transfer network and its compatibility with other potential targets. Transfer accessibility, orbital proximity, community structure, and network centrality can consequently contribute to mission-level target selection alongside conventional debris-prioritization criteria. Second, the framework introduces a computational decomposition between target-subset discovery and trajectory optimization. Rather than solving the full combinatorial target-selection and trajectory-design problem as a single monolithic optimization, the proposed architecture uses GBS-based sampling to identify a limited set of candidate subsets and subsequently applies multi-objective evolutionary optimization to the resulting reduced search spaces. This decoupling reduces the effective dimensionality and combinatorial burden of the downstream trajectory-optimization stage while preserving the ability to evaluate detailed physical mission trade-offs for the selected candidates.\\

The broader significance of the framework lies in connecting environmental structure, combinatorial target selection, and physical trajectory optimization within a unified computational workflow. GBS provides the probabilistic sampling mechanism, the orbital-transfer graph provides the structural representation, and evolutionary optimization resolves the detailed mission-design problem. The resulting decomposition provides a computational paradigm in which a large combinatorial candidate space can be filtered using relational structure before expensive trajectory optimization is undertaken. In this sense, the quantum stage and the classical optimization stage address different aspects of the overall problem: the former operates on the structured target-selection space, while the latter concentrates computational effort on the physical execution of candidate missions.\\

More generally, the results demonstrate the potential of combining quantum sampling with domain-specific graph representations for large-scale space-network optimization. The framework is not limited to a particular ADR catalogue or optimization configuration, but provides a basis for investigating how quantum sampling can be coupled with physically meaningful network structure while separating large-scale combinatorial candidate generation from computationally intensive mission design. This perspective may also support future extensions to orbital servicing, constellation maintenance, in-orbit logistics, and other large-scale space-domain optimization problems.

\clearpage
\bibliographystyle{elsarticle-num-names}
\bibliography{references}
\end{document}